\documentclass[twocolumn]{aastex701}
\usepackage{subcaption}
\usepackage{booktabs}

\DeclareRobustCommand{\okina}{%
  \raisebox{\dimexpr\fontcharht\font`A-\height}{%
    \scalebox{0.8}{`}%
  }%
}

\linenumbers

\begin{document}

\defcitealias{tully_galaxy_2015}{T15}

\title{Supernova Cousins: The Intrinsic Dispersion and Environmental Dependence of Type Ia Supernovae in Galaxy Groups with TITAN}

\author[0009-0000-9959-5216]{Allison Blum}
\affiliation{Institute for Astronomy, University of Hawai\okina i, 2680 Woodlawn Drive, Honolulu, HI 96822, USA}
\email[show]{ablum@hawaii.edu}

\author[0000-0001-8596-4746]{Erik R. Peterson}
\affiliation{Department of Physics, University of Michigan, Ann Arbor, MI 48109, USA}
\affiliation{Society of Fellows, University of Michigan, Ann Arbor, MI 48109, USA}
\email{erikpete@umich.edu}

\author[0000-0002-6230-0151]{David~O.~Jones}
\affiliation{Institute for Astronomy, University of Hawai\okina i, 640 N. A'ohoku Pl., Hilo, HI 96720, USA}
\email{dojones@hawaii.edu}

\author[0000-0002-9291-1981]{R.\ Brent Tully}
\affiliation{Institute for Astronomy, University of Hawai\okina i, 2680 Woodlawn Drive, Honolulu, HI 96822, USA}
\email{tully@ifa.hawaii.edu}

\author[0009-0003-4631-3184]{Elijah G. Marlin}
\email{emarlin@bu.edu}
\affiliation{Departments of Astronomy and Physics, Boston University, Boston MA 02215}

\author[0000-0002-8342-3804]{Yukei S. Murakami}
\email{ymuraka2@jhu.edu}
\affiliation{Department of Physics and Astronomy, Johns Hopkins University, Baltimore, MD 21218, USA}

\author[0009-0004-5681-545X]{Jack W. Tweddle}
\email{jack.tweddle@physics.ox.ac.uk}
\affiliation{Astrophysics sub-Department, Department of Physics, University of Oxford, Keble Road, Oxford, OX1 3RH, UK}

\author[0000-0001-5201-8374]{Dillon Brout}
\email{dbrout@bu.edu}
\affiliation{Departments of Astronomy and Physics, Boston University, Boston MA 02215}

\author[0000-0003-0928-0494]{Mitchell Dixon}
\affiliation{Institute for Astronomy, University of Hawai\okina i, 640 N. A'ohoku Pl., Hilo, HI 96720, USA}
\email{mtdixon@hawaii.edu}

\author[0000-0002-8229-1731]{Stephen J. Smartt}
\email{stephen.smartt@physics.ox.ac.uk}
\affiliation{Astrophysics sub-Department, Department of Physics, University of Oxford, Keble Road, Oxford, OX1 3RH, UK}
\affiliation{Astrophysics Research Centre, School of Mathematics and Physics, Queen’s University Belfast, BT7 1NN, UK}




\begin{abstract}

At low redshift, scatter in the Type Ia supernova (SN~Ia) Hubble diagram is increased by peculiar velocities (PVs) and correlations between SN distances and their host-galaxy environments, limiting the precision of cosmological distance measurements. We investigate these effects using 1086 spectroscopically-confirmed SNe~Ia from the Type Ia Supernova Trove from ATLAS in the Nearby Universe (TITAN) sample in the redshift range $0.01<z<0.05$. We identify 314 group-associated SNe, including 80 SN ``cousins" occurring in different galaxies within 32 common groups. We first examine whether using the group-averaged redshifts of SNe in galaxy groups can mitigate PV-induced scatter; however, we find that the Hubble residual scatter increases slightly from $0.184^{+0.017}_{-0.009}$~mag using individual host-galaxy redshifts to $0.200^{+0.016}_{-0.015}$~mag using group-averaged redshifts. For the sample of SN cousins, we next analyze pairwise differences in their distances and compare their scatter to the Hubble residual scatter of the full SN sample.  For analysis variants that limit line-of-sight differences in distance modulus between cousins, we find $\sim$30--40\% lower scatter from SN cousins at significances of 2.3--2.8$\sigma$.
We also find tentative evidence that these differences in scatter may be driven by the youngest groups, with a 2$\sigma$ reduction in scatter for groups with high fractions of spiral galaxies.  
These results suggest that SN cousins may provide a novel route for increasing SN~Ia distance precision and understanding the role of host-galaxy and group properties on SN~Ia distance measurements.

\end{abstract}



\section{Introduction} 
Type Ia supernovae (SNe Ia) are among the most precise extragalactic distance indicators. They play a central role in constraining key cosmological parameters, including the Hubble constant, H$_0$ \citep{planck_collaboration_planck_2020, riess_comprehensive_2021}, the dark energy equation-of-state parameter, $w$ \citep[e.g.,][]{brout_pantheon_2022,abbott_dark_2024}, and the growth-of-structure parameter, $f\sigma_8$ \citep[e.g.,][]{boruah_cosmic_2020,stahl_peculiar-velocity_2021, carreres_growth-rate_2023,dixon_peculiar_2026}. Their utility as standardizable candles arises from empirical correlations between their light-curve shape, peak luminosity, and color, which reduce intrinsic scatter in their distance estimates to $\sim$5--7\% \citep[e.g.,][]{riess_precise_1996, guy_supernova_2010, burns_carnegie_2011, kenworthy_salt3_2021, mandel_hierarchical_2022}. 

A major source of uncertainty in local SN Ia distances within $z<0.05$ comes from peculiar velocities (PVs), which are caused by deviations of galaxy motions from the smooth Hubble flow due to gravitational interactions with nearby structures \citep[e.g.,][]{peterson_pantheon_2022, carreres_ztf_2025}. PVs can contribute $\sim$250--300 km s$^{-1}$ of uncertainty to the measured redshift, adding scatter to the Hubble diagram and reducing the precision of H$_0$ and $w$ measurements. Although higher-redshift SNe Ia distances are largely unaffected by PVs because the same velocity perturbation represents a smaller fraction of their recession velocity, local SN Ia samples are critical for most cosmological parameter measurements \citep[e.g.,][]{krisciunas_carnegie_2017, foley_foundation_2018}.

Another contributor to scatter and systematic biases biases in SN Ia distances is the so-called host-galaxy mass step: an empirical effect in which SNe Ia in high-stellar-mass galaxies (typically $\log (\frac{M_\star}{M_\odot}) \gtrsim 10$) appear $\sim$0.05--0.1~mag brighter after standardization than those in lower-mass hosts \citep{kelly_hubble_2010, lampeitl_effect_2010,sullivan_dependence_2010}. This effect likely reflects underlying differences in progenitor age, metallicity, or dust properties between SNe Ia in different host galaxies \citep{childress_host_2013, hayden_fundamental_2013,jones_should_2018,rigault_strong_2020,uddin_carnegie_2020,kelsey_effect_2021,brout_its_2021,wiseman_galaxy-driven_2022,meldorf_dark_2023,popovic_pantheon_2023,peterson_dehvils_2024,toy_reduction_2025,murakami_old_2026}.

Previous analyses have compared SN ``siblings," SNe occurring in the same host galaxy, which effectively remove the effects of small-scale peculiar velocities as well as correlations between distance measurements and global host-galaxy environment \citep[e.g.,][]{Burns_2020, wiseman_rates_2021, Hoogendam_2022, kelsey_archival_2024}. Early sibling studies were limited by very small samples, with \citet{Burns_2020} finding 12 sibling SN~Ia pairs in the literature and \citet{Scolnic_2020} identifying only eight pairs in the Dark Energy Survey. More recently, \citet{kelsey_archival_2024} compiled a substantially larger archival sample of 327 SNe~Ia across 158 host galaxies, although only a much smaller cosmology-quality subsample of 44 objects currently has the homogeneous light-curve quality and distance measurements needed for precision cosmological analyses. As a result, the statistical leverage currently available from sibling systems remains limited.

\begin{figure}
\centering
\includegraphics[width=\linewidth]{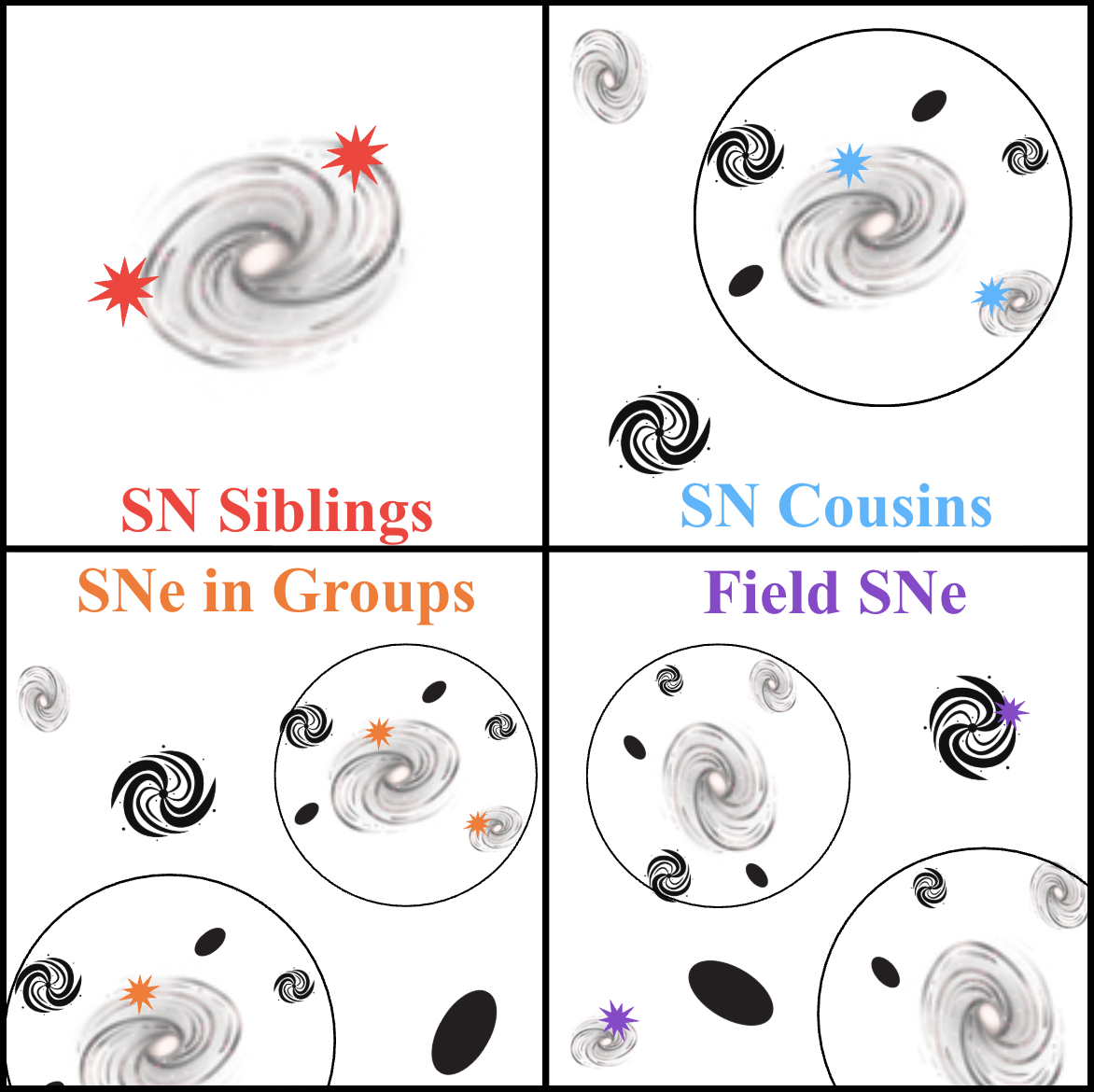}
\caption{Graphic depicting (upper left) SN siblings which are SNe with the same host galaxy, (upper right) SN cousins which are SNe in the same galaxy group, (lower left) SNe in galaxy groups, and (lower right) field SNe which are SNe not associated with a galaxy group.}
\label{fig:graphic}
\end{figure}

Although SN siblings are intrinsically rare, a similar analysis can be extended to larger-scale environments through SN ``cousins" --- SNe~Ia occurring in different galaxies within the same galaxy group or cluster. Unlike siblings, cousins do not share the same host galaxy, but they reside within a common group-scale environment where both galaxy properties and peculiar velocities remain correlated (see Figure~\ref{fig:graphic}). Consequently, cousins retain many of the advantages of sibling analyses while substantially increasing the available sample size.

Galaxy groups and clusters provide a natural framework for reducing the impact of small-scale peculiar velocities arising from the internal motions of galaxies within gravitationally bound systems. Galaxies in the same group share a common bulk motion, so comparing SNe within these environments cancels a substantial fraction of the velocity-induced scatter associated with internal group dynamics \citep{peterson_pantheon_2022}. This is distinct from intermediate- and large-scale bulk flows, which arise from coherent motions of entire galaxy groups over much larger scales and require separate corrections \citep[e.g.,][]{boruah_cosmic_2020,stahl_peculiar-velocity_2021,carreres_growth-rate_2023}. Early work using galaxy-group catalogs demonstrated that assigning SNe to groups improves Hubble residuals and reduces their $\chi^2$ values by up to 37\% for nearby samples \citep{peterson_pantheon_2022}, while updated analyses have shown that group redshifts can reduce the contribution of small-scale peculiar velocities by approximately $135~{\rm km~s^{-1}}$ \citep{peterson_improving_2025}. However, current group catalogs remain incomplete and magnitude limited, so only a fraction of nearby SNe can presently be associated with known groups \citep[e.g., 30\% of the Pantheon$+$ sample compared to a predicted 73\% from N-body simulations;][]{peterson_pantheon_2022,peterson_improving_2025}.

Beyond reducing small-scale peculiar-velocity scatter, cousins also provide a means of partially controlling for environmental effects. SNe Ia in dense environments may themselves be more homogeneous; for example, cluster centers preferentially host low-$x_1$ SNe Ia \citep{larison_environmental_2024}. Galaxies in groups and clusters exhibit correlated properties, including enhanced passive fractions, suppressed star formation, and the well-established morphology-density relation, in which denser environments preferentially host larger fractions of early-type galaxies relative to the field \citep{dressler_galaxy_1980,helsdon_morphology-density_2003,peng_mass_2010}. Comparing standardized distances among cousins therefore enables simultaneous tests of both correlated peculiar velocities and environmental influences on SN~Ia luminosities, extending the concept of sibling analyses to group-scale environments.

The rapid growth of time-domain sky surveys has dramatically expanded the available low-redshift SN Ia sample for such analyses including the All-Sky Automated Survey for Supernovae (ASAS-SN; \citealt{shappee_man_2014}), the Zwicky Transient Facility (ZTF; \citealt{bellm_zwicky_2019}), and the Asteroid Terrestrial-impact Last Alert System (ATLAS; \citealt{tonry_atlas_2018}). In particular, the ATLAS survey via the Type Ia Supernova Trove from ATLAS in the Nearby Universe (TITAN) compilation (\citealp{marlin_titan_2025, murakami_old_2026}; Tweddle et al.\ 2026, \textit{under review}) now provides a homogeneous set of $\sim$3000 spectroscopically-classified SNe, one of the largest low-$z$ samples suitable for cosmological analysis. TITAN presents an opportunity to identify a significantly larger number of SNe residing in galaxy groups and to systematically search for both SN siblings and SN cousins.

In this work, we use the TITAN SN sample along with the \citet[hereafter \citetalias{tully_galaxy_2015}]{tully_galaxy_2015} galaxy-group catalog to (1) assign group memberships to known SNe, (2) identify SNe that share the same groups or clusters, and (3) quantify how group associations reduce peculiar-velocity contributions and reduce distance scatter.  The goal of this work is to identify new pathways for improving the precision of cosmological parameter measurements. In Section \ref{sec:data} we describe the data acquisition and data samples, and in Section \ref{sec:methods} we describe the analysis methods performed on such data. Our data and sample selection are detailed in Section \ref{sec:sample}. Results from the cousins and galaxy-group redshifts are included in Section \ref{sec:results}. Finally, in Sections \ref{sec:disc} and \ref{sec:conclusions} we present our discussions and conclusions. Throughout this work, we assume a flat $\Lambda$CDM cosmology with $\Omega_m = 0.3$ and $H_0 = 70~{\rm km~s^{-1}~Mpc^{-1}}$.

\section{Data} \label{sec:data}

\subsection{The TITAN Supernova Sample} \label{sec:atlas-gold}

We draw our SN sample from the  TITAN compilation of SNe with light curves from the ATLAS survey. ATLAS \citep{tonry_atlas_2018} is an all-sky, high-cadence optical survey consisting of four 0.5-m robotic telescopes distributed between the Northern and Southern Hemispheres. The system images the accessible sky nightly to a typical $5\sigma$ depth of $m_{\rm AB}\sim19.5$\,mag in two broad optical bands (``cyan" and ``orange", roughly equivalent to $g+r$ and $r+i$ bands), enabling efficient discovery and follow-up of nearby SNe across a wide range of host-galaxy environments \citep{smith_20202020PASP..132h5002S}.

Our analysis uses the TITAN ``gold" sample, which consists of spectroscopically-confirmed SNe Ia with spectroscopic host-galaxy redshifts reported to the Transient Name Server (TNS\footnote{\url{https://www.wis-tns.org/}.})
between February 2017 and March 2025. Only normal and 91T-like SNe~Ia are included, while peculiar subtypes such as 91bg-like or 02cx-like events are excluded, to limit the sample to SNe for which reliable distances can be obtained. Spectral classifications from publicly available spectra reported to TNS are obtained via the SNID-SAGE pipeline \citep{blondin_determining_2007,stoppa_snidsage_2026}. SNID-SAGE cross-correlates observed spectra against a library of SN templates for classification\footnote{SNID-SAGE can also estimate redshifts, though all redshifts in this work are separately computed from host-galaxy spectra.} and accounts for host-galaxy spectral features when present. 

The light-curve data used here are from the TITAN Data Release 1 (DR1; Murakami et al. in prep.), which originates from 
 the ATLAS force-photometry server \citep{Shingles21}.\footnote{\url{https://fallingstar-data.com/forcedphot/}.}
Briefly, forced point-spread-function (PSF)-fitting photometry is measured on ATLAS difference images in both survey bands.  The photometry is then calibrated using the Pan-STARRS-based Refcat2 tertiary star catalog \citep{tonry_vizier_2021}, together with the TITAN DR1 recalibration presented by \citet{marlin_titan_2025}. The TITAN recalibration identified and corrected spatially dependent zeropoint offsets across ATLAS detectors, reducing calibration residuals to the $\sim$5--10~millimag level and improved photometric uniformity across the survey. The recalibration additionally accounts for filter transmission-function color terms and validates the final photometry against external standard-star catalogs and SN Ia distance measurements \citep{marlin_titan_2025}.

Each SN light curve is fit with the SALT3 model \citep{kenworthy_salt3_2021}, providing standardized peak magnitudes, stretch, and color parameters used to compute distance moduli. 
We then apply standard light-curve quality cuts similar to those applied in the Pantheon+ and DES analyses \citep[e.g.,][]{brout_pantheon_2022,abbott_dark_2024}. To be included in the TITAN gold sample, only SNe with SALT3 stretch and color parameters within $-3 < x_1 < 3$ and $-0.3 < c < 0.3$ are included, with uncertainties $\sigma_{x_1} < 1.0$, $\sigma_c < 0.1$, and a well-constrained time of maximum light, $\sigma_{t_0} \leq 1.0$ day. Additionally, each light curve must have at least seven observations in the phase range $-10 \leq t-t_0 \leq 50$ days, including at least two detections before and two after maximum light. Both the cyan and orange bands must be represented within this phase range to ensure robust determination of light-curve colors.

We also define a higher-fidelity subset, the ``platinum" sample. This sample further restricts the light-curve parameters to $-2<x_1<2$ and $-0.15<c<0.15$, while tightening the color uncertainty requirement to $\sigma_c<0.05$. It also applies stricter spectroscopic and fit-quality criteria, requiring a minimum SNID-SAGE match statistic (H$\sigma$LAP-CCC $>6$) and a SALT fit probability greater than $0\%$. The H$\sigma$LAP-CCC statistic combines the cross-correlation peak height, fractional wavelength overlap, concordance correlation coefficient, and a penalty for spectral sharpness mismatch to rank the quality of template matches \citep{stoppa_snidsage_2026}. The SALT fit probability quantifies the statistical likelihood that the observed light-curve data are well represented by the fitted SALT model, with low values indicating poor agreement between the model and observations. These cuts were identified by examining Hubble residual outliers across the full sample and were motivated by the observation that fast-declining SNe, events with redder-than-average colors, and SNe with uncertain spectroscopic classifications were disproportionately associated with large Hubble residuals.


Lastly, we restrict our analysis to the redshift range $0.01 < z < 0.05$. The lower limit reduces the impact of large fractional uncertainties from peculiar velocities at very low redshift, while the upper limit ensures reliable group associations, as described below.


\subsection{Host Galaxy Association}

Host galaxies for the TITAN SN sample are identified and characterized in a companion analysis (Tweddle et al.\ 2026, \textit{under review}). Candidate hosts are assigned using a combination of the directional light radius (DLR) formalism \citep{Sullivan2006, Gupta2016}, redshift-matching between the SN and galaxy, and visual inspection, achieving a secure host galaxy association for 7856 SNe (93.8\% of the full TITAN sample). 

Spectroscopic redshifts are compiled from a hierarchical cross-match against multiple public catalogs, yielding a secure host redshift for 5522 SNe (65.9\%). Using the \texttt{HostPhot} package \citep{HostPhot}, uniform host photometry spanning the far-ultraviolet through the mid-infrared is compiled for the associated hosts and used to derive global and local host-galaxy properties --- including stellar mass, star-formation rate, dust attenuation, and mass-weighted stellar age --- via spectral energy distribution (SED) fitting with \texttt{Bagpipes} \citep{Bagpipes}. We draw on this TITAN host-galaxy catalog throughout this work, both for the galaxy properties examined in Sec. \ref{sec:host_props}, and as the input host positions and redshifts used to construct the group associations described in the following section.

\subsection{Galaxy Group Catalog and Association} \label{sec:galaxy-groups}

We associate SN host galaxies with groups and clusters using the nearby group catalog of \citetalias{tully_galaxy_2015}. This catalog is constructed from the 2MASS Redshift Survey \citep[2MRS; ][]{huchra_2mass_2012}, which provides near-infrared-selected positions and spectroscopic redshifts for galaxies with $K_s < 11.75$ mag over nearly the full sky. Selection in the near-infrared reduces sensitivity to dust extinction and to transient brightening from young, star-forming stellar populations that can bias optical luminosities. As a result, the $K_s$ galaxy luminosities more closely trace total stellar mass, yielding a sample that better reflects the underlying mass distribution in the local universe.

Galaxy groups in \citetalias{tully_galaxy_2015} are identified using an iterative association procedure based on the expected scaling relations of dark matter halos. Candidate groups are seeded from the most luminous galaxies in the 2MRS sample, and neighboring galaxies are associated if they lie within the projected second-turnaround radius ($R_{\rm 2t}$, a proxy for the virial radius) and within the expected velocity range of the halo. As additional members are assigned, the total group luminosity, inferred halo mass, and corresponding values of $R_{\rm 2t}$ and velocity dispersion are updated iteratively until no further candidate members satisfy the association criteria. The halo masses are estimated from the total $K_s$-band luminosities using an empirical mass-to-light relation calibrated from nearby groups and clusters.

The resulting catalog contains 24,044 galaxies in the redshift range relevant to this work, of which 13,900 (57.8\%) are assigned to groups containing two or more members. For each group \citetalias{tully_galaxy_2015} provides, among other quantities, a mean redshift, velocity dispersion, the number of identified member galaxies ($N_g$), total $K$-band luminosity, and estimated halo mass, enabling quantitative characterization of the local gravitational environment. The parameter $N_g$ represents the number of galaxies in a group brighter than the $K_s=11.75$ magnitude limit of the 2MRS \citep{huchra_2mass_2012}. Because the survey is flux limited, $N_g$ depends not only on the intrinsic richness of a group but also on its distance, with more distant groups containing fewer detected members. The estimated halo mass accounts for this selection effect and therefore provides a more robust measure of the underlying group properties than $N_g$ alone.


Since the 2MRS catalog is magnitude-limited, \citetalias{tully_galaxy_2015} applies a correction for missing luminosity from galaxies that fall below the survey detection threshold at larger distances. Nevertheless, incompleteness in group assignments remains an important limitation for low-redshift SN samples. For example, only $\sim$30\% of SN host galaxies in the Pantheon+ low-redshift sample are associated with galaxy groups when using existing magnitude-limited catalogs, while the true group membership fraction is likely $\sim$73\%  \citep{peterson_improving_2025}.

For the TITAN SN~Ia sample, we cross-match each SN host galaxy to the \citetalias{tully_galaxy_2015} catalog using both sky position and spectroscopic redshift. A host galaxy is associated with a group when its projected separation from the group center is smaller than $1.5R_{\rm 2t}$ and its redshift lies within 2$\sigma_p$, where $\sigma_p$ is the velocity dispersion of the group. When multiple candidate associations exist, the nearest group is adopted. For hosts assigned to groups, we use the group-averaged redshift and dynamical properties for subsequent analyses.

We restrict our analysis to nearby systems with $z<0.05$. At this redshift, the luminosity correction applied to account for galaxies that fall below the 2MRS flux limit remains below approximately a factor of seven. At larger redshifts, an increasing fraction of faint galaxies are not detected, requiring substantially larger luminosity corrections. These larger corrections make galaxy-group identification and the inferred group properties progressively less reliable.

We further require that the maximum second-turnaround radius ($R_{2t}$), which characterizes the physical extent of each group, satisfies $\frac{R_{2t}}{D_{L}} < 0.03$, where $D_{L}$ is the luminosity distance to the observer. This cut removes systems for which the projected angular extent of the group becomes sufficiently large that the physical separation of cousins from each other is comparable to the distance modulus measurement for these SNe.
We note, however, that the effect of distance scatter even with this cut may still be substantial; for example, SN cousins that are separated from each other by 3\% in distance (a separation of $1 \times R_{2t}$ at the maximum allowed $\frac{R_{2t}}{D_{L}}$ cut) would have an inherent 0.06 mag difference in distance modulus.  The maximum potential $\frac{R_{2t}}{D_{L}}$ effect could be $\sim$twice as large, as two SN cousins at $+1.5R_{2t}$ and $-1.5R_{2t}$ along the line of sight, for a $\frac{R_{2t}}{D_{L}} = 3$\% group, would differ by 0.12~mag.

We can further reduce this effect by tightening the requirement to $R_{2t}/D_{L}<0.02$, but find that this reduces the cousins sample size from 80 to 54 SNe and substantially increases statistical uncertainties. Relaxing the cut to $R_{2t}/D_{\rm L}<0.05$ produced nearly identical scatter measurements to the 3\% case. We adopt the 3\% criterion for our baseline as a compromise between maximizing sample size and reducing intra-group scatter.

\subsection{Supernova Cousins and Siblings} \label{sec:cousins-siblings}

Using the \citetalias{tully_galaxy_2015} group associations, we construct subsamples designed to isolate different contributions to SN~Ia distance scatter. We first identify SN siblings; because siblings share the same global host stellar population, metallicity, and peculiar velocity, they provide a nearly direct probe of intrinsic SN~Ia luminosity scatter by removing the effects of the global host environment and peculiar velocities \citep{wiseman_rates_2021, kelsey_archival_2024, dwomoh_evaluating_2024}. However, siblings can still experience different local environments within their host galaxies, including differences in dust extinction and star-formation conditions. We also identify SN cousins; for each SN host assigned to a group following the methods discussed in Section \ref{sec:galaxy-groups}, all SNe in the same group are considered cousins.

After applying the redshift and group-selection criteria described previously, our final sample contains 314 SN~Ia events associated with galaxy groups and 772 field SNe without group associations, drawn from the full TITAN gold sample within our redshift cuts of 1086 objects. The cousins subsample, consisting of multiple SNe occurring within the same galaxy group, contains 80 SNe distributed across 32 distinct systems. We additionally identify 12 SN siblings occurring in six host galaxies that each hosted two SNe~Ia. A summary of the samples used throughout this work is provided in Table~\ref{tab:stats}.

\begin{table*}
\centering
\caption{Summary of the SN~Ia samples used in this work after applying all selection criteria including the group-extent requirement.}
\label{tab:sample_summary}
\begin{tabular}{lcccc}
\toprule
Sample & Number of SNe & Fraction of Total & Median Redshift & Number of Systems\\
\midrule
\multicolumn{5}{c}{\textit{Gold Sample}}\\ \midrule

TITAN Gold Sample ($0.01 < z < 0.05$) & 1086 & 100\% & 0.0350 & --\\
Group-associated SNe & 314 & 28.9\% & 0.0330 & --\\
Field (non-group) SNe & 772 & 71.1\% & 0.0362 & --\\
Sibling Sample & 12 & 1.1\% & 0.0576 & 6 hosts\\
Cousin Sample & 80 & 7.4\% & 0.0349 & 32 groups\\
\midrule 
\multicolumn{5}{c}{\textit{Platinum Sample}}\\ \midrule
TITAN Platinum Sample ($0.01 < z < 0.05$) & 396 & 100\% & 0.0324 & --\\
Group-associated SNe & 122 & 30.8\% & 0.0285 & --\\
Field (non-group) SNe & 274 & 69.2\% & 0.0345 & -- \\
Sibling Sample & -- & -- & -- & --\\
Cousin Sample & 20 & 5.1\% & 0.0239 & 9 groups\\
\bottomrule
\label{tab:stats}
\end{tabular}
\end{table*}


\section{Analysis}
\label{sec:methods}

\subsection{Measuring Hubble Residual Dispersion}

We measure the dispersion of Hubble residuals using three samples, each of which has different sensitivity to peculiar velocities. Unless otherwise noted, we quantify the dispersion using the robust standard deviation (RelSD), defined as $1.48\times\mathrm{MAD}$, where $\mathrm{MAD}=\mathrm{median}(|x_i-\mathrm{median}(x)|)$ is the median absolute deviation \citep{Hoaglin00}. This estimator is less sensitive to outliers than the standard deviation although it is equivalent to the standard deviation when using data drawn from a normal distribution. Hubble residuals are computed by comparing the observed SN distance modulus to the distance modulus predicted by the fiducial $\Lambda$CDM cosmology at the measured redshift of each SN or host group, depending on the method used. Uncertainties on RelSD are estimated via bootstrap resampling of the data. All redshifts are transformed to the cosmic microwave background (CMB) frame and no external peculiar-velocity corrections are applied. The same analysis procedure is applied to both the TITAN gold and platinum datasets, allowing direct comparison between different data-quality selections throughout the paper. The three samples considered in this work are:

\begin{itemize}

\item {\bf SNe in Groups}. For the sample of SNe associated with \citetalias{tully_galaxy_2015} groups, we compute the RelSD from the Hubble residuals using either the redshift measured from the SN host galaxy or the average redshift of the group from \citetalias{tully_galaxy_2015}.

\item {\bf Field SNe}. For SNe for which a group association has not been identified, dispersion is computed from the Hubble residuals using the individual host-galaxy redshift of each SN.


\item {\bf SN Cousins:} We compute the RelSD separately for each group containing two or more SNe and average over groups. The dispersion for SN cousins is computed on a group-by-group basis. Within each galaxy group containing two or more SNe, the RelSD is computed from the distances in each group individually and then averaged over the full set of groups.  This approach differs slightly from that used for the previous two samples, as groups with more than two SNe are weighted equally to groups with just two SNe; however, in practice only 18.8\% of cousins belong to groups with more than two SNe, and we find that modified weighting schemes produce negligible changes in the results. 

Additionally, because the RelSD measured from small samples is biased low,\footnote{This effect also occurs in standard deviation measurements, and can be fixed by Bessel's correction.}  we correct this bias using a Monte Carlo approach. For each cousins sample, we generate random realizations of that sample drawn from a normal distribution with width equal to the RelSD of the full TITAN gold sample. We then apply the recovered RelSD divided by the input RelSD as a correction factor to each cousins RelSD measurement.


\end{itemize}

The resulting dispersion measurements for the three samples are presented in Section \ref{sec:results} as a function of several physical parameters describing the SN environment.

\subsection{Peculiar Velocity and Intrinsic Scatter Modeling}
\label{sec:pvmodel}

As a separate method to constrain the contribution of peculiar velocities to the observed Hubble residual scatter, we model the total dispersion as arising from three independent components: measurement uncertainties, intrinsic SN scatter, and an additional contribution from peculiar velocities. The intrinsic scatter term represents the irreducible dispersion of standardized SN~Ia luminosities after light-curve corrections and is assumed to be independent of photometric measurement uncertainty and redshift-dependent velocity effects.

We assume that Hubble residuals follow a Gaussian distribution with variance given by the quadrature sum of these contributions. Measurement uncertainties are taken from the reported distance modulus errors prior to including any intrinsic-scatter contribution. The intrinsic scatter is treated as a free redshift-independent parameter, while the peculiar-velocity contribution is parameterized through a characteristic velocity amplitude ($\sigma_v$) propagated into distance-modulus space, following \citet{Davis_2011}:

\begin{equation}
    \sigma_{\mu} = \frac{\sigma_v}{c}\frac{5}{\ln(10)}\frac{1+z}{z(1+z/2)},
\end{equation}

\noindent where $z$ is the CMB-frame redshift and $c$ is the speed of light.

We maximize the resulting Gaussian likelihood and estimate the free parameters using Markov Chain Monte Carlo sampling with the \texttt{emcee} ensemble sampler \citep{foreman-mackey_emcee_2013}. Constraints on all parameters are derived from the posterior distributions.

For the cousins sample, the likelihood framework is slightly modified: for each group containing two or more SNe~Ia, we compute pairwise differences in distance modulus between all SN pairs within the group and define a characteristic intra-group scatter statistic, $\Delta\mu_i$, from the root-mean-square of these differences. The likelihood is then evaluated over groups rather than individual SNe using the representative group redshift and the average distance-modulus uncertainty of the SNe in each group.  We note that because cousins occupy the same galaxy group or cluster environment, we expect that their PV scatter will be consistent with zero.

The same modeling framework is applied to both the gold and platinum samples and to all environmental subsamples considered later in this work, including subdivisions by group mass and host-galaxy properties.

\section{Sample Overview}
\label{sec:sample}

\begin{figure}
\centering
\includegraphics[width=\linewidth]{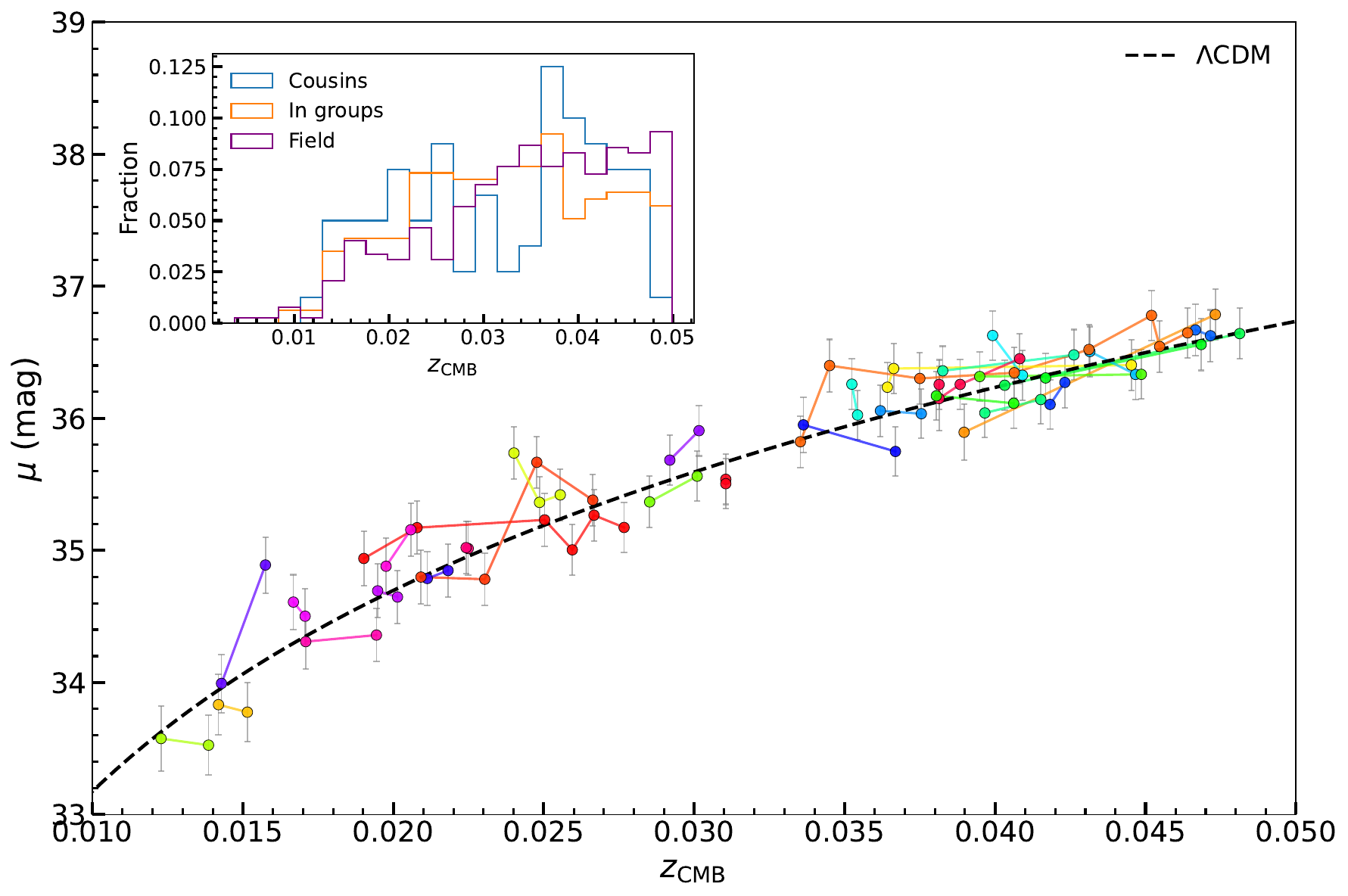}
\caption{Hubble diagram of the cousins sample in the TITAN gold catalog, restricted to $0.01 < z_{\rm CMB} < 0.05$. Cousins sharing a common galaxy group are shown with the same color and connected by solid lines. The x-axis shows the CMB-frame redshift for each SN, and the increased scatter toward low redshift is consistent with the expected contribution from peculiar velocities, while the overall distribution follows the fiducial $\Lambda$CDM model (dashed line). The inset shows the normalized CMB-frame redshift distributions of the cousins sample (blue), all group-associated SNe (orange), and field SNe without group associations (purple), demonstrating that the cousins sample occupies a similar redshift distribution as the full group-associated sample.}
\label{fig:hubblediagram_hist}
\end{figure}

The SN samples used in this analysis are summarized in Table \ref{tab:sample_summary} and the cousins sample is illustrated in Figure~\ref{fig:hubblediagram_hist}.
The overall group-association fraction of our sample is of 28.9\%, consistent with the fraction found by \citet{peterson_pantheon_2022} for SN host galaxies in groups but significantly below the fraction of galaxies found in groups in \citetalias{tully_galaxy_2015} (57.8\%). The fraction of SNe associated with galaxy groups decreases with increasing redshift because the underlying 2MRS group catalog is magnitude limited and becomes progressively incomplete for lower-luminosity galaxies at larger distances. Restricting to the nearby sample ($z<0.03$), we identify 126 group-associated SNe out of 358 total objects, corresponding to a group fraction of 35.2\%. This fraction decreases steadily with redshift, from $\sim36.8\%$ at $0.02<z<0.03$ to $\sim23.1\%$ at $0.04<z<0.05$. This behavior is qualitatively consistent with previous low-redshift studies using similar catalogs (e.g., $\sim$30\% in \citealt{peterson_pantheon_2022}), and reflects the decreasing completeness of magnitude-limited galaxy-group catalogs at larger distances (for example, \citealp{peterson_improving_2025} find that 73\% of SNe\,Ia should be found in groups in an unbiased sample).


The median redshift of the entire $z < 0.05$ TITAN gold sample is $\tilde{z}_{\rm gold}=0.0350$, which includes both group and field SNe. The median redshift of group-associated SNe is $\tilde{z}_{\rm group}=0.0330$, slightly lower than the median value for field SNe, $\tilde{z}_{\rm field}=0.0362$, reflecting the reduced completeness of group identifications at larger distances. The cousins subsample has a median redshift of $\tilde{z}_{\rm cousins}=0.0349$, slightly higher than $\tilde{z}_{\rm group}$ and comparable to the overall gold sample. Given the small offset relative to the group sample, we do not interpret this difference further.  

The 80 cousins SNe are distributed across 32 galaxy groups spanning a range of multiplicities. Most systems contain only two SNe, with 26 groups (81.2\%) containing exactly two events. Two systems contain three SNe, two contain four SNe, one contains six SNe (the Coma cluster), and one contains eight SNe (the Hercules complex). Thus, 18.8\% of cousins groups contain three or more SNe, while 12.5\% contain four or more events. 

\begin{figure}
\centering
\includegraphics[width=\columnwidth]{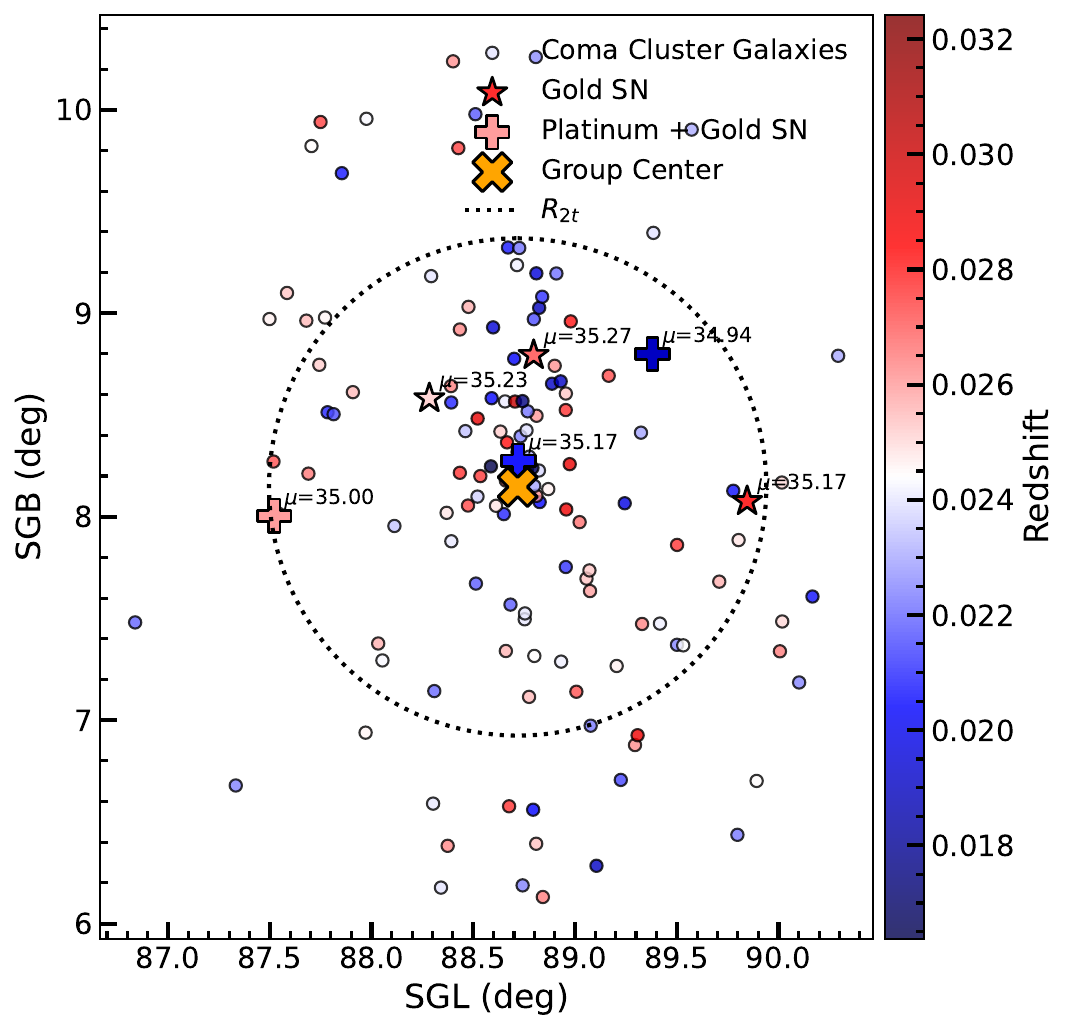}
\caption{
Spatial distribution of SNe in the Coma cluster in supergalactic coordinates. The cross marks the group center from \citetalias{tully_galaxy_2015} and the dashed circle indicates the projected second-turnaround radius $R_{2t}$. SNe are shown as stars and labeled by distance modulus. Host galaxies and SNe are colored according to redshift. Three events additionally satisfy the platinum sample selection criteria and are indicated by colored cross symbols.
}
\label{fig:coma_sgl_sgb}
\end{figure}

The Coma cluster is a useful example of one of the rich, dynamically evolved galaxy groups in our sample, shown in Figure \ref{fig:coma_sgl_sgb}. All six SNe lie within the projected second-turnaround radius $R_{2t}$, with a median projected separation of $0.45R_{2t}$. The measured cousins distance scatter is ${\rm RelSD}=0.356$ mag, substantially larger than the overall cousins sample (in part, this may be because it lies near the threshold of our $R_{2t}/D_L$ cut), and the cousins in the cluster exhibit a large velocity range spanning approximately $2600~{\rm km~s^{-1}}$. Coma is also dominated by early type galaxies, with a group spiral fraction of only 0.14. A more in-depth discussion of SNe in the Coma cluster and Hercules complex is presented in Appendix \ref{appendix:coma_hercules}.

\subsection{Group Properties}

The full sample of 314 group-associated SNe spans a broad range of galaxy environments, with group richness ranging from sparse systems containing only two identified galaxies to rich clusters with as many as 136 members.  The distribution is strongly weighted toward low-richness systems, with $64\%$ of group-associated SNe residing in groups with $N_g\leq6$, while only $13.1\%$ occur in richer systems with $N_g>25$. 
Compared to the sample of 314 group-associated SNe, the cousins subsample is preferentially found in richer and more massive galaxy groups; the median cousins SN resides in a system with richness $N_g=17$, compared with $N_g=5$ for the full group-associated SN sample.
 These differences show that the cousins sample is drawn from systematically richer environments.  We note that because the observed richness depends on both intrinsic group mass and the magnitude-limited nature of the group catalog, we use it primarily as a descriptive quantity rather than a direct physical measure of environment.

Group masses span nearly four orders of magnitude, from $1.3\times10^{12}M_\odot$ to $1.1\times10^{16}M_\odot$, with a median group mass of $1.32\times10^{14}M_\odot$ for the full group-associated SN sample. The median group mass is slightly higher for the cousins sample, at $4.27\times10^{14}M_\odot$. This environmental bias is expected because more massive groups generally contain larger galaxy populations and stellar masses, increasing the probability of hosting multiple observable SN~Ia events.

\begin{figure}
\centering
\includegraphics[width=\linewidth]{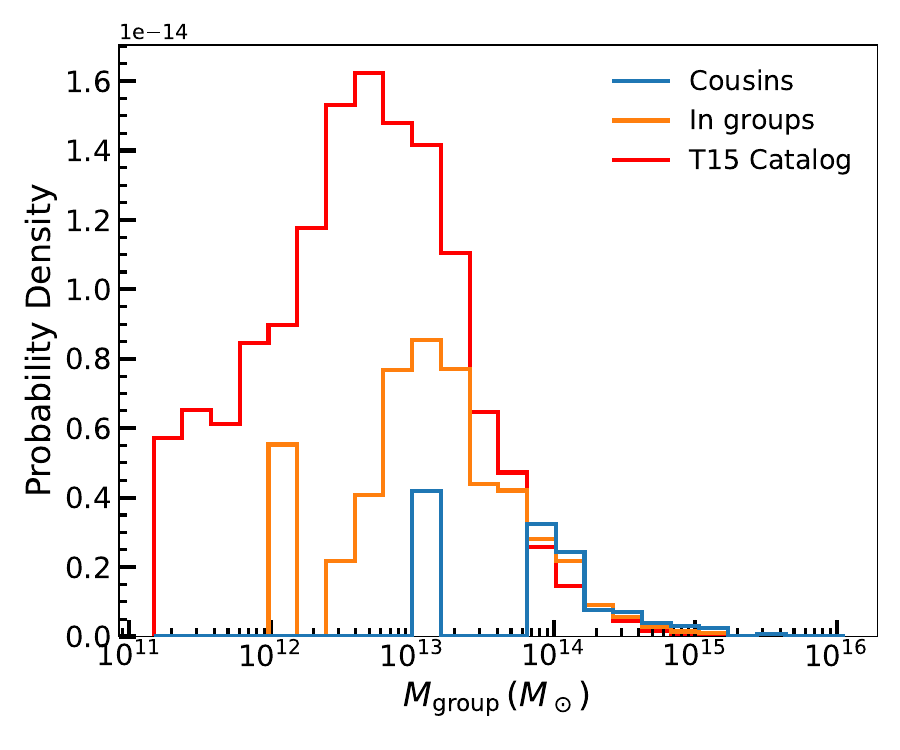}
\caption{Distribution of group masses for the \citetalias{tully_galaxy_2015} catalog at $z<0.05$ (red), all group-associated SNe (orange), and the cousins subsample (blue). The SN-associated group populations are preferentially weighted toward higher-mass systems relative to the underlying group catalog, with the cousins sample showing the strongest preference for massive groups.}
\label{fig:group_mass_hist}
\end{figure}

We compare the distribution of group masses for the \citetalias{tully_galaxy_2015} catalog within $z < 0.05$, the full group-associated SN sample, and the cousins subsample in Figure~\ref{fig:group_mass_hist}. The SN sample is preferentially weighted toward more massive groups relative to the underlying catalog population, while the cousins sample is shifted even further toward high-mass systems (unsurprising given that the SN~Ia rate is generally observed to trace stellar mass; \citealp[e.g.,][]{nugent_characterizing_2026}).

Cousins systems also span a substantial range of group velocity dispersion. The median difference in redshift of cousins within individual systems is $\Delta z=0.00156$, corresponding to a characteristic velocity span of approximately $468~{\rm km~s^{-1}}$, while the mean values are $\Delta z=0.00296$ and $887~{\rm km~s^{-1}}$, respectively. These values reflect that cousins sample extends from relatively compact systems to systems with extended and complex structure.

\begin{figure*}
\centering
\includegraphics[width=\linewidth]{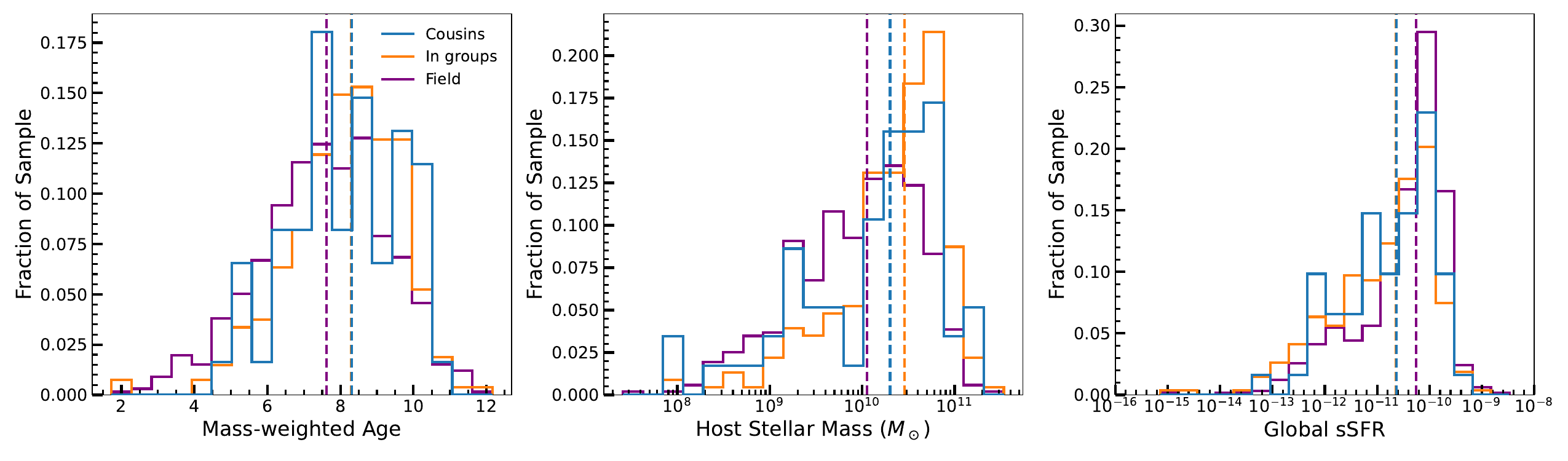}
\caption{Distributions of host-galaxy properties for group-associated SNe (orange), field SNe (purple), and the cousins subsample (blue). Panels show mass-weighted stellar age (left), host stellar mass (center), and specific star formation rate (right). Dashed vertical lines indicate the median value for each sample respectively. Field SNe preferentially occur in younger and lower-mass host galaxies, whereas group-associated and cousins SNe are more strongly associated with older, more massive systems.}
\label{fig:host_prop}
\end{figure*}

\begin{figure*}
    \centering
    \includegraphics[width=0.6\linewidth]{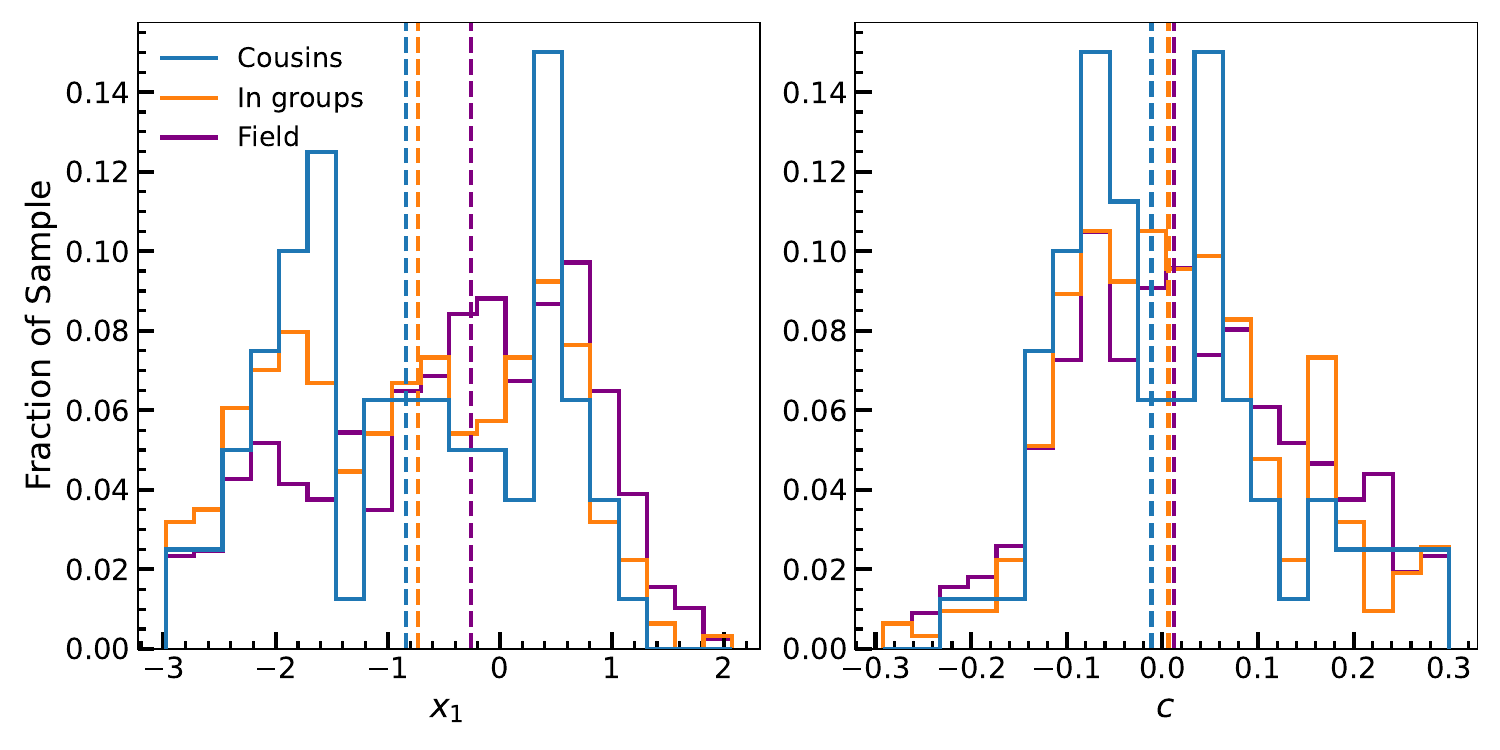}
    \caption{Distributions of SALT3 stretch parameter $x_1$ (center) and color parameter $c$ (right) for the TITAN gold sample with $z < 0.05$. Group-associated SNe are shown in orange, field SNe without group associations are in purple, and cousins SNe are in blue. Dashed vertical lines indicate the median value for each sample respectively.}
    \label{fig:hist}
\end{figure*}

\subsection{Supernova Light-Curve and Host-Galaxy Properties} \label{sec:host_props}

We compare the distributions of host-galaxy mass-weighted stellar age, stellar mass, and specific star formation rate (sSFR) for group-associated SNe, field SNe, and the cousins subsample in Figure~\ref{fig:host_prop}. Field SNe preferentially occur in younger and lower-mass host galaxies compared to both group-associated and cousins SNe. 
Similarly, the sSFR distributions show that the host galaxies of field SNe exhibit a somewhat narrower distribution and have fewer passive hosts relative to the group-associated populations. This likely reflects environmental differences in galaxy populations, as denser environments preferentially host a larger fraction of massive, quiescent galaxies \citep{dressler_galaxy_1980, peng_mass_2010}.  To quantify these differences, we use two-sample Kolmogorov--Smirnov (KS) tests, where the KS statistic $D$ measures the maximum separation between the cumulative distributions and the corresponding $p$-value quantifies the statistical significance of the difference. The group-associated and field SN host-property distributions differ significantly for all three parameters, with $D=0.185$ ($p=3.6\times10^{-6}$) for mass-weighted stellar age, $D=0.290$ ($p=2.8\times10^{-12}$) for stellar mass, and $D=0.223$ ($p=8.7\times10^{-9}$) for global sSFR. Together, Figures~\ref{fig:group_mass_hist} and \ref{fig:host_prop} demonstrate that SNe in groups occupy systematically different group and host-galaxy environments than the full distribution of galaxies.

Figure~\ref{fig:hist} compares the distributions of SALT3 stretch ($x_1$) and color ($c$). The stretch distribution differs significantly between group-associated and field SNe (a KS test returns $D=0.204$ and a $p$-value of $p=3.0\times10^{-6}$), with group environments exhibiting a more pronounced, bimodal low-$x_1$ population, whereas the color distributions are statistically consistent ($D=0.060$, $p=0.582$). The cousins subsample shows a similar tendency toward lower stretch values, although the smaller sample size makes the comparison less statistically significant. 


The overabundance of fast-declining SNe in group environments is consistent with previous studies demonstrating that the SN~Ia stretch distribution depends on host-galaxy properties. Lower mean stretch is preferentially observed in more massive and older galaxies, and the stretch distribution itself can exhibit bimodality and environmental dependence in large, volume-limited samples \citep{sullivan_dependence_2010,ginolin_ztf_2025}. 




\section{Results}
\label{sec:results}

We investigate whether galaxy-group information can improve SN Ia distance measurements using two complementary approaches that are carried through each of the following subsections. First, we examine the cousins sample, in which multiple SNe occurring within the same galaxy group are compared directly. Because these SNe are expected to share the same large-scale peculiar velocity and correlated environmental properties, this analysis tests whether a common group environment reduces Hubble residual scatter. Second, we evaluate the use of group-averaged redshifts for the full sample of group-associated SNe, testing whether replacing individual host-galaxy redshifts with the mean group redshift improves distance precision. For most subsections below, we first summarize the analysis and then discuss the cousins and group-redshift analyses separately.

\subsection{Distance Measurements and Hubble Residual Dispersion}
\label{sec:hr_scatter}

We begin by comparing the overall Hubble residual dispersion measured for the principal samples considered in this work. Table~\ref{tab:global_RelSD} summarizes the global RelSD measurements for the gold and platinum samples, including field SNe, group-associated SNe analyzed using both individual and group-averaged redshifts, the cousins sample, and the sibling sample.

The full gold sample has a scatter of $0.170^{+0.007}_{-0.007}$~mag, while the platinum sample's scatter is 13\% lower at $0.148^{+0.010}_{-0.007}$~mag. As a whole, SNe associated with groups have slightly higher scatter than SNe in the field in both the platinum and gold samples; while this difference is not highly significant (1.9$\sigma$ in the gold sample), it may reflect either increased peculiar velocities in group environments or slightly higher SN distance scatter in early-type hosts.


\begin{table}
\centering
\caption{Hubble residual dispersion measurements. For the cousins samples, the 3\% and 2\% variants denote maximum allowed values of $R_{2t}/D_{\rm L}$ of 3\% and 2\%, respectively. The 3\% variant is adopted as the baseline selection, while the 2\% variant provides a more restrictive association criterion to assess the sensitivity of the results to projection effects.}
\label{tab:global_RelSD}
\begin{tabular}{lrrr}
\hline
Sample & Redshift Used & RelSD (mag) & $N_{\rm SN}$\\
\hline
\multicolumn{4}{c}{\textit{Gold Sample}}\\
\hline

All SNe in groups & Group redshift &
$0.200^{+0.016}_{-0.015}$ & 314\\

All SNe in groups & Indiv.~redshift &
$0.184^{+0.017}_{-0.009}$ & 314\\

Field SNe & Indiv.~redshift &
$0.164^{+0.005}_{-0.008}$ & 772\\

Full sample & Indiv.~redshift &
$0.170^{+0.007}_{-0.007}$ & 1086\\

Cousins (3\% Var.) & No redshift &
$0.177^{+0.042}_{-0.030}$ & 80\\

Cousins (2\% Var.) & No redshift &
$0.122^{+0.020}_{-0.017}$ & 54\\

Siblings & Indiv.~redshift &
$0.275^{+0.108}_{-0.068}$ & 12\\

\hline
\multicolumn{4}{c}{\textit{Platinum Sample$^{a}$}}\\
\hline

All SNe in groups & Group redshift &
$0.161^{+0.015}_{-0.023}$ & 122\\

All SNe in groups & Indiv.~redshift &
$0.158^{+0.033}_{-0.017}$ & 122\\

Field SNe & Indiv.~redshift &
$0.144^{+0.011}_{-0.008}$ & 274\\

Full sample & Indiv.~redshift &
$0.148^{+0.010}_{-0.007}$ & 396\\

Cousins (3\% Var.) & No redshift &
$0.093^{+0.026}_{-0.022}$ & 20\\

Cousins (2\% Var.) & No redshift$^{b}$ &
$0.088^{+0.029}_{-0.023}$ & 17\\

\hline
\end{tabular}
\tablenotetext{a}{No sibling systems satisfy the stricter platinum sample selection criteria.}
\tablenotetext{b}{For the cousins samples, ``No redshift'' indicates that the analysis compares relative SN Ia distances within each group rather than using individual or group redshifts.}
\end{table}

\subsubsection{Distance Dispersion in the Cousins Sample}

For the cousins sample, we measure a Hubble residual dispersion of $0.177^{+0.042}_{-0.029}$~mag (Table~\ref{tab:global_RelSD}). This is consistent with the dispersion of the full TITAN gold sample, $0.170^{+0.006}_{-0.007}$~mag. Thus, using the cousins distance comparison does not produce a statistically significant reduction in distance scatter relative to the full SN sample. As a complementary comparison, analyzing these same cousins SNe using their individual host-galaxy redshifts gives a nearly identical scatter of $0.180^{+0.028}_{-0.029}$~mag.

We do, however, find a larger reduction in scatter for the platinum sample. The full platinum sample has a Hubble residual dispersion of $0.148^{+0.010}_{-0.007}$~mag, while the platinum cousins subsample has a dispersion of $0.093^{+0.026}_{-0.022}$~mag for the baseline $R_{2t}/D_{\rm L} < 3\%$ selection cut. Relative to the scatter from the full platinum sample, this corresponds to a $37.0^{+14.7}_{-17.7}\%$ reduction in scatter at $2.3\sigma$ significance. For the same set of platinum cousins SNe, the scatter computed from their Hubble residuals is $0.209^{+0.073}_{-0.043}$~mag,  $55.5^{+20.1}_{-33.7}\%$ larger than the scatter computed from the difference in their pairwise distances ($1.4\sigma$ significance).

\begin{figure}
    \centering
    \includegraphics[width=\columnwidth]{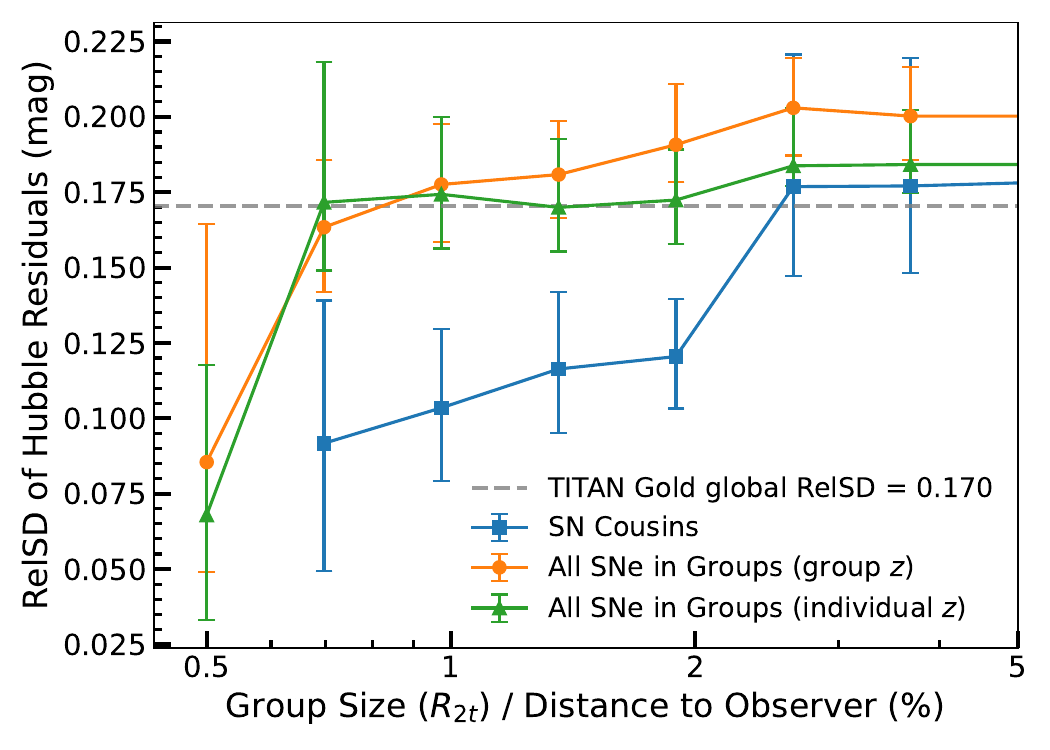}
    \caption{Global Hubble residual scatter as a function of the maximum allowed ratio of group extent to observer distance, $R_{2t}/D_{\rm L}$. The adopted threshold is $R_{2t}/D_{\rm L}=3\%$. The scatter becomes approximately stable near this value, while more restrictive cuts substantially reduce the available sample size.}
    \label{fig:r2t_cut}
\end{figure}

Additionally, Figure~\ref{fig:r2t_cut} illustrates how the Hubble residual dispersion measured from SN cousins depends on our cut on the maximum allowed $R_{2t}/D_{\rm L}$. For SNe located a line-of-sight distance of $R_{2t}$ from the group center, a 3\% cut corresponds to a maximum distance-modulus difference of approximately 0.06~mag between the SN location and the group center. We note, however, that SN cousins can be separated by up to $2\times R_{2t}$ even while both are still inside the second turnaround radius. In addition, 19.1\% of group-associated SNe lie beyond one projected $R_{2t}$, indicating that residual line-of-sight separation between cousins may occasionally contribute more than 0.06~mag to the measured dispersion. We adopt a 3\% threshold as our baseline while considering a 2\% threshold as a more restrictive variant that reduces the potential impact of projection effects.

For the gold sample, tightening the $R_{2t}/D_{\rm L}$ threshold from 3\% to 2\% reduces the number of group-associated SNe from 314 to 282 and the cousins sample from 80 SNe in 32 groups to 54 SNe in 25 groups. The cousins dispersion decreases substantially from $0.177^{+0.042}_{-0.029}$~mag to $0.122^{+0.020}_{-0.017}$~mag. Relative to the full TITAN gold sample, this corresponds to a $28.7^{+10.8}_{-10.4}\%$ reduction in scatter at $2.8\sigma$. For the same 54 SNe analyzed using the Hubble residuals from their individual host-galaxy redshifts, the scatter is $0.162^{+0.031}_{-0.028}$~mag, corresponding to a $24.9^{+15.4}_{-19.4}\%$ reduction --- for the same SNe using their individual redshifts --- at $1.2\sigma$ significance.

The platinum sample shows a similar trend. The 2\% $R_{2t}/D_{\rm L}$ cut reduces the platinum cousins dispersion from $0.093^{+0.026}_{-0.022}$~mag to $0.088^{+0.029}_{-0.023}$~mag, which changes the scatter relative to the full platinum sample from a $37.0^{+14.7}_{-17.7}\%$ ($2.3\sigma$) reduction to a $40.5^{+17.6}_{-17.8}\%$ reduction ($2.3\sigma$). The more restrictive cut reduces the platinum cousins sample from 20 to 17~SNe, and reduces the number of groups from 9 to 8. Additionally, when we analyze the same objects using Hubble residuals with respect to their individual host-galaxy redshifts, the $R_{2t}/D_{\rm L} < 2\%$ sample gives  a $35.2^{+23.0}_{-53.7}\%$ increase relative to the pairwise distance measurement ($0.9\sigma$ significance). Therefore, both the platinum sample and the gold sample show a consistent trend in which a more restrictive $R_{2t}/D_{\rm L}$ cut produces lower cousins scatter, although this comes at the cost of a smaller sample. We note that we do not consider tightening the cut further to 1\%, which would leave only 20 SNe in 10 groups for the gold sample.

Lastly, while the sibling sample provides a useful limiting case because multiple SNe within the same host galaxy share identical global host properties and the same peculiar velocity, our statistical leverage is limited. We measure a median pairwise Hubble residual dispersion of $0.275^{+0.108}_{-0.068}$~mag for the six sibling pairs in our sample when all sibling events are considered collectively. These values are statistically consistent with both the overall TITAN gold sample scatter and the cousins scatter.

\begin{figure*}
    \centering
    \includegraphics[width=0.49\linewidth]{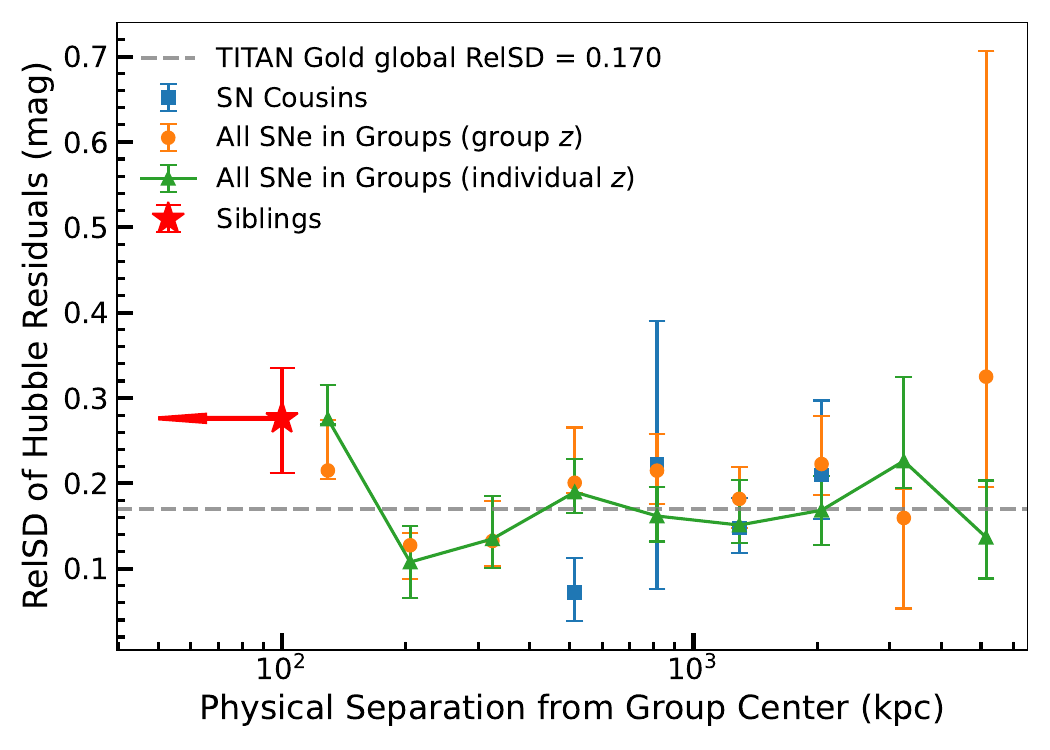}
    \hfill
    \includegraphics[width=0.49\linewidth]{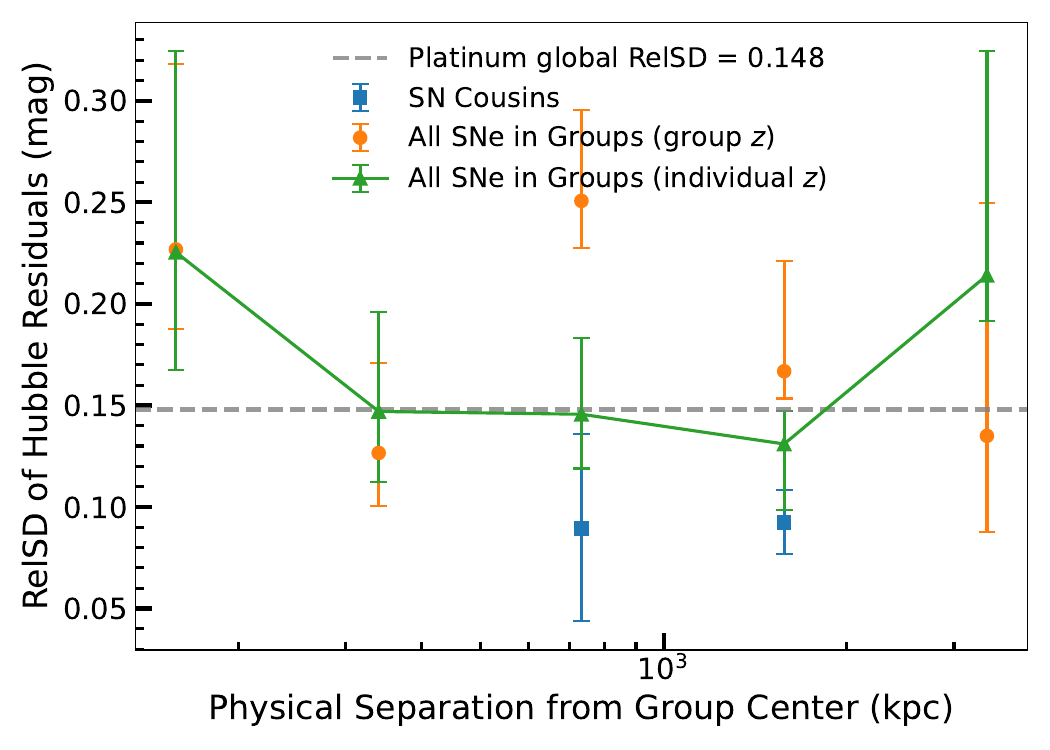}
    \caption{
    Hubble residual dispersion (RelSD) as a function of projected physical separation from the galaxy-group center. The TITAN gold sample is on the left and the platinum sample is on the right. Blue points show the cousins sample, orange points show all SNe in groups using group-averaged redshifts, and green points show all SNe in groups using individual host-galaxy redshifts. Horizontal dashed lines indicate the global RelSD of the corresponding parent sample. The red star denotes the sibling sample, plotted at an arbitrary horizontal position with an arrow indicating that its true mean separation ($6.37$~kpc) lies to the left of the plotted range.
    }
    \label{fig:disp_vs_sep}
\end{figure*}

\subsubsection{Distance Dispersion using Group Redshifts}

We next investigate whether replacing individual host-galaxy redshifts with galaxy-group averaged redshifts improves the precision of SN~Ia distance measurements for the full sample of group-associated SNe.

 For the gold sample, SNe associated with galaxy groups have a Hubble residual dispersion of $0.200^{+0.016}_{-0.015}$~mag when group redshifts are used, compared with $0.184^{+0.017}_{-0.009}$~mag when using individual host-galaxy redshifts (Table~\ref{tab:global_RelSD}). Rather than reducing the scatter, adopting group-averaged redshifts produces a marginal increase in the Hubble residual dispersion, although the difference is significant at only the $\sim$1.0$\sigma$ level. A similar trend is observed for the platinum sample, where the scatter increases from $0.158^{+0.033}_{-0.017}$~mag using individual host-galaxy redshifts to $0.161^{+0.015}_{-0.023}$~mag when group redshifts are adopted. Thus, we find no evidence that replacing individual galaxy redshifts with group-averaged values systematically improves SN~Ia distance precision.

As an additional robustness test, we repeated the analysis after applying large-scale bulk-flow corrections to both the individual host-galaxy and group-averaged redshifts. Following \citet{peterson_pantheon_2022}, we applied these corrections using the \texttt{pvhub}\footnote{\url{https://github.com/KSaid-1/pvhub}.} package, which derives peculiar velocity corrections from the reconstructed density and velocity fields of the 2M++ galaxy redshift survey \citep{lavaux_2m_2011}. These corrections are intended to account for coherent motions induced by large-scale structure beyond the local group environment. We find that the resulting Hubble residual dispersions and separation-dependent trends remain consistent with those presented above;  no statistically significant reduction in scatter emerges when group-averaged redshifts are used. This indicates that our conclusions are not driven by the treatment of large-scale bulk motions and instead primarily reflect the local group environments and the quality of the underlying group assignments.

\subsection{Dependence on Distance from Group Center}

\begin{figure}
\centering
\includegraphics[width=\linewidth]{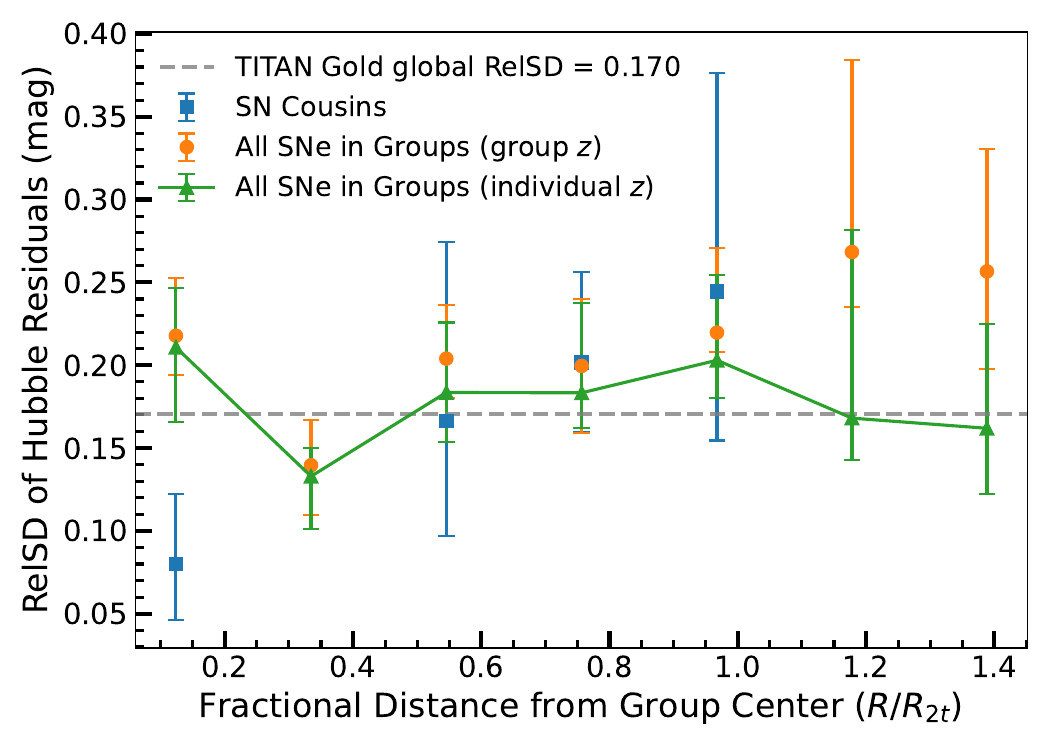}
\caption{Hubble residual dispersion as a function of projected separation normalized by the group second-turnaround radius, $R_{2t}$. The plot shows three samples: the cousins sample (blue), all SNe in groups using group redshifts (orange), and all SNe in groups using individual redshifts (green). The global RelSD for the TITAN gold sample is indicated by a dashed horizontal line.}
\label{fig:disp_vs_fracradius}
\end{figure}

We next investigate whether the location of a SN within its host-galaxy group influences the Hubble residual dispersion. Figure~\ref{fig:disp_vs_sep} shows the dispersion as a function of projected physical separation from the group center for both the TITAN gold and platinum samples, while Figure~\ref{fig:disp_vs_fracradius} presents the same analysis using the projected separation normalized by the group second-turnaround radius, $R_{2t}$, to measure differences as a fraction of group extent.

\subsubsection{Dispersion Trends on Distance from Group Center for Cousins}

For the cousins sample, the projected separation is defined as the median projected distance of the cousins from the center of their host-galaxy group. Thus, unlike the group-redshift and individual-redshift samples, where we measure the separation of each SN from the group center, each cousins system is assigned a single representative separation corresponding to the median of its constituent SNe. 
While the cousins Hubble residual scatter measurements are broadly consistent as a function of projected separation, we note that the lowest-separation bin for the cousins sample in the gold sample, which contains only four groups with two SNe each, may have reduced projection effects and exhibits a particularly low mean RelSD of $0.072^{+0.040}_{-0.034}$~mag. Because this bin contains only a small number of systems, however, the bootstrap uncertainties are themselves poorly constrained and we therefore do not interpret this apparent reduction as a highly significant effect.

To quantify the dependence on projected separation, we divide the cousins sample at the median projected distance from the group center ($1.26$~Mpc). Low-separation systems exhibit a scatter of $0.140^{+0.053}_{-0.042}$~mag, while high-separation systems have a larger scatter of $0.213^{+0.051}_{-0.047}$~mag. This corresponds to an increase of $0.075^{+0.067}_{-0.073}$~mag ($\sim 1.0\sigma$ significance). Although not statistically significant, this trend is consistent with the expectation that SNe farther from the group center may also be farther away along the line of sight, or have a more uncertain group membership. The platinum sample exhibits the same qualitative behavior, with lower absolute scatter but no statistically significant change in the dependence on projected separation.

A similar trend is observed when the projected separation is normalized by the group second-turnaround radius, $R_{2t}$ (Figure~\ref{fig:disp_vs_fracradius}), thereby accounting for the different physical sizes of galaxy groups. Cousins located at smaller fractional radii ($R/R_{2t}<0.744$) exhibit a scatter of $0.130^{+0.052}_{-0.042}$~mag, compared with $0.227^{+0.057}_{-0.044}$~mag at larger fractional radii, corresponding to a marginal significance of approximately $1.4\sigma$. Overall, both analyses suggest that the cousins method performs best for SNe located closer to the centers of their host groups, although the statistical significance remains modest.

\subsubsection{Dispersion Trends on Distance from Group Center for Galaxy Group}

The sample of SNe using group-averaged redshifts shows that, relative to the same sample using individual host-galaxy redshifts, scatter increases as projected separation increases (Figure \ref{fig:disp_vs_sep}). Splitting the full group-associated sample at the median separation of $0.74$~Mpc, we find that at small separations the two methods are statistically consistent. At larger separations, group-averaged redshifts produce larger scatter, with $\Delta {\rm RelSD}=0.049^{+0.025}_{-0.025}$~mag. The difference between the low- and high-separation regimes is $0.041^{+0.028}_{-0.033}$~mag, significant at approximately $1.9\sigma$, indicating that the degradation in distance precision associated with group-averaged redshifts becomes increasingly important for SNe farther from the group center.
The platinum sample follows a similar qualitative behavior, albeit with lower scatter. No separation bin shows a clear improvement when using group-averaged redshifts instead of individual host-galaxy redshifts.

We find a similar result when the projected separation is normalized by $R_{2t}$. At smaller fractional radii, the difference between the group-redshift and individual-redshift scatter is only $\Delta{\rm RelSD}=0.003$~mag, but at larger fractional radii, the group-redshift scatter becomes larger ($\Delta{\rm RelSD}=0.038$~mag).

Lastly, we examine whether the effectiveness of group-averaged redshifts depends on the galaxy-group velocity dispersion, $\sigma_p$. One might expect group-averaged redshifts to provide the greatest improvement for groups with large velocity dispersions, where individual host-galaxy redshifts are expected to have the largest deviations from the group mean. Instead, we find that the Hubble residual scatter is smallest for both the individual-redshift and group-redshift samples in groups with low $\sigma_p$, with no evidence that group-averaged redshifts become more advantageous at larger velocity dispersions. 


\begin{table*}[t]
\centering
\caption{Cousins distance scatter as a function of SN, host-galaxy, and group properties. The table lists the sample median for each parameter, the Hubble residual dispersion (RelSD) below and above this median, the difference between the two, and the corresponding significance.}
\label{tab:median_split}
\begin{tabular}{lrrrrr}
\hline
Parameter & Median & RelSD$_{\rm low}$ & RelSD$_{\rm high}$ & Difference & Significance \\
\hline
\multicolumn{6}{l}{\textbf{SN Property}}\\
\hline
$x_1$
& $-0.610$
& $0.143^{+0.054}_{-0.040}$
& $0.211^{+0.060}_{-0.045}$
& $0.069^{+0.070}_{-0.071}$
& $0.98\sigma$ \\
\hline
\multicolumn{6}{l}{\textbf{Host-Galaxy Properties}}\\
\hline
Host mass ($\log M_\star/M_\odot$)
& $9.751$
& $0.173^{+0.051}_{-0.044}$
& $0.178^{+0.051}_{-0.046}$
& $0.005^{+0.074}_{-0.062}$
& $0.07\sigma$ \\
sSFR ($\log_{10},\mathrm{yr}^{-1}$)
& $-10.604$
& $0.194^{+0.057}_{-0.049}$
& $0.179^{+0.054}_{-0.046}$
& $-0.015^{+0.074}_{-0.071}$
& $0.20\sigma$ \\
Age (Gyr)
& $8.494$
& $0.211^{+0.062}_{-0.049}$
& $0.161^{+0.053}_{-0.041}$
& $-0.050^{+0.070}_{-0.073}$
& $0.71\sigma$ \\
\hline
\multicolumn{6}{l}{\textbf{Galaxy-Group Properties}}\\
\hline
Group spiral fraction
& $0.400$
& $0.251^{+0.069}_{-0.059}$
& $0.112^{+0.024}_{-0.024}$
& $-0.139^{+0.070}_{-0.063}$
& $2.06\sigma$ \\
Group mass ($10^{12}M_\odot$)
& $297$
& $0.135^{+0.055}_{-0.040}$
& $0.221^{+0.058}_{-0.046}$
& $0.086^{+0.068}_{-0.074}$
& $1.18\sigma$ \\
Projected separation (kpc)
& $1256$
& $0.142^{+0.052}_{-0.041}$
& $0.213^{+0.052}_{-0.046}$
& $0.071^{+0.073}_{-0.067}$
& $1.03\sigma$ \\
\hline
\end{tabular}
\end{table*}

\subsection{Distance Scatter versus Host-Galaxy and Galaxy Group Properties}
\label{sec:group_props}

\begin{figure}
\centering
\includegraphics[width=\linewidth]{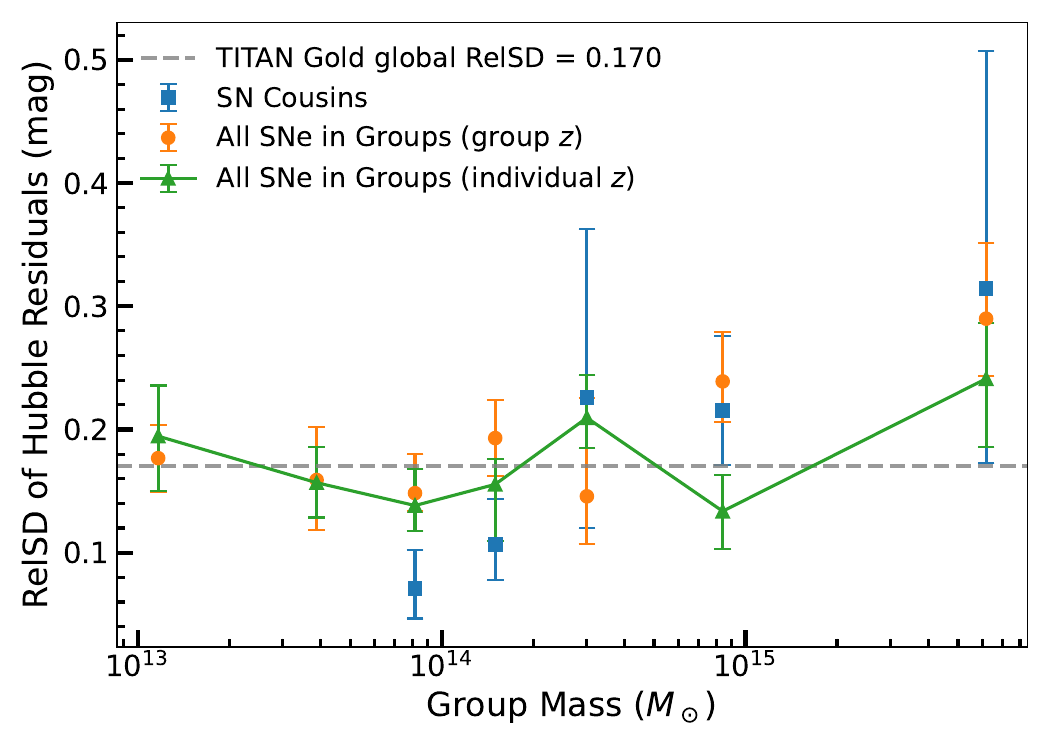}
\caption{RelSD of SN Ia Hubble residuals as a function of group mass. The plot shows three samples: the cousins sample (blue), all SNe in groups using group redshifts (orange), and all SNe in groups using individual redshifts (green). The global RelSD for the full sample is indicated by a dashed horizontal line.}
\label{fig:disp_vs_mass}
\end{figure}

While the projected location of a SN within its host-galaxy group shows only marginal evidence for influencing the Hubble residual scatter, the global properties of the host galaxy group may play a more important role. We therefore examine how the scatter depends on both galaxy-group and host-galaxy properties. Figure~\ref{fig:disp_vs_mass} shows the Hubble residual dispersion as a function of group halo mass, while Tables~\ref{tab:median_split} and \ref{tab:median_split_group} summarize the dependence of distance dispersion on  the cousins and full group-associated samples, respectively.

\subsubsection{Dispersion Dependence on Galactic Properties for Cousins}


To investigate whether  specific SN, host-galaxy, or group properties can reduce cousins scatter, we measured the scatter below and above the median of several such properties; we summarize this analysis in Table~\ref{tab:median_split} and show the scatter as a function of group mass in Figure \ref{fig:disp_vs_mass}. For the SN $x_1$ parameter and the host-galaxy parameters (stellar mass, stellar age, and sSFR), we use the average value from all SNe within that group to determine whether a given group is above or below the sample median.  This is necessary due to the nature of the cousins distance comparison; choosing only groups where all SNe have higher than median $x_1$, for example, would substantially reduce the sample size available for comparisons.

\begin{table*}[t]
\centering
\caption{Hubble residual scatter for SNe associated with galaxy groups as a function of SN, host-galaxy, and group properties. The table lists the sample median for each parameter, the Hubble residual dispersion (RelSD) below and above this median, the difference between the two, and the corresponding significance.}
\label{tab:median_split_group}
\begin{tabular}{lrrrrr}
\hline
Parameter & Median & RelSD$_{\rm low}$ & RelSD$_{\rm high}$ & Difference & Significance \\
\hline

\multicolumn{6}{l}{\textbf{SN Property}}\\
\hline
$x_1$
& $-0.732$
& $0.164^{+0.016}_{-0.015}$
& $0.186^{+0.026}_{-0.020}$
& $0.022^{+0.030}_{-0.025}$
& $0.77\sigma$ \\

\hline
\multicolumn{6}{l}{\textbf{Host-Galaxy Properties}}\\
\hline
Host mass ($\log M_\star/M_\odot$)
& $10.171$
& $0.203^{+0.021}_{-0.022}$
& $0.141^{+0.027}_{-0.021}$
& $-0.062^{+0.033}_{-0.030}$
& $1.99\sigma$ \\

sSFR ($\log_{10}\,\mathrm{yr}^{-1}$)
& $-10.638$
& $0.160^{+0.024}_{-0.016}$
& $0.163^{+0.017}_{-0.015}$
& $0.003^{+0.026}_{-0.027}$
& $0.11\sigma$ \\

Age (Gyr)
& $8.298$
& $0.147^{+0.032}_{-0.024}$
& $0.193^{+0.018}_{-0.023}$
& $0.045^{+0.032}_{-0.035}$
& $1.36\sigma$ \\

\hline
\multicolumn{6}{l}{\textbf{Galaxy-Group Properties}}\\
\hline
Group spiral fraction
& $0.441$
& $0.189^{+0.020}_{-0.025}$
& $0.178^{+0.020}_{-0.022}$
& $-0.011^{+0.031}_{-0.031}$
& $0.35\sigma$ \\

Group mass ($10^{12}M_\odot$)
& $133$
& $0.176^{+0.025}_{-0.023}$
& $0.195^{+0.018}_{-0.024}$
& $0.020^{+0.033}_{-0.030}$
& $0.62\sigma$ \\

Projected separation (kpc)
& $737$
& $0.177^{+0.022}_{-0.029}$
& $0.185^{+0.019}_{-0.026}$
& $0.008^{+0.036}_{-0.033}$
& $0.24\sigma$ \\

\hline
\end{tabular}
\end{table*}

The only effect showing modest significance is that groups with higher spiral fractions have lower distance scatter than groups dominated by early type galaxies, with the scatter decreasing from $0.251^{+0.069}_{-0.059}$ to $0.112^{+0.024}_{-0.024}$~mag, significant at the $2.1\sigma$ level. A weaker but likely correlated trend is also present with group halo mass, where lower-mass systems yield a smaller scatter than higher-mass groups by $0.086^{+0.068}_{-0.074}$~mag ($1.2\sigma$), in agreement with the behavior shown in Figure~\ref{fig:disp_vs_mass}. SNe with higher $x_1$ values, which are expected to be more prevalent in younger groups/galaxies, also have lower scatter but at just 1.0$\sigma$ significance.



\subsubsection{Dispersion Dependence on Galactic Properties for Galaxy Groups}

We repeat this same analysis for the full sample of 314 group-associated SNe using their individual redshifts, although here we treat each SN as an independent data point rather than assigning equal weight to each galaxy group. The results are summarized in Table~\ref{tab:median_split_group}. Unlike the cousins sample, the full group-associated population does not clearly follow the same trends as the cousins sample.

We find the largest difference in scatter when comparing low versus high host-galaxy masses, with lower-mass hosts exhibiting a larger Hubble residual scatter than higher-mass hosts by $0.062^{+0.033}_{-0.030}$~mag, corresponding to $2.0\sigma$. We also observe larger scatter among SNe in older hosts --- which would be higher mass on average ---  at the $1.4\sigma$ level. We see no statistically significant differences in scatter as a function of the remaining parameters, including group mass, projected separation, group spiral fraction, and sSFR.

These trends differ from those found for the cousins sample. In particular, the marginal dependence on group halo mass and spiral fraction seen for the cousins sample is absent in the full group-associated sample.  These differences may be due to our relatively small sample sizes, and larger group-hosted SN samples may be needed to determine which of these trends are robust.


\subsection{The Effect of Group Redshifts on Peculiar Velocities}

\begin{figure}
    \centering
    \includegraphics[width=\columnwidth]{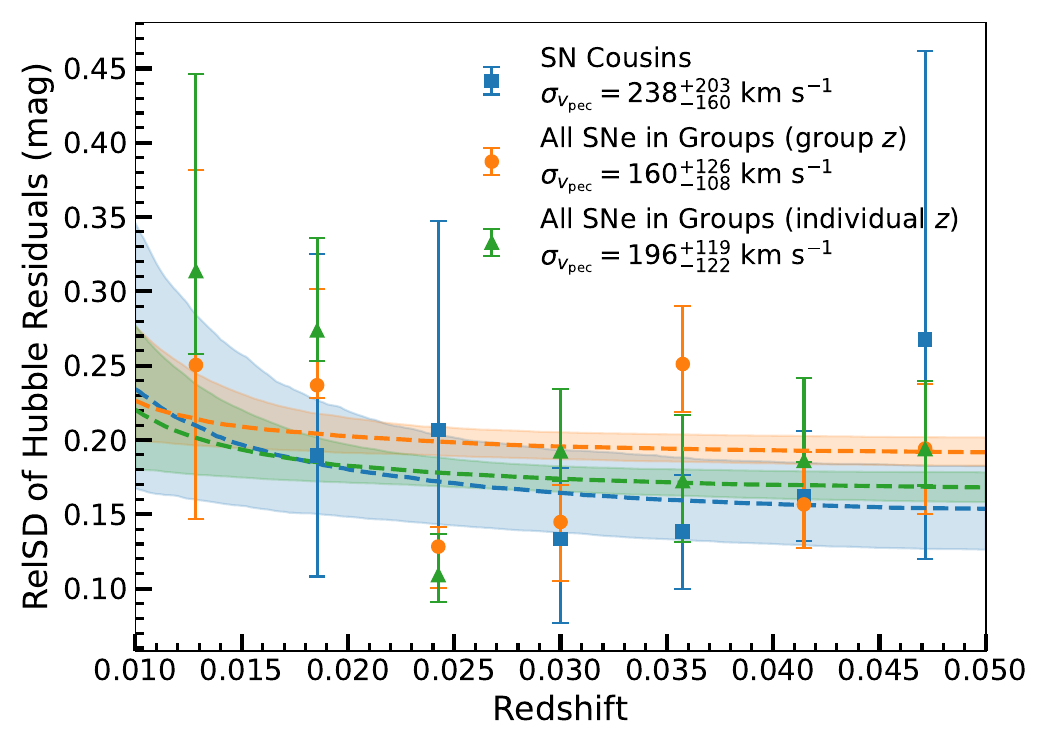}
    \caption{RelSD of SN Ia Hubble residuals as a function of redshift. The figure includes three samples: cousins in blue, all SNe in groups using group redshifts in orange, and  individual galaxy redshifts in green. Dashed lines indicate the best fit peculiar velocity contribution for each sample and are included in the legend, with the uncertainties shown as the shaded regions.}
    \label{fig:disp_vs_z}
\end{figure}

To constrain the characteristic peculiar velocity dispersion amplitude, which represents the root-mean-square amplitude of random galaxy motions about the Hubble flow, together with the intrinsic SN scatter in each subsample, we use the likelihood framework described in Section~\ref{sec:pvmodel}. We perform this analysis for three samples: (1) SNe in groups using individual CMB-frame host-galaxy redshifts, (2) SNe in groups using group-averaged redshifts, and (3) the cousins sample, which is expected to be insensitive to PVs. The resulting constraints are summarized in Table~\ref{tab:pv} and Figure~\ref{fig:disp_vs_z}.

\begin{table}
\caption{Constraints on peculiar velocity amplitude and intrinsic scatter.}
\label{tab:pv}
\begin{tabular}{lcc}
\hline
Sample & $\sigma_{v_{\mathrm{pec}}}\ ({\rm km~s}^{-1}$) & $\sigma_{\mathrm{int}}$ (mag) \\
\hline
Cousins & $239^{+203}_{-160}$ & $0.146^{+0.032}_{-0.032}$ \\
Group redshift & $160^{+126}_{-108}$ & $0.190^{+0.011}_{-0.011}$ \\
Indiv. redshift & $196^{+119}_{-122}$ & $0.165^{+0.011}_{-0.013}$ \\
\hline
\end{tabular}
\end{table}

The inferred peculiar velocity amplitudes are broadly consistent across the three samples, with relatively large uncertainties that limit strong conclusions about differences in the velocity contribution. For the cousins sample, we recover a peculiar velocity scatter of $\sigma_{v_{\rm pec}}=239^{+203}_{-160}$~km~s$^{-1}$. In the idealized case, cousins should share the same large-scale group motion, so little residual peculiar velocity contribution is expected. Our measurement  is consistent with zero peculiar velocity scatter at 1.5$\sigma$ significance; the larger average $R_{2t}/D_L$ at low redshifts may explain this marginal trend in cousins scatter with redshift.

Using individual host-galaxy redshifts, we infer a peculiar velocity scatter of $\sigma_{v_{\rm pec}}=196^{+119}_{-122}$~km~s$^{-1}$, while adopting group-averaged redshifts yields a slightly smaller value of $\sigma_{v_{\rm pec}}=160^{+126}_{-108}$~km~s$^{-1}$. Although these measurements are fully consistent within their uncertainties, the lower central value obtained using group redshifts is qualitatively consistent with the expectation that averaging the redshifts of multiple group members partially suppresses random galaxy-scale velocity components by providing a better estimate of the group's bulk cosmological motion.

More significant differences are found in the inferred intrinsic scatter. The cousins and individual-redshift samples yield similar values of $\sigma_{\rm int}=0.146^{+0.032}_{-0.032}$ and $0.165^{+0.011}_{-0.013}$~mag, respectively, whereas for the group-redshift sample we measure a substantially larger intrinsic scatter of $\sigma_{\rm int}=0.190\pm0.011$~mag. Possible contributors to this higher scatter include uncertainties in group membership, imperfectly determined group redshifts, departures from dynamical equilibrium within galaxy groups, or other limitations of the group catalog itself. Thus, while group averaging may partially reduce random peculiar motions, it does not improve the overall precision of SN~Ia distance measurements in the current sample.

The platinum sample exhibits qualitatively similar behavior, with lower overall Hubble residual scatter but no evidence that adopting group-averaged redshifts improves the total distance precision. This suggests that the conclusions presented here are not primarily driven by SN photometric quality, but instead reflect limitations associated with assigning group environments and accurately determining group redshifts.


\section{Discussion}
\label{sec:disc}

\subsection{Insights from Supernova Cousins}

Although our baseline analysis does not demonstrate better distance precision from SN cousins, we explore several analysis variants that show improved distances from cousins may be possible.  First, with the higher-quality platinum sample, the distance dispersion decreases to $0.093^{+0.026}_{-0.022}$~mag for cousins compared with $0.148^{+0.010}_{-0.007}$~mag from the full sample (2.3$\sigma$ significance).  Compared to the Hubble residual dispersion from that same sample of platinum cousins, the pairwise distance dispersion is $55.5^{+20.1}_{-33.7}$\% lower.

Second, more restrictive 
 $R_{2t}/D_L$ cuts may reduce  distance dispersion along the line of sight between cousins;  two SN cousins on opposite sides of a group at the threshold of our baseline cut ($R_{2t}/D_L = 3$\%) could have as much as $\sim$0.12~mag difference in distance modulus between them.  For a tighter cut of $R_{2t}/D_L < 2$\%, distance scatter in the gold sample decreases to $0.122$~mag (2.8$\sigma$ significance) and distance scatter in the platinum sample decreases to 0.088~mag (2.3$\sigma$ significance). Taken together, these results suggest that SN cousins may have improved distance precision under tighter selection criteria.

The reduction in cousins scatter seen in the more restrictive samples is not simply explained by the removal of peculiar-velocity contributions. We find that even when simultaneously modeling the peculiar velocity and intrinsic scatter of our Hubble residuals, the intrinsic scatter is dominant.  Figure \ref{fig:disp_vs_z} shows that  Hubble residual scatter in the gold sample is roughly constant until  $z < 0.02$, after which it increases by $\lesssim0.04$~mag for the 11\% of our sample at $0.01 < z < 0.02$.

On the other hand, our analysis of group and host-galaxy environments provides tentative evidence that the observed scatter within the cousins sample may correlate with the broader properties of the host-galaxy group. The strongest trend is that groups with higher spiral fractions have lower cousins scatter at $\sim$2$\sigma$ significance. A weaker --- and likely correlated --- trend is observed with group halo mass, with lower-mass groups exhibiting lower scatter than higher-mass groups. Cousins located farther from group centers also tend to exhibit larger scatter, although this trend is only marginally significant.   However, when analyzing the Hubble residuals of group-hosted SNe as a whole, we do not recover the same marginally significant trends with environment that we see in the cousins sample.  We also do not see significant differences in scatter when considering the individual host-galaxy properties of the SN cousins instead of the group-level properties.


Our data are too limited to constrain whether SN siblings in our sample have comparably low distance scatter compared to SN cousins, as we find a poorly constrained scatter of $0.275^{+0.108}_{-0.068}$~mag for the six pairs of SNe in our sample.  Measurements of SN siblings in the literature have a broad range in scatter, including \citet{kelsey_archival_2024}, who find 0.185~mag, \citet{dwomoh_evaluating_2024}, who find 0.177--0.186~mag, and \citet{scolnic_pantheon_2022}, who find 0.32~mag.  Alternatively, \citet{Burns_2020} find intrinsic scatter of $\sim$0.06~mag in literature samples, \citet{dhawan_ztf_2025} find $<0.097$~mag intrinsic scatter in ZTF, and \citet{ward_relative_2023} find near-zero intrinsic dispersion for the three SN siblings in NGC~3147. Thus, our cousins scatter --- depending on our sample cuts --- is consistent with the wide range of values reported for SN siblings, and more data will be needed to provide consensus on how SN cousins and siblings compare.

We also find a scatter comparable to that achieved in some SN ``twins" analyses, which compare spectroscopically similar SNe Ia to improve distance precision \citep{fakhouri_improving_2015}. The most precise distance measurements from twin-based approaches to date are from ``Twins Embedding" \citep{boone_twins_2021}, which finds a $\sim$40\% reduction in distance scatter, corresponding to 0.073--0.101~mag depending on the assumed peculiar-velocity contributions and the size of the reference sample. Our baseline cousins scatter of $0.177$~mag is higher than this range, although the $0.093$~mag scatter measured for our platinum cousins sample falls within it. This comparison suggests that the distance precision achieved by SN cousins could approach that obtained by methods that exploit spectroscopic similarity.

Given that SN cousins are less similar in host environment than siblings, and likely have less similarity in their SEDs than twins, it is somewhat surprising that several of our analysis variants exhibit such low scatter. However, we caution that the apparent improvement may be influenced by the limited sample size. Larger cousins samples and group catalogs will be needed to determine whether shared galaxy-group environments can provide a robust improvement in SN Ia distance precision and what selection criteria may be needed to realize this effect. If such an improvement is confirmed, SN cousins could provide a complementary way to both improve distance estimates for SNe Ia and better understand the factors impacting SN Ia distance scatter.

\subsection{Why Do Group Redshifts Not Consistently Improve Distances?}

A central motivation of this work is the expectation that galaxy groups provide a physically meaningful pathway to mitigating PV-induced scatter in SN~Ia distance measurements. Galaxies residing within the same gravitationally bound structure are expected to experience correlated velocity fields, implying that averaging over multiple group members could yield a redshift estimate closer to the underlying Hubble flow than the redshift of any individual galaxy.
Numerical simulations and theoretical studies generally predict that group-averaged redshifts should provide more accurate estimates of cosmological recession velocities than individual galaxy redshifts \citep{peterson_improving_2025}.

Surprisingly, we find no evidence that group-averaged redshifts improve the precision of SN~Ia distances for the full group-associated sample. For the gold sample, the Hubble residual scatter increases from $0.184^{+0.017}_{-0.009}$~mag when individual host-galaxy redshifts are used to $0.200^{+0.016}_{-0.015}$~mag when group-averaged redshifts are adopted, corresponding to a marginal increase at approximately $1.0\sigma$ significance. The same qualitative behavior is observed in the higher-quality platinum sample, where the scatter increases from $0.158^{+0.033}_{-0.017}$~mag using individual galaxy redshifts to $0.161^{+0.015}_{-0.023}$~mag using group redshifts. The absence of a clear improvement is also reflected in the inferred peculiar velocity contributions; we find $\sigma_{v_{\mathrm{pec}}}=196^{+119}_{-122}$~km~s$^{-1}$ for the individual-redshift sample and a fully consistent $160^{+126}_{-108}$~km~s$^{-1}$ for the group-redshift sample. Thus, although group-averaged redshifts are theoretically expected to reduce the contribution of individual galaxy peculiar velocities, they do not produce the anticipated reduction in Hubble residual scatter. Instead, we observe a small, statistically insignificant increase in scatter, indicating that group averaging provides no measurable improvement in distance precision for the full group-associated sample.  

We also see that group averaging may become less effective away from group centers. For the full group-associated sample, the scatter difference between group and individual redshifts becomes larger at projected separations above the median of $0.74$~Mpc. At large separations, group-redshift measurements have $0.049^{+0.025}_{-0.025}$~mag more scatter than the corresponding individual-redshift measurements, while the difference between the low- and high-separation regimes is $0.041^{+0.028}_{-0.033}$~mag at approximately $1.86\sigma$. 
These trends are suggestive but remain below the threshold for a statistically significant detection.

Several factors may contribute to the lack of a clear improvement from group redshifts. First, many systems in the \citetalias{tully_galaxy_2015} catalog are low-richness groups with relatively small internal velocity dispersions. In such environments, the group-averaged and individual host-galaxy redshifts are often very similar, leaving little opportunity for averaging over group members to produce a more accurate estimate of the underlying cosmological redshift.
Conversely, some of the larger and more complex identified groups may not be fully virialized systems. Ongoing accretion, substructure, and departures from dynamical equilibrium can cause the cataloged group velocity to differ from the velocity field experienced by an individual SN host galaxy. 

Second, uncertainties in group membership may become increasingly important toward the outskirts of groups. A substantial fraction of the group-associated sample lies at large projected separations from the nominal group center, and the separation-dependent analyses suggest that the performance of group-averaged redshifts may degrade at larger group-centric radii. 

Finally, the completeness of currently available group catalogs may limit the achievable improvement. Theoretical expectations generally assume complete membership assignments and well-determined group properties. In practice, observational catalogs are incomplete, especially at higher redshift, and are constructed from the most luminous galaxies. As a result, real-world group redshifts may not yet achieve the precision assumed in simulations.



These considerations suggest that group-averaged redshifts should not be viewed as a universal replacement for individual host-galaxy redshifts. Rather, their usefulness likely depends on the dynamical structure and membership quality of the underlying galaxy group. Our results provide no statistically significant evidence that group redshifts improve distance precision for the full group-associated SN sample. The data do, however, show tentative indications that group-redshift performance may be more comparable to individual-redshift performance for SNe located near group centers.  They may also offer improvement over redshifts derived from SN spectra, which are substantially less precise than those derived from SN host galaxies.

\subsection{Future Directions and Improved Group Catalogs}

The results presented here highlight the importance of improving both SN samples and galaxy-group catalogs. While the TITAN gold sample provides one of the largest low-redshift and cosmology grade SN datasets currently available, the statistical power of the cousins and sibling analyses remains limited by the relatively small number of systems containing multiple events.

Equally important are improvements in group identification. The \citetalias{tully_galaxy_2015} catalog represents one of the most comprehensive nearby group catalogs currently available, but it remains fundamentally limited by the depth and completeness of the underlying galaxy survey. Work with future and ongoing spectroscopic surveys, such as 
FAST \citep{zhang_drafts_2025}, WALLABY \citep{koribalski_wallaby_2020}, and DESI \citep{desi_collaboration_desi_2016} will identify substantially more low-luminosity galaxies and provide more accurate group memberships and velocity dispersions out to higher redshifts.

Combining these improved catalogs with forthcoming SN samples from surveys such as the Vera C.~Rubin Observatory Legacy Survey of Space and Time (LSST) will dramatically increase the number of known cousin and sibling systems. Such datasets will enable direct tests of how peculiar velocities, group dynamics, and environmental properties contribute to the scatter in SN Ia distance measurements.


\section{Conclusions}
\label{sec:conclusions}
In this work, we investigate how galaxy-group environments affect SN Ia distance measurements.  We use the gold sample of SNe~Ia from TITAN and a subsample of multiple SNe occurring within the same galaxy group, which we refer to as ``SN cousins." By combining SN~Ia distance measurements with galaxy-group information from the \citetalias{tully_galaxy_2015} catalog, we examine whether galaxy-group information can be used to yield improved precision in SN~Ia distances.

For our baseline analysis of SN cousins, we measure a Hubble residual scatter of $0.177^{+0.042}_{-0.029}$~mag from pairwise comparison of their distances, nearly identical to the $0.180^{+0.028}_{-0.029}$~mag we measure for the same SNe analyzed using individual host-galaxy redshifts. However, when restricting to a higher-quality ``platinum" sample we find that SN cousins yield a scatter that is $37.0^{+14.7}_{-17.7}\%$ lower than that of the full platinum sample ($2.3\sigma$ significance). Additionally, when applying more restrictive cuts to reduce the effect of line-of-sight differences in distance modulus among the cousins, we find that the cousins again have lower scatter, a reduction of $\sim$30--40\% at 2.8$\sigma$ and 2.3$\sigma$ significance for the gold and platinum samples, respectively.


We also find tentative evidence that the observed scatter of SN cousins varies with some properties of the host-galaxy group. We see that cousins located farther from group centers have marginal evidence for higher scatter, and that groups with higher spiral fractions have lower scatter than groups with lower spiral fractions at 2$\sigma$ significance. By comparison, we see no relationship between the scatter in SN cousins and the host-galaxy properties of the SNe themselves. These results suggest that the scatter among cousins may vary with the broader group environment, although our results are limited by the available sample size.

When analyzing the effect of measuring Hubble residuals from  group-averaged redshifts across our full group-associated SN sample, we do not find evidence that they improve SN~Ia distance precision. For the gold sample, the scatter {\it increases} slightly from $0.184^{+0.017}_{-0.009}$~mag using individual host-galaxy redshifts to $0.200^{+0.016}_{-0.015}$~mag using group-averaged redshifts. The platinum sample shows the same qualitative behavior. Although group averaging may reduce some contribution from individual galaxy peculiar velocities, our results indicate that this potential benefit is offset by additional sources of scatter that may be caused by group membership uncertainties, group-redshift estimates, and the dynamical complexity of real galaxy groups.

Overall, our results suggest that the theoretical advantages of group-based velocity averaging are difficult to realize using current group catalogs, but that galaxy-group environments may nevertheless provide a promising avenue for improving SN~Ia distance precision.  Larger low-redshift SN samples, combined with deeper and more complete galaxy-group catalogs, will be essential for testing whether the trends observed here can be reproduced and systematically exploited.  Fortunately, ongoing SN surveys from the Rubin Observatory alongside current and upcoming redshift surveys should provide a wealth of new data for understanding the properties and distance precision of SNe~Ia in group environments.

\begin{acknowledgments}

A.B.\ and D.O.J.\ acknowledge support from NSF grants AST-2407632, AST-2429450, and AST-2510993, NASA grants 80NSSC24M0023 and 80NSSC24K0353, and HST/JWST grants HST-GO-17128.028 and JWST-GO-05324.031, awarded by the Space Telescope Science Institute (STScI), which is operated by the Association of Universities for Research in Astronomy, Inc., for NASA, under contract NAS5-26555.  This work is also funded in part by the Gordon and Betty Moore Foundation through Grant GBMF13900 to D.O.J.

This work has made use of data from the Asteroid Terrestrial-impact Last Alert System (ATLAS) project. The Asteroid Terrestrial-impact Last Alert System (ATLAS) project is primarily funded to search for near earth asteroids through NASA grants NN12AR55G, 80NSSC18K0284, and 80NSSC18K1575; byproducts of the NEO search include images and catalogs from the survey area. This work was partially funded by Kepler/K2 grant J1944/80NSSC19K0112 and HST GO-15889, and STFC grants ST/T000198/1 and ST/S006109/1. The ATLAS science products have been made possible through the contributions of the University of Hawaii Institute for Astronomy, the Queen’s University Belfast, the Space Telescope Science Institute, the South African Astronomical Observatory, and The Millennium Institute of Astrophysics (MAS) Chile, the Instituto de Astrofísica de Canarias, and the University of Oxford. 

S.J.S.\ acknowledges funding from STFC Grants ST/Y001605/1, ST/X001253/1, a Royal Society Research Professorship and the Hintze Family Charitable Foundation.

\end{acknowledgments}

%



\appendix

\section{Case Studies of the Coma Cluster and Hercules Complex}
\label{appendix:coma_hercules}

\begin{figure}
\centering
\includegraphics[width=0.6\columnwidth]{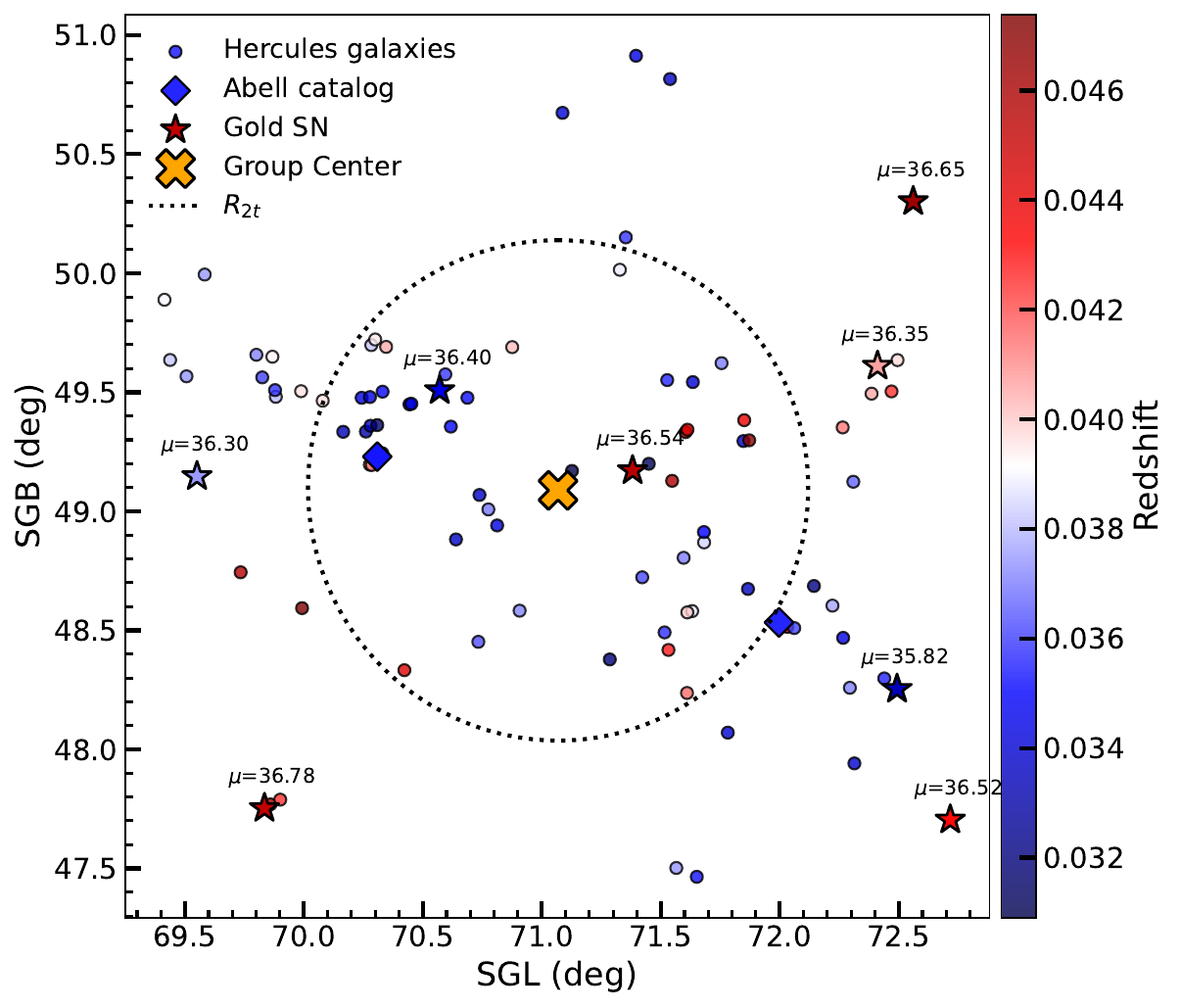}
\caption{
Spatial distribution of SNe in the Hercules complex in supergalactic coordinates. The cross marks the group center from \citetalias{tully_galaxy_2015} and the dashed circle indicates the projected second-turnaround radius $R_{2t}$. SNe are shown as stars colored according to redshift and labeled with their distance moduli. Diamond symbols identify galaxies associated with the Abell catalog.
}
\label{fig:hercules_sgl_sgb}
\end{figure}

As discussed briefly in Section~\ref{sec:sample}, the richest systems in the sample provide useful examples of the environments represented within the cousins population. Here we examine the Coma cluster and Hercules complex in greater detail. Figures~\ref{fig:coma_sgl_sgb} and \ref{fig:hercules_sgl_sgb} show the projected distributions of SNe in these systems.

The Coma cluster contains six SNe, three of which additionally satisfy the platinum selection criteria. As noted in Section~\ref{sec:sample}, all six events lie within the projected second-turnaround radius $R_{2t}$ and the system exhibits a relatively large Hubble residual scatter (${\rm RelSD}=0.356$ mag).

We additionally examine whether the SN light-curve properties are related to the measured scatter within Coma. Splitting the sample at the median SALT3 stretch ($x_1=-0.86$) yields a lower Hubble residual scatter for the higher-stretch subsample (${\rm RelSD}=0.041$~mag) than for the lower-stretch subsample ($0.156$~mag), corresponding to a difference of $0.115^{+0.041}_{-0.122}$~mag ($1.7\sigma$). Coma is also strongly dominated by early-type galaxies, with a group spiral fraction of only 0.14 in the \citetalias{tully_galaxy_2015} catalog. Although no firm conclusions can be drawn from a single system, its relatively large Hubble residual scatter is qualitatively consistent with the broader trend identified in this work that spiral-poor galaxy groups exhibit larger distance  scatter in the cousins subset.

The Hercules complex contains eight SNe, making it the richest cousins system in our sample. In contrast to Coma, the Hercules distribution is substantially more spatially extended, with five of eight SNe located beyond $R_{2t}$, although all remain within the $1.5R_{2t}$ projected group-association criterion adopted in this work. The median projected separation is $1.06R_{2t}$, and the most distant event lies at $1.44R_{2t}$. Despite this larger spatial extent, the measured Hubble residual scatter is lower than in Coma, with ${\rm RelSD}=0.140$ mag, although the difference is not necessarily statistically meaningful given the small number of events.

Unlike Coma, the Hercules complex shows no obvious trend between Hubble residuals and either projected separation or redshift. Figure~\ref{fig:hercules_sgl_sgb} additionally shows galaxies associated with the \citet{abell_catalog_1989} catalog, a catalog of rich galaxy clusters identified as significant overdensities of galaxies on the sky. The presence of multiple Abell clusters within the Hercules complex indicates that it comprises several massive substructures rather than a single, relaxed galaxy group. The cousins in the system span a redshift of $\Delta z=0.0129$, corresponding to approximately $3861~{\rm km~s^{-1}}$, substantially larger than the velocity range observed for Coma. This suggests that Hercules does not represent a single uniform gravitational potential, but instead a dynamically complex environment containing multiple cluster-scale overdensities. The large number of associated SNe located beyond $R_{2t}$ additionally raises the possibility that the system extends beyond the nominal group boundary.

A median split by SALT3 stretch ($x_1=-1.33$) likewise yields a lower scatter for the higher-stretch SNe (${\rm RelSD}=0.101$~mag) than for the lower-stretch events ($0.184$~mag) in Hercules. Hercules also has a substantially larger group spiral fraction (0.31) than Coma (0.14), indicating a more mixed galaxy population. While based on only two systems, the lower Hubble residual scatter observed in Hercules is qualitatively consistent with the broader statistical result that spiral-rich galaxy groups tend to exhibit lower cousins distance scatter.

The contrasting properties of Coma and Hercules illustrate the diversity of environments represented within the cousins sample. Although neither system individually provides statistically significant constraints, these case studies qualitatively mirror the broader trends identified in the main analysis, suggesting that the effectiveness of the cousins method depends more strongly on the overall properties of the host galaxy group than on its spatial extent alone.

\bibliography{699_Proposal}{}

@article{brout_pantheon_2022,
	title = {The {Pantheon}+ {Analysis}: {Cosmological} {Constraints}},
	volume = {938},
	issn = {0004-637X, 1538-4357},
	shorttitle = {The {Pantheon}+ {Analysis}},
	url = {https://iopscience.iop.org/article/10.3847/1538-4357/ac8e04},
	doi = {10.3847/1538-4357/ac8e04},
	number = {2},
	urldate = {2025-12-03},
	journal = {ApJ},
	author = {Brout, Dillon and Scolnic, Dan and Popovic, Brodie and Riess, Adam G. and Carr, Anthony and Zuntz, Joe and Kessler, Rick and Davis, Tamara M. and Hinton, Samuel and Jones, David and Kenworthy, W. D’Arcy and Peterson, Erik R. and Said, Khaled and Taylor, Georgie and Ali, Noor and Armstrong, Patrick and Charvu, Pranav and Dwomoh, Arianna and Meldorf, Cole and Palmese, Antonella and Qu, Helen and Rose, Benjamin M. and Sanchez, Bruno and Stubbs, Christopher W. and Vincenzi, Maria and Wood, Charlotte M. and Brown, Peter J. and Chen, Rebecca and Chambers, Ken and Coulter, David A. and Dai, Mi and Dimitriadis, Georgios and Filippenko, Alexei V. and Foley, Ryan J. and Jha, Saurabh W. and Kelsey, Lisa and Kirshner, Robert P. and Möller, Anais and Muir, Jessie and Nadathur, Seshadri and Pan, Yen-Chen and Rest, Armin and Rojas-Bravo, Cesar and Sako, Masao and Siebert, Matthew R. and Smith, Mat and Stahl, Benjamin E. and Wiseman, Phil},
	month = oct,
	year = {2022},
	pages = {110},
}

@article{carreres_growth-rate_2023,
	title = {Growth-rate measurement with type-{Ia} supernovae using {ZTF} survey simulations},
	volume = {674},
	copyright = {https://creativecommons.org/licenses/by/4.0},
	issn = {0004-6361, 1432-0746},
	url = {https://www.aanda.org/10.1051/0004-6361/202346173},
	doi = {10.1051/0004-6361/202346173},
	urldate = {2025-12-03},
	journal = {A\&A},
	author = {Carreres, Bastien and Bautista, Julian E. and Feinstein, Fabrice and Fouchez, Dominique and Racine, Benjamin and Smith, Mathew and Amenouche, Melissa and Aubert, Marie and Dhawan, Suhail and Ginolin, Madeleine and Goobar, Ariel and Gris, Philippe and Lacroix, Leander and Nuss, Eric and Regnault, Nicolas and Rigault, Mickael and Robert, Estelle and Rosnet, Philippe and Sommer, Kelian and Dekany, Richard and Groom, Steven L. and Sravan, Niharika and Masci, Frank J. and Purdum, Josiah},
	month = jun,
	year = {2023},
	pages = {A197},
}

@article{brout_its_2021,
	title = {It’s {Dust}: {Solving} the {Mysteries} of the {Intrinsic} {Scatter} and {Host}-galaxy {Dependence} of {Standardized} {Type} {Ia} {Supernova} {Brightnesses}},
	volume = {909},
	issn = {0004-637X, 1538-4357},
	shorttitle = {It’s {Dust}},
	url = {https://iopscience.iop.org/article/10.3847/1538-4357/abd69b},
	doi = {10.3847/1538-4357/abd69b},
	number = {1},
	urldate = {2025-12-03},
	journal = {ApJ},
	author = {Brout, Dillon and Scolnic, Daniel},
	month = mar,
	year = {2021},
	pages = {26},
}

@article{riess_comprehensive_2021,
	title = {A {Comprehensive} {Measurement} of the {Local} {Value} of the {Hubble} {Constant} with 1 km/s/{Mpc} {Uncertainty} from the {Hubble} {Space} {Telescope} and the {SH0ES} {Team}},
	copyright = {arXiv.org perpetual, non-exclusive license},
	url = {https://arxiv.org/abs/2112.04510},
	doi = {10.48550/ARXIV.2112.04510},
	urldate = {2025-12-03},
	publisher = {arXiv},
	author = {Riess, Adam G. and Yuan, Wenlong and Macri, Lucas M. and Scolnic, Dan and Brout, Dillon and Casertano, Stefano and Jones, David O. and Murakami, Yukei and Breuval, Louise and Brink, Thomas G. and Filippenko, Alexei V. and Hoffmann, Samantha and Jha, Saurabh W. and Kenworthy, W. D'arcy and Anand, Gagandeep and Mackenty, John and Stahl, Benjamin E. and Zheng, Weikang},
	year = {2021},
	note = {Version Number: 3},
}

@article{peterson_pantheon_2022,
	title = {The {Pantheon}+ {Analysis}: {Evaluating} {Peculiar} {Velocity} {Corrections} in {Cosmological} {Analyses} with {Nearby} {Type} {Ia} {Supernovae}},
	volume = {938},
	issn = {0004-637X, 1538-4357},
	shorttitle = {The {Pantheon}+ {Analysis}},
	url = {https://iopscience.iop.org/article/10.3847/1538-4357/ac4698},
	doi = {10.3847/1538-4357/ac4698},
	number = {2},
	urldate = {2025-12-03},
	journal = {ApJ},
	author = {Peterson, Erik R. and Kenworthy, W. D’Arcy and Scolnic, Daniel and Riess, Adam G. and Brout, Dillon and Carr, Anthony and Courtois, Hélène and Davis, Tamara and Dwomoh, Arianna and Jones, David O. and Popovic, Brodie and Rose, Benjamin M. and Said, Khaled},
	month = oct,
	year = {2022},
	pages = {112},
}

@ARTICLE{Scolnic_2020,
       author = {{Scolnic}, D. and {Smith}, M. and {Massiah}, A. and {Wiseman}, P. and {Brout}, D. and {Kessler}, R. and {Davis}, T.~M. and {Foley}, R.~J. and {Galbany}, L. and {Hinton}, S.~R. and {Hounsell}, R. and {Kelsey}, L. and {Lidman}, C. and {Macaulay}, E. and {Morgan}, R. and {Nichol}, R.~C. and {M{\"o}ller}, A. and {Popovic}, B. and {Sako}, M. and {Sullivan}, M. and {Thomas}, B.~P. and {Tucker}, B.~E. and {Abbott}, T.~M.~C. and {Aguena}, M. and {Allam}, S. and {Annis}, J. and {Avila}, S. and {Bechtol}, K. and {Bertin}, E. and {Brooks}, D. and {Burke}, D.~L. and {Rosell}, A. Carnero and {Carollo}, D. and {Kind}, M. Carrasco and {Carretero}, J. and {Costanzi}, M. and {da Costa}, L.~N. and {De Vicente}, J. and {Desai}, S. and {Diehl}, H.~T. and {Doel}, P. and {Drlica-Wagner}, A. and {Eckert}, K. and {Eifler}, T.~F. and {Everett}, S. and {Flaugher}, B. and {Fosalba}, P. and {Frieman}, J. and {Garc{\'\i}a-Bellido}, J. and {Gaztanaga}, E. and {Gerdes}, D.~W. and {Glazebrook}, K. and {Gruen}, D. and {Gruendl}, R.~A. and {Gschwend}, J. and {Gutierrez}, G. and {Hartley}, W.~G. and {Hollowood}, D.~L. and {Honscheid}, K. and {James}, D.~J. and {Kuehn}, K. and {Kuropatkin}, N. and {Lewis}, G.~F. and {Li}, T.~S. and {Lima}, M. and {Maia}, M.~A.~G. and {Marshall}, J.~L. and {Menanteau}, F. and {Miquel}, R. and {Palmese}, A. and {Paz-Chinch{\'o}n}, F. and {Plazas}, A.~A. and {Pursiainen}, M. and {Sanchez}, E. and {Scarpine}, V. and {Schubnell}, M. and {Serrano}, S. and {Sevilla-Noarbe}, I. and {Sommer}, N.~E. and {Suchyta}, E. and {Swanson}, M.~E.~C. and {Tarle}, G. and {Varga}, T.~N. and {Walker}, A.~R. and {Wilkinson}, R. and {DES Collaboration}},
        title = "{Supernova Siblings: Assessing the Consistency of Properties of Type Ia Supernovae that Share the Same Parent Galaxies}",
      journal = {\apjl},
         year = 2020,
        month = jun,
       volume = {896},
       number = {1},
          eid = {L13},
        pages = {L13},
          doi = {10.3847/2041-8213/ab8735},
archivePrefix = {arXiv},
       eprint = {2002.00974},
 primaryClass = {astro-ph.GA},
       adsurl = {https://ui.adsabs.harvard.edu/abs/2020ApJ...896L..13S}
}

@ARTICLE{Burns_2020,
       author = {{Burns}, Christopher R. and {Ashall}, Chris and {Contreras}, Carlos and {Brown}, Peter and {Stritzinger}, Maximilian and {Phillips}, M.~M. and {Flores}, Ricardo and {Suntzeff}, Nicholas B. and {Hsiao}, Eric Y. and {Uddin}, Syed and {Simon}, Joshua D. and {Krisciunas}, Kevin and {Campillay}, Abdo and {Foley}, Ryan J. and {Freedman}, Wendy L. and {Galbany}, Llu{\'\i}s and {Gonz{\'a}lez}, Consuelo and {Hoeflich}, Peter and {Holmbo}, S. and {Kilpatrick}, Charles D. and {Kirshner}, Robert P. and {Morrell}, Nidia and {Mu{\~n}oz-Elgueta}, Nahir and {Piro}, Anthony L. and {Rojas-Bravo}, C{\'e}sar and {Sand}, David and {Vargas-Gonz{\'a}lez}, Jaime and {Ulloa}, Natalie and {Vilchez}, Jorge Anais},
        title = "{SN 2013aa and SN 2017cbv: Two Sibling Type Ia Supernovae in the Spiral Galaxy NGC 5643}",
      journal = {\apj},
         year = 2020,
        month = jun,
       volume = {895},
       number = {2},
          eid = {118},
        pages = {118},
          doi = {10.3847/1538-4357/ab8e3e},
archivePrefix = {arXiv},
       eprint = {2004.13069},
 primaryClass = {astro-ph.SR},
       adsurl = {https://ui.adsabs.harvard.edu/abs/2020ApJ...895..118B}
}

@ARTICLE{Davis_2011,
       author = {{Davis}, Tamara M. and {Hui}, Lam and {Frieman}, Joshua A. and {Haugb{\o}lle}, Troels and {Kessler}, Richard and {Sinclair}, Benjamin and {Sollerman}, Jesper and {Bassett}, Bruce and {Marriner}, John and {M{\"o}rtsell}, Edvard and {Nichol}, Robert C. and {Richmond}, Michael W. and {Sako}, Masao and {Schneider}, Donald P. and {Smith}, Mathew},
        title = "{The Effect of Peculiar Velocities on Supernova Cosmology}",
      journal = {\apj},
         year = 2011,
        month = nov,
       volume = {741},
       number = {1},
          eid = {67},
        pages = {67},
          doi = {10.1088/0004-637X/741/1/67},
archivePrefix = {arXiv},
       eprint = {1012.2912},
 primaryClass = {astro-ph.CO},
       adsurl = {https://ui.adsabs.harvard.edu/abs/2011ApJ...741...67D}
}

@article{carreres_ztf_2025,
	title = {{ZTF} {SN} {Ia} {DR2}: {Peculiar} velocities’ impact on the {Hubble} diagram},
	volume = {694},
	copyright = {https://creativecommons.org/licenses/by/4.0},
	issn = {0004-6361, 1432-0746},
	shorttitle = {{ZTF} {SN} {Ia} {DR2}},
	url = {https://www.aanda.org/10.1051/0004-6361/202450389},
	doi = {10.1051/0004-6361/202450389},
	urldate = {2025-12-03},
	journal = {A\&A},
	author = {Carreres, B. and Rosselli, D. and Bautista, J. E. and Feinstein, F. and Fouchez, D. and Racine, B. and Ravoux, C. and Sanchez, B. and Dimitriadis, G. and Goobar, A. and Johansson, J. and Nordin, J. and Rigault, M. and Smith, M. and Amenouche, M. and Aubert, M. and Barjou-Delayre, C. and Burgaz, U. and D’Arcy Kenworthy, W. and De Jaeger, T. and Dhawan, S. and Galbany, L. and Ginolin, M. and Kuhn, D. and Kowalski, M. and Müller-Bravo, T. E. and Nugent, P. E. and Popovic, B. and Rosnet, P. and Ruppin, F. and Sollerman, J. and Terwel, J. H. and Townsend, A. and Groom, S. L. and Kulkarni, S. R. and Purdum, J. and Rusholme, B. and Sravan, N.},
	month = feb,
	year = {2025},
	pages = {A8},
}

@ARTICLE{Shingles21,
       author = {{Shingles}, L. and {Smith}, K.~W. and {Young}, D.~R. and {Smartt}, S.~J. and {Tonry}, J. and {Denneau}, L. and {Heinze}, A. and {Weiland}, H. and {Flewelling}, H. and {Stalder}, B. and {Clocchiatti}, A. and {F{\"o}rster}, F. and {Pignata}, G. and {Rest}, A. and {Anderson}, J. and {Stubbs}, C. and {Erasmus}, N.},
        title = "{Release of the ATLAS Forced Photometry server for public use}",
      journal = {Transient Name Server AstroNote},
         year = 2021,
        month = jan,
       volume = {7},
        pages = {1-7},
       adsurl = {https://ui.adsabs.harvard.edu/abs/2021TNSAN...7....1S}
}

@article{kelly_hubble_2010,
	title = {Hubble {Residuals} of {Nearby} {Type} {Ia} {Supernovae} are {Correlated} with {Host} {Galaxy} {Masses}},
	volume = {715},
	issn = {0004-637X},
	url = {https://ui.adsabs.harvard.edu/abs/2010ApJ...715..743K},
	doi = {10.1088/0004-637X/715/2/743},
	urldate = {2025-12-03},
	journal = {The Astrophysical Journal},
	publisher = {IOP},
	author = {Kelly, Patrick L. and Hicken, Malcolm and Burke, David L. and Mandel, Kaisey S. and Kirshner, Robert P.},
	month = jun,
	year = {2010},
	note = {ADS Bibcode: 2010ApJ...715..743K},
	pages = {743--756},
}

@ARTICLE{Hoogendam_2022,
       author = {{Hoogendam}, W.~B. and {Ashall}, C. and {Galbany}, L. and {Shappee}, B.~J. and {Burns}, C.~R. and {Lu}, J. and {Phillips}, M.~M. and {Baron}, E. and {Holmbo}, S. and {Hsiao}, E.~Y. and {Morrell}, N. and {Stritzinger}, M.~D. and {Suntzeff}, N.~B. and {Taddia}, F. and {Young}, D.~R. and {Lyman}, J.~D. and {Benetti}, S. and {Mazzali}, P.~A. and {Delgado Manche{\~n}o}, M. and {D{\'\i}az}, R. Gonz{\'a}lez and {Torres}, S. Mu{\~n}oz},
        title = "{A Tale of Two Type Ia Supernovae: The Fast-declining Siblings SNe 2015bo and 1997cn}",
      journal = {\apj},
         year = 2022,
        month = apr,
       volume = {928},
       number = {2},
          eid = {103},
        pages = {103},
          doi = {10.3847/1538-4357/ac54aa},
archivePrefix = {arXiv},
       eprint = {2109.14644},
 primaryClass = {astro-ph.HE},
       adsurl = {https://ui.adsabs.harvard.edu/abs/2022ApJ...928..103H}
}

@article{sullivan_dependence_2010,
	title = {The dependence of {Type} {Ia} {Supernovae} luminosities on their host galaxies},
	volume = {406},
	issn = {0035-8711},
	url = {https://ui.adsabs.harvard.edu/abs/2010MNRAS.406..782S},
	doi = {10.1111/j.1365-2966.2010.16731.x},
	urldate = {2025-12-03},
	journal = {Monthly Notices of the Royal Astronomical Society},
	publisher = {OUP},
	author = {Sullivan, M. and Conley, A. and Howell, D. A. and Neill, J. D. and Astier, P. and Balland, C. and Basa, S. and Carlberg, R. G. and Fouchez, D. and Guy, J. and Hardin, D. and Hook, I. M. and Pain, R. and Palanque-Delabrouille, N. and Perrett, K. M. and Pritchet, C. J. and Regnault, N. and Rich, J. and Ruhlmann-Kleider, V. and Baumont, S. and Hsiao, E. and Kronborg, T. and Lidman, C. and Perlmutter, S. and Walker, E. S.},
	month = aug,
	year = {2010},
	note = {ADS Bibcode: 2010MNRAS.406..782S},
	pages = {782--802},
}

@article{wiseman_rates_2021,
	title = {Rates and delay times of type {Ia} supernovae in the {Dark} {Energy} {Survey}},
	copyright = {https://academic.oup.com/journals/pages/open\_access/funder\_policies/chorus/standard\_publication\_model},
	issn = {0035-8711, 1365-2966},
	url = {https://academic.oup.com/mnras/advance-article/doi/10.1093/mnras/stab1943/6318383},
	doi = {10.1093/mnras/stab1943},
	language = {en},
	urldate = {2025-12-03},
	journal = {Monthly Notices of the Royal Astronomical Society},
	author = {Wiseman, P and Sullivan, M and Smith, M and Frohmaier, C and Vincenzi, M and Graur, O and Popovic, B and Armstrong, P and Brout, D and Davis, T M and Galbany, L and Hinton, S R and Kelsey, L and Kessler, R and Lidman, C and Möller, A and Nichol, R C and Rose, B and Scolnic, D and Toy, M and Zontou, Z and Asorey, J and Carollo, D and Glazebrook, K and Lewis, G F and Tucker, B E and Abbott, T M C and Aguena, M and Allam, S and Andrade-Oliveira, F and Annis, J and Bacon, D and Bertin, E and Brooks, D and Buckley-Geer, E and Burke, D L and Rosell, A Carnero and Kind, M Carrasco and Carretero, J and Costanzi, M and Da Costa, L N and Pereira, M E S and Desai, S and Diehl, H T and Doel, P and Everett, S and Ferrero, I and Flaugher, B and Fosalba, P and Frieman, J and García-Bellido, J and Gaztanaga, E and Giannantonio, T and Gruen, D and Gruendl, R A and Gschwend, J and Gutierrez, G and Hollowood, D L and Honscheid, K and Hoyle, B and James, D J and Krause, E and Kuehn, K and Kuropatkin, N and Maia, M A G and Marshall, J L and Martini, P and Menanteau, F and Miquel, R and Morgan, R and Ogando, R L C and Palmese, A and Paz-Chinchón, F and Petravick, D and Pieres, A and Malagón, A A Plazas and Romer, A K and Sanchez, E and Scarpine, V and Schubnell, M and Serrano, S and Sevilla-Noarbe, I and Soares-Santos, M and Suchyta, E and Swanson, M E C and Tarle, G and Thomas, D and To, C and Varga, T N and Walker, A R},
	month = jul,
	year = {2021},
	pages = {stab1943},
}

@article{scolnic_pantheon_2022,
	title = {The {Pantheon}+ {Analysis}: {The} {Full} {Data} {Set} and {Light}-curve {Release}},
	volume = {938},
	issn = {0004-637X, 1538-4357},
	shorttitle = {The {Pantheon}+ {Analysis}},
	url = {https://iopscience.iop.org/article/10.3847/1538-4357/ac8b7a},
	doi = {10.3847/1538-4357/ac8b7a},
	number = {2},
	urldate = {2025-12-03},
	journal = {ApJ},
	author = {Scolnic, Dan and Brout, Dillon and Carr, Anthony and Riess, Adam G. and Davis, Tamara M. and Dwomoh, Arianna and Jones, David O. and Ali, Noor and Charvu, Pranav and Chen, Rebecca and Peterson, Erik R. and Popovic, Brodie and Rose, Benjamin M. and Wood, Charlotte M. and Brown, Peter J. and Chambers, Ken and Coulter, David A. and Dettman, Kyle G. and Dimitriadis, Georgios and Filippenko, Alexei V. and Foley, Ryan J. and Jha, Saurabh W. and Kilpatrick, Charles D. and Kirshner, Robert P. and Pan, Yen-Chen and Rest, Armin and Rojas-Bravo, Cesar and Siebert, Matthew R. and Stahl, Benjamin E. and Zheng, WeiKang},
	month = oct,
	year = {2022},
	pages = {113},
}

@article{larison_environmental_2024,
	title = {Environmental {Dependence} of {Type} {Ia} {Supernovae} in {Low}-redshift {Galaxy} {Clusters}},
	volume = {961},
	issn = {0004-637X, 1538-4357},
	url = {https://iopscience.iop.org/article/10.3847/1538-4357/ad0e0f},
	doi = {10.3847/1538-4357/ad0e0f},
	number = {2},
	urldate = {2025-12-03},
	journal = {ApJ},
	author = {Larison, Conor and Jha, Saurabh W. and Kwok, Lindsey A. and Camacho-Neves, Yssavo},
	month = feb,
	year = {2024},
	pages = {185},
}

@article{tonry_atlas_2018,
	title = {{ATLAS}: {A} {High}-cadence {All}-sky {Survey} {System}},
	volume = {130},
	issn = {0004-6280},
	shorttitle = {{ATLAS}},
	url = {https://ui.adsabs.harvard.edu/abs/2018PASP..130f4505T},
	doi = {10.1088/1538-3873/aabadf},
	urldate = {2025-12-03},
	journal = {Publications of the Astronomical Society of the Pacific},
	publisher = {IOP},
	author = {Tonry, J. L. and Denneau, L. and Heinze, A. N. and Stalder, B. and Smith, K. W. and Smartt, S. J. and Stubbs, C. W. and Weiland, H. J. and Rest, A.},
	month = jun,
	year = {2018},
	note = {ADS Bibcode: 2018PASP..130f4505T},
	pages = {064505},
}

@article{tully_galaxy_2015,
	title = {{GALAXY} {GROUPS}: {A} {2MASS} {CATALOG}},
	volume = {149},
	copyright = {http://iopscience.iop.org/info/page/text-and-data-mining},
	issn = {1538-3881},
	shorttitle = {{GALAXY} {GROUPS}},
	url = {https://iopscience.iop.org/article/10.1088/0004-6256/149/5/171},
	doi = {10.1088/0004-6256/149/5/171},
	number = {5},
	urldate = {2025-12-03},
	journal = {AJ},
	author = {Tully, R. Brent},
	month = apr,
	year = {2015},
	pages = {171},
}

@article{kenworthy_salt3_2021,
	title = {{SALT3}: {An} {Improved} {Type} {Ia} {Supernova} {Model} for {Measuring} {Cosmic} {Distances}},
	volume = {923},
	issn = {0004-637X, 1538-4357},
	shorttitle = {{SALT3}},
	url = {https://iopscience.iop.org/article/10.3847/1538-4357/ac30d8},
	doi = {10.3847/1538-4357/ac30d8},
	number = {2},
	urldate = {2025-12-03},
	journal = {ApJ},
	author = {Kenworthy, W. D. and Jones, D. O. and Dai, M. and Kessler, R. and Scolnic, D. and Brout, D. and Siebert, M. R. and Pierel, J. D. R. and Dettman, K. G. and Dimitriadis, G. and Foley, R. J. and Jha, S. W. and Pan, Y.-C. and Riess, A. and Rodney, S. and Rojas-Bravo, C.},
	month = dec,
	year = {2021},
	pages = {265},
}

@article{lampeitl_effect_2010,
	title = {The {Effect} of {Host} {Galaxies} on {Type} {Ia} {Supernovae} in the {SDSS}-{II} {Supernova} {Survey}},
	volume = {722},
	issn = {0004-637X},
	url = {https://ui.adsabs.harvard.edu/abs/2010ApJ...722..566L},
	doi = {10.1088/0004-637X/722/1/566},
	urldate = {2025-12-03},
	journal = {The Astrophysical Journal},
	publisher = {IOP},
	author = {Lampeitl, Hubert and Smith, Mathew and Nichol, Robert C. and Bassett, Bruce and Cinabro, David and Dilday, Benjamin and Foley, Ryan J. and Frieman, Joshua A. and Garnavich, Peter M. and Goobar, Ariel and Im, Myungshin and Jha, Saurabh W. and Marriner, John and Miquel, Ramon and Nordin, Jakob and Östman, Linda and Riess, Adam G. and Sako, Masao and Schneider, Donald P. and Sollerman, Jesper and Stritzinger, Maximilian},
	month = oct,
	year = {2010},
	note = {ADS Bibcode: 2010ApJ...722..566L},
	pages = {566--576},
}

@article{guy_supernova_2010,
	title = {The {Supernova} {Legacy} {Survey} 3-year sample: {Type} {Ia} supernovae photometric distances and cosmological constraints},
	volume = {523},
	issn = {0004-6361, 1432-0746},
	shorttitle = {The {Supernova} {Legacy} {Survey} 3-year sample},
	url = {http://www.aanda.org/10.1051/0004-6361/201014468},
	doi = {10.1051/0004-6361/201014468},
	urldate = {2025-12-09},
	journal = {A\&A},
	author = {Guy, J. and Sullivan, M. and Conley, A. and Regnault, N. and Astier, P. and Balland, C. and Basa, S. and Carlberg, R. G. and Fouchez, D. and Hardin, D. and Hook, I. M. and Howell, D. A. and Pain, R. and Palanque-Delabrouille, N. and Perrett, K. M. and Pritchet, C. J. and Rich, J. and Ruhlmann-Kleider, V. and Balam, D. and Baumont, S. and Ellis, R. S. and Fabbro, S. and Fakhouri, H. K. and Fourmanoit, N. and González-Gaitán, S. and Graham, M. L. and Hsiao, E. and Kronborg, T. and Lidman, C. and Mourao, A. M. and Perlmutter, S. and Ripoche, P. and Suzuki, N. and Walker, E. S.},
	month = nov,
	year = {2010},
	pages = {A7},
}

@article{planck_collaboration_planck_2020,
	title = {\textit{{Planck}} 2018 results: {VI}. {Cosmological} parameters},
	volume = {641},
	copyright = {https://www.edpsciences.org/en/authors/copyright-and-licensing},
	issn = {0004-6361, 1432-0746},
	shorttitle = {\textit{{Planck}} 2018 results},
	url = {https://www.aanda.org/10.1051/0004-6361/201833910},
	doi = {10.1051/0004-6361/201833910},
	urldate = {2025-12-09},
	journal = {A\&A},
	author = {{Planck Collaboration} and Aghanim, N. and Akrami, Y. and Ashdown, M. and Aumont, J. and Baccigalupi, C. and Ballardini, M. and Banday, A. J. and Barreiro, R. B. and Bartolo, N. and Basak, S. and Battye, R. and Benabed, K. and Bernard, J.-P. and Bersanelli, M. and Bielewicz, P. and Bock, J. J. and Bond, J. R. and Borrill, J. and Bouchet, F. R. and Boulanger, F. and Bucher, M. and Burigana, C. and Butler, R. C. and Calabrese, E. and Cardoso, J.-F. and Carron, J. and Challinor, A. and Chiang, H. C. and Chluba, J. and Colombo, L. P. L. and Combet, C. and Contreras, D. and Crill, B. P. and Cuttaia, F. and De Bernardis, P. and De Zotti, G. and Delabrouille, J. and Delouis, J.-M. and Di Valentino, E. and Diego, J. M. and Doré, O. and Douspis, M. and Ducout, A. and Dupac, X. and Dusini, S. and Efstathiou, G. and Elsner, F. and Enßlin, T. A. and Eriksen, H. K. and Fantaye, Y. and Farhang, M. and Fergusson, J. and Fernandez-Cobos, R. and Finelli, F. and Forastieri, F. and Frailis, M. and Fraisse, A. A. and Franceschi, E. and Frolov, A. and Galeotta, S. and Galli, S. and Ganga, K. and Génova-Santos, R. T. and Gerbino, M. and Ghosh, T. and González-Nuevo, J. and Górski, K. M. and Gratton, S. and Gruppuso, A. and Gudmundsson, J. E. and Hamann, J. and Handley, W. and Hansen, F. K. and Herranz, D. and Hildebrandt, S. R. and Hivon, E. and Huang, Z. and Jaffe, A. H. and Jones, W. C. and Karakci, A. and Keihänen, E. and Keskitalo, R. and Kiiveri, K. and Kim, J. and Kisner, T. S. and Knox, L. and Krachmalnicoff, N. and Kunz, M. and Kurki-Suonio, H. and Lagache, G. and Lamarre, J.-M. and Lasenby, A. and Lattanzi, M. and Lawrence, C. R. and Le Jeune, M. and Lemos, P. and Lesgourgues, J. and Levrier, F. and Lewis, A. and Liguori, M. and Lilje, P. B. and Lilley, M. and Lindholm, V. and López-Caniego, M. and Lubin, P. M. and Ma, Y.-Z. and Macías-Pérez, J. F. and Maggio, G. and Maino, D. and Mandolesi, N. and Mangilli, A. and Marcos-Caballero, A. and Maris, M. and Martin, P. G. and Martinelli, M. and Martínez-González, E. and Matarrese, S. and Mauri, N. and McEwen, J. D. and Meinhold, P. R. and Melchiorri, A. and Mennella, A. and Migliaccio, M. and Millea, M. and Mitra, S. and Miville-Deschênes, M.-A. and Molinari, D. and Montier, L. and Morgante, G. and Moss, A. and Natoli, P. and Nørgaard-Nielsen, H. U. and Pagano, L. and Paoletti, D. and Partridge, B. and Patanchon, G. and Peiris, H. V. and Perrotta, F. and Pettorino, V. and Piacentini, F. and Polastri, L. and Polenta, G. and Puget, J.-L. and Rachen, J. P. and Reinecke, M. and Remazeilles, M. and Renzi, A. and Rocha, G. and Rosset, C. and Roudier, G. and Rubiño-Martín, J. A. and Ruiz-Granados, B. and Salvati, L. and Sandri, M. and Savelainen, M. and Scott, D. and Shellard, E. P. S. and Sirignano, C. and Sirri, G. and Spencer, L. D. and Sunyaev, R. and Suur-Uski, A.-S. and Tauber, J. A. and Tavagnacco, D. and Tenti, M. and Toffolatti, L. and Tomasi, M. and Trombetti, T. and Valenziano, L. and Valiviita, J. and Van Tent, B. and Vibert, L. and Vielva, P. and Villa, F. and Vittorio, N. and Wandelt, B. D. and Wehus, I. K. and White, M. and White, S. D. M. and Zacchei, A. and Zonca, A.},
	month = sep,
	year = {2020},
	pages = {A6},
}

@article{childress_host_2013,
	title = {Host {Galaxy} {Properties} and {Hubble} {Residuals} of {Type} {Ia} {Supernovae} from the {Nearby} {Supernova} {Factory}},
	volume = {770},
	issn = {0004-637X},
	url = {https://ui.adsabs.harvard.edu/abs/2013ApJ...770..108C},
	doi = {10.1088/0004-637X/770/2/108},
	urldate = {2025-12-09},
	journal = {The Astrophysical Journal},
	publisher = {IOP},
	author = {Childress, M. and Aldering, G. and Antilogus, P. and Aragon, C. and Bailey, S. and Baltay, C. and Bongard, S. and Buton, C. and Canto, A. and Cellier-Holzem, F. and Chotard, N. and Copin, Y. and Fakhouri, H. K. and Gangler, E. and Guy, J. and Hsiao, E. Y. and Kerschhaggl, M. and Kim, A. G. and Kowalski, M. and Loken, S. and Nugent, P. and Paech, K. and Pain, R. and Pecontal, E. and Pereira, R. and Perlmutter, S. and Rabinowitz, D. and Rigault, M. and Runge, K. and Scalzo, R. and Smadja, G. and Tao, C. and Thomas, R. C. and Weaver, B. A. and Wu, C.},
	month = jun,
	year = {2013},
	note = {ADS Bibcode: 2013ApJ...770..108C},
	pages = {108},
}

@article{peterson_improving_2025,
	title = {Improving the {Determination} of {Supernova} {Cosmological} {Redshifts} by {Using} {Galaxy} {Groups}},
	volume = {980},
	issn = {0004-637X, 1538-4357},
	url = {https://iopscience.iop.org/article/10.3847/1538-4357/ada285},
	doi = {10.3847/1538-4357/ada285},
	number = {1},
	urldate = {2025-12-09},
	journal = {ApJ},
	author = {Peterson, Erik R. and Carreres, Bastien and Carr, Anthony and Scolnic, Daniel and Bailey, Ava and Davis, Tamara M. and Brout, Dillon and Howlett, Cullan and Jones, David O. and Riess, Adam G. and Said, Khaled and Taylor, Georgie},
	month = feb,
	year = {2025},
	pages = {21},
}

@article{ginolin_ztf_2025,
	title = {{ZTF} {SN} {Ia} {DR2}: {Environmental} dependencies of stretch and luminosity for a volume-limited sample of 1000 type {Ia} supernovae},
	volume = {695},
	copyright = {https://creativecommons.org/licenses/by/4.0},
	issn = {0004-6361, 1432-0746},
	shorttitle = {{ZTF} {SN} {Ia} {DR2}},
	url = {https://www.aanda.org/10.1051/0004-6361/202450378},
	doi = {10.1051/0004-6361/202450378},
	urldate = {2026-01-27},
	journal = {A\&A},
	author = {Ginolin, M. and Rigault, M. and Smith, M. and Copin, Y. and Ruppin, F. and Dimitriadis, G. and Goobar, A. and Johansson, J. and Maguire, K. and Nordin, J. and Amenouche, M. and Aubert, M. and Barjou-Delayre, C. and Betoule, M. and Burgaz, U. and Carreres, B. and Deckers, M. and Dhawan, S. and Feinstein, F. and Fouchez, D. and Galbany, L. and Ganot, C. and Harvey, L. and De Jaeger, T. and Kenworthy, W. D. and Kim, Y.-L. and Kowalski, M. and Kuhn, D. and Lacroix, L. and Müller-Bravo, T. E. and Nugent, P. and Popovic, B. and Racine, B. and Rosnet, P. and Rosselli, D. and Sollerman, J. and Terwel, J. H. and Townsend, A. and Brugger, J. and Bellm, E. C. and Kasliwal, M. M. and Kulkarni, S. and Laher, R. R. and Masci, F. J. and Riddle, R. L. and Sharma, Y.},
	month = mar,
	year = {2025},
	pages = {A140},
}

@article{dhawan_ztf_2025,
	title = {{ZTF} {SN} {Ia} {DR2}: {Cosmology}-independent constraints on {Type} {Ia} supernova standardisation from supernova siblings},
	volume = {702},
	issn = {0004-6361},
	shorttitle = {{ZTF} {SN} {Ia} {DR2}},
	url = {https://ui.adsabs.harvard.edu/abs/2025A&A...702A.190D},
	doi = {10.1051/0004-6361/202450392},
	urldate = {2026-02-11},
	journal = {Astronomy and Astrophysics},
	publisher = {EDP},
	author = {Dhawan, S. and Mortsell, E. and Johansson, J. and Goobar, A. and Rigault, M. and Smith, M. and Maguire, K. and Nordin, J. and Dimitriadis, G. and Nugent, P. E. and Galbany, L. and Sollerman, J. and Kenworthy, W. D. and de Jaeger, T. and Terwel, J. H. and Kim, Y.-L. and Burgaz, U. and Rosnet, P. and Helou, G. and Purdum, J. and Groom, S. L. and Laher, R. and Healy, B.},
	month = oct,
	year = {2025},
	note = {ADS Bibcode: 2025A\&A...702A.190D},
	pages = {A190},
}

@article{dwomoh_evaluating_2024,
	title = {Evaluating the {Consistency} of {Cosmological} {Distances} {Using} {Supernova} {Siblings} in the {Near}-infrared},
	volume = {965},
	issn = {0004-637X, 1538-4357},
	url = {https://iopscience.iop.org/article/10.3847/1538-4357/ad1ff5},
	doi = {10.3847/1538-4357/ad1ff5},
	number = {1},
	urldate = {2026-02-14},
	journal = {ApJ},
	author = {Dwomoh, Arianna M. and Peterson, Erik R. and Scolnic, Daniel and Ashall, Chris and DerKacy, James M. and Do, Aaron and Johansson, Joel and Jones, David O. and Riess, Adam G. and Shappee, Benjamin J.},
	month = apr,
	year = {2024},
	pages = {90},
}

@article{tonry_vizier_2021,
	title = {{VizieR} {Online} {Data} {Catalog}: {ATLAS} all-sky stellar ref. catalog, {ATLAS}-{REFCAT2} ({Tonry}+, 2018)},
	volume = {186},
	shorttitle = {{VizieR} {Online} {Data} {Catalog}},
	url = {https://ui.adsabs.harvard.edu/abs/2021yCat..18670105T},
	doi = {10.26093/cds/vizier.18670105},
	urldate = {2026-02-19},
	journal = {VizieR Online Data Catalog},
	author = {Tonry, J. L. and Denneau, L. and Flewelling, H. and Heinze, A. N. and Onken, C. A. and Smartt, S. J. and Stalder, B. and Weiland, H. J. and Wolf, C.},
	month = mar,
	year = {2021},
	note = {ADS Bibcode: 2021yCat..18670105T},
	pages = {J/ApJ/867/105},
}

@article{huchra_2mass_2012,
	title = {The {2MASS} {Redshift} {Survey}—{Description} and {Data} {Release}},
	volume = {199},
	issn = {0067-0049},
	url = {https://ui.adsabs.harvard.edu/abs/2012ApJS..199...26H},
	doi = {10.1088/0067-0049/199/2/26},
	urldate = {2026-02-19},
	journal = {The Astrophysical Journal Supplement Series},
	publisher = {IOP},
	author = {Huchra, John P. and Macri, Lucas M. and Masters, Karen L. and Jarrett, Thomas H. and Berlind, Perry and Calkins, Michael and Crook, Aidan C. and Cutri, Roc and Erdoǧdu, Pirin and Falco, Emilio and George, Teddy and Hutcheson, Conrad M. and Lahav, Ofer and Mader, Jeff and Mink, Jessica D. and Martimbeau, Nathalie and Schneider, Stephen and Skrutskie, Michael and Tokarz, Susan and Westover, Michael},
	month = apr,
	year = {2012},
	note = {ADS Bibcode: 2012ApJS..199...26H},
	pages = {26},
}

@article{dressler_galaxy_1980,
	title = {Galaxy morphology in rich clusters - {Implications} for the formation and evolution of galaxies},
	volume = {236},
	issn = {0004-637X, 1538-4357},
	url = {http://adsabs.harvard.edu/doi/10.1086/157753},
	doi = {10.1086/157753},
	language = {en},
	urldate = {2026-05-27},
	journal = {ApJ},
	author = {Dressler, A.},
	month = mar,
	year = {1980},
	pages = {351},
}

@misc{marlin_titan_2025,
	title = {{TITAN} {DR1}: {An} {Improved}, {Validated}, and {Systematically}-{Controlled} {Recalibration} of {ATLAS} {Photometry} toward {Type} {Ia} {Supernova} {Cosmology}},
	copyright = {arXiv.org perpetual, non-exclusive license},
	shorttitle = {{TITAN} {DR1}},
	url = {https://arxiv.org/abs/2512.21903},
	doi = {10.48550/ARXIV.2512.21903},
	urldate = {2026-05-28},
	publisher = {arXiv},
	author = {Marlin, Elijah G. and Murakami, Yukei S. and Brout, Dillon and Tweddle, Jack W. and Popovic, Brodie and Smith, Ken W. and Smartt, Stephen J. and Scolnic, Daniel M. and Jones, David and Peterson, Erik R. and Riess, Adam G. and Vincenzi, Maria and Sherman, Nora F. and Acevedo, Maria and Milstein, Jasper and Dixon, Mitchell and Rest, Armin},
	year = {2025},
	note = {Version Number: 1},
}

@article{blondin_determining_2007,
	title = {Determining the {Type}, {Redshift}, and {Age} of a {Supernova} {Spectrum}},
	volume = {666},
	issn = {0004-637X, 1538-4357},
	url = {http://arxiv.org/abs/0709.4488},
	doi = {10.1086/520494},
	number = {2},
	urldate = {2026-05-28},
	journal = {ApJ},
	author = {Blondin, Stéphane and Tonry, John L.},
	month = sep,
	year = {2007},
	note = {arXiv:0709.4488 [astro-ph]},
	pages = {1024--1047},
}

@article{foreman-mackey_emcee_2013,
	title = {emcee: {The} {MCMC} {Hammer}},
	volume = {125},
	issn = {00046280, 15383873},
	shorttitle = {emcee},
	url = {http://arxiv.org/abs/1202.3665},
	doi = {10.1086/670067},
	number = {925},
	urldate = {2026-05-28},
	journal = {Publications of the Astronomical Society of the Pacific},
	author = {Foreman-Mackey, Daniel and Hogg, David W. and Lang, Dustin and Goodman, Jonathan},
	month = mar,
	year = {2013},
	note = {arXiv:1202.3665 [astro-ph.IM]},
	pages = {306--312},
}

@article{abbott_dark_2024,
	title = {The {Dark} {Energy} {Survey}: {Cosmology} {Results} with ∼1500 {New} {High}-redshift {Type} {Ia} {Supernovae} {Using} the {Full} 5 yr {Data} {Set}},
	volume = {973},
	issn = {2041-8205, 2041-8213},
	shorttitle = {The {Dark} {Energy} {Survey}},
	url = {https://iopscience.iop.org/article/10.3847/2041-8213/ad6f9f},
	doi = {10.3847/2041-8213/ad6f9f},
	number = {1},
	urldate = {2026-06-11},
	journal = {ApJL},
	author = { {DES Collaboration} and Abbott, Des Collaboration: T. M. C. and Acevedo, M. and Aguena, M. and Alarcon, A. and Allam, S. and Alves, O. and Amon, A. and Andrade-Oliveira, F. and Annis, J. and Armstrong, P. and Asorey, J. and Avila, S. and Bacon, D. and Bassett, B. A. and Bechtol, K. and Bernardinelli, P. H. and Bernstein, G. M. and Bertin, E. and Blazek, J. and Bocquet, S. and Brooks, D. and Brout, D. and Buckley-Geer, E. and Burke, D. L. and Camacho, H. and Camilleri, R. and Campos, A. and Carnero Rosell, A. and Carollo, D. and Carr, A. and Carretero, J. and Castander, F. J. and Cawthon, R. and Chang, C. and Chen, R. and Choi, A. and Conselice, C. and Costanzi, M. and Da Costa, L. N. and Crocce, M. and Davis, T. M. and DePoy, D. L. and Desai, S. and Diehl, H. T. and Dixon, M. and Dodelson, S. and Doel, P. and Doux, C. and Drlica-Wagner, A. and Elvin-Poole, J. and Everett, S. and Ferrero, I. and Ferté, A. and Flaugher, B. and Foley, R. J. and Fosalba, P. and Friedel, D. and Frieman, J. and Frohmaier, C. and Galbany, L. and García-Bellido, J. and Gatti, M. and Gaztanaga, E. and Giannini, G. and Glazebrook, K. and Graur, O. and Gruen, D. and Gruendl, R. A. and Gutierrez, G. and Hartley, W. G. and Herner, K. and Hinton, S. R. and Hollowood, D. L. and Honscheid, K. and Huterer, D. and Jain, B. and James, D. J. and Jeffrey, N. and Kasai, E. and Kelsey, L. and Kent, S. and Kessler, R. and Kim, A. G. and Kirshner, R. P. and Kovacs, E. and Kuehn, K. and Lahav, O. and Lee, J. and Lee, S. and Lewis, G. F. and Li, T. S. and Lidman, C. and Lin, H. and Malik, U. and Marshall, J. L. and Martini, P. and Mena-Fernández, J. and Menanteau, F. and Miquel, R. and Mohr, J. J. and Mould, J. and Muir, J. and Möller, A. and Neilsen, E. and Nichol, R. C. and Nugent, P. and Ogando, R. L. C. and Palmese, A. and Pan, Y.-C. and Paterno, M. and Percival, W. J. and Pereira, M. E. S. and Pieres, A. and Plazas Malagón, A. A. and Popovic, B. and Porredon, A. and Prat, J. and Qu, H. and Raveri, M. and Rodríguez-Monroy, M. and Romer, A. K. and Roodman, A. and Rose, B. and Sako, M. and Sanchez, E. and Sanchez Cid, D. and Schubnell, M. and Scolnic, D. and Sevilla-Noarbe, I. and Shah, P. and Smith, J. Allyn. and Smith, M. and Soares-Santos, M. and Suchyta, E. and Sullivan, M. and Suntzeff, N. and Swanson, M. E. C. and Sánchez, B. O. and Tarle, G. and Taylor, G. and Thomas, D. and To, C. and Toy, M. and Troxel, M. A. and Tucker, B. E. and Tucker, D. L. and Uddin, S. A. and Vincenzi, M. and Walker, A. R. and Weaverdyck, N. and Wechsler, R. H. and Weller, J. and Wester, W. and Wiseman, P. and Yamamoto, M. and Yuan, F. and Zhang, B. and Zhang, Y.},
	month = sep,
	year = {2024},
	pages = {L14},
}

@ARTICLE{smith_20202020PASP..132h5002S,
       author = {{Smith}, K.~W. and {Smartt}, S.~J. and {Young}, D.~R. and {Tonry}, J.~L. and {Denneau}, L. and {Flewelling}, H. and {Heinze}, A.~N. and {Weiland}, H.~J. and {Stalder}, B. and {Rest}, A. and {Stubbs}, C.~W. and {Anderson}, J.~P. and {Chen}, T.-W. and {Clark}, P. and {Do}, A. and {F{\"o}rster}, F. and {Fulton}, M. and {Gillanders}, J. and {McBrien}, O.~R. and {O'Neill}, D. and {Srivastav}, S. and {Wright}, D.~E.},
        title = "{Design and Operation of the ATLAS Transient Science Server}",
      journal = {\pasp},
         year = 2020,
        month = aug,
       volume = {132},
       number = {1014},
          eid = {085002},
        pages = {085002},
          doi = {10.1088/1538-3873/ab936e},
archivePrefix = {arXiv},
       eprint = {2003.09052},
 primaryClass = {astro-ph.IM},
       adsurl = {https://ui.adsabs.harvard.edu/abs/2020PASP..132h5002S}
}

@article{peng_mass_2010,
	title = {{MASS} {AND} {ENVIRONMENT} {AS} {DRIVERS} {OF} {GALAXY} {EVOLUTION} {IN} {SDSS} {AND} {zCOSMOS} {AND} {THE} {ORIGIN} {OF} {THE} {SCHECHTER} {FUNCTION}},
	volume = {721},
	issn = {0004-637X, 1538-4357},
	url = {https://iopscience.iop.org/article/10.1088/0004-637X/721/1/193},
	doi = {10.1088/0004-637X/721/1/193},
	number = {1},
	urldate = {2026-06-12},
	journal = {ApJ},
	author = {Peng, Ying-Jie and Lilly, Simon J. and Kovač, Katarina and Bolzonella, Micol and Pozzetti, Lucia and Renzini, Alvio and Zamorani, Gianni and Ilbert, Olivier and Knobel, Christian and Iovino, Angela and Maier, Christian and Cucciati, Olga and Tasca, Lidia and Carollo, C. Marcella and Silverman, John and Kampczyk, Pawel and De Ravel, Loic and Sanders, David and Scoville, Nicholas and Contini, Thierry and Mainieri, Vincenzo and Scodeggio, Marco and Kneib, Jean-Paul and Le Fèvre, Olivier and Bardelli, Sandro and Bongiorno, Angela and Caputi, Karina and Coppa, Graziano and De La Torre, Sylvain and Franzetti, Paolo and Garilli, Bianca and Lamareille, Fabrice and Le Borgne, Jean-Francois and Le Brun, Vincent and Mignoli, Marco and Montero, Enrique Perez and Pello, Roser and Ricciardelli, Elena and Tanaka, Masayuki and Tresse, Laurence and Vergani, Daniela and Welikala, Niraj and Zucca, Elena and Oesch, Pascal and Abbas, Ummi and Barnes, Luke and Bordoloi, Rongmon and Bottini, Dario and Cappi, Alberto and Cassata, Paolo and Cimatti, Andrea and Fumana, Marco and Hasinger, Gunther and Koekemoer, Anton and Leauthaud, Alexei and Maccagni, Dario and Marinoni, Christian and McCracken, Henry and Memeo, Pierdomenico and Meneux, Baptiste and Nair, Preethi and Porciani, Cristiano and Presotto, Valentina and Scaramella, Roberto},
	month = sep,
	year = {2010},
	pages = {193--221},
}

@article{helsdon_morphology-density_2003,
	title = {The morphology-density relation in {X}-ray-bright galaxy groups},
	volume = {339},
	issn = {0035-8711},
	url = {https://ui.adsabs.harvard.edu/abs/2003MNRAS.339L..29H},
	doi = {10.1046/j.1365-8711.2003.06300.x},
	urldate = {2026-06-12},
	journal = {Monthly Notices of the Royal Astronomical Society},
	publisher = {OUP},
	author = {Helsdon, Stephen F. and Ponman, Trevor J.},
	month = mar,
	year = {2003},
	note = {ADS Bibcode: 2003MNRAS.339L..29H},
	pages = {L29--L32},
}

@article{abell_catalog_1989,
	title = {A {Catalog} of {Rich} {Clusters} of {Galaxies}},
	volume = {70},
	issn = {0067-0049},
	url = {https://ui.adsabs.harvard.edu/abs/1989ApJS...70....1A},
	doi = {10.1086/191333},
	urldate = {2026-06-12},
	journal = {The Astrophysical Journal Supplement Series},
	publisher = {IOP},
	author = {Abell, George O. and Corwin, Jr., Harold G. and Olowin, Ronald P.},
	month = may,
	year = {1989},
	note = {ADS Bibcode: 1989ApJS...70....1A},
	pages = {1},
}

@article{shappee_man_2014,
	title = {The {Man} behind the {Curtain}: {X}-{Rays} {Drive} the {UV} through {NIR} {Variability} in the 2013 {Active} {Galactic} {Nucleus} {Outburst} in {NGC} 2617},
	volume = {788},
	issn = {0004-637X},
	shorttitle = {The {Man} behind the {Curtain}},
	url = {https://ui.adsabs.harvard.edu/abs/2014ApJ...788...48S},
	doi = {10.1088/0004-637X/788/1/48},
	urldate = {2026-06-12},
	journal = {The Astrophysical Journal},
	publisher = {IOP},
	author = {Shappee, B. J. and Prieto, J. L. and Grupe, D. and Kochanek, C. S. and Stanek, K. Z. and De Rosa, G. and Mathur, S. and Zu, Y. and Peterson, B. M. and Pogge, R. W. and Komossa, S. and Im, M. and Jencson, J. and Holoien, T. W.-S. and Basu, U. and Beacom, J. F. and Szczygieł, D. M. and Brimacombe, J. and Adams, S. and Campillay, A. and Choi, C. and Contreras, C. and Dietrich, M. and Dubberley, M. and Elphick, M. and Foale, S. and Giustini, M. and Gonzalez, C. and Hawkins, E. and Howell, D. A. and Hsiao, E. Y. and Koss, M. and Leighly, K. M. and Morrell, N. and Mudd, D. and Mullins, D. and Nugent, J. M. and Parrent, J. and Phillips, M. M. and Pojmanski, G. and Rosing, W. and Ross, R. and Sand, D. and Terndrup, D. M. and Valenti, S. and Walker, Z. and Yoon, Y.},
	month = jun,
	year = {2014},
	note = {ADS Bibcode: 2014ApJ...788...48S},
	pages = {48},
}

@article{bellm_zwicky_2019,
	title = {The {Zwicky} {Transient} {Facility}: {System} {Overview}, {Performance}, and {First} {Results}},
	volume = {131},
	issn = {0004-6280, 1538-3873},
	shorttitle = {The {Zwicky} {Transient} {Facility}},
	url = {http://arxiv.org/abs/1902.01932},
	doi = {10.1088/1538-3873/aaecbe},
	number = {995},
	urldate = {2026-06-12},
	journal = {PASP},
	author = {Bellm, Eric C. and Kulkarni, Shrinivas R. and Graham, Matthew J. and Dekany, Richard and Smith, Roger M. and Riddle, Reed and Masci, Frank J. and Helou, George and Prince, Thomas A. and Adams, Scott M. and Barbarino, C. and Barlow, Tom and Bauer, James and Beck, Ron and Belicki, Justin and Biswas, Rahul and Blagorodnova, Nadejda and Bodewits, Dennis and Bolin, Bryce and Brinnel, Valery and Brooke, Tim and Bue, Brian and Bulla, Mattia and Burruss, Rick and Cenko, S. Bradley and Chang, Chan-Kao and Connolly, Andrew and Coughlin, Michael and Cromer, John and Cunningham, Virginia and De, Kishalay and Delacroix, Alex and Desai, Vandana and Duev, Dmitry A. and Eadie, Gwendolyn and Farnham, Tony L. and Feeney, Michael and Feindt, Ulrich and Flynn, David and Franckowiak, Anna and Frederick, S. and Fremling, C. and Gal-Yam, Avishay and Gezari, Suvi and Giomi, Matteo and Goldstein, Daniel A. and Golkhou, V. Zach and Goobar, Ariel and Groom, Steven and Hacopians, Eugean and Hale, David and Henning, John and Ho, Anna Y. Q. and Hover, David and Howell, Justin and Hung, Tiara and Huppenkothen, Daniela and Imel, David and Ip, Wing-Huen and Ivezić, Željko and Jackson, Edward and Jones, Lynne and Juric, Mario and Kasliwal, Mansi M. and Kaspi, S. and Kaye, Stephen and Kelley, Michael S. P. and Kowalski, Marek and Kramer, Emily and Kupfer, Thomas and Landry, Walter and Laher, Russ R. and Lee, Chien-De and Lin, Hsing Wen and Lin, Zhong-Yi and Lunnan, Ragnhild and Giomi, Matteo and Mahabal, Ashish and Mao, Peter and Miller, Adam A. and Monkewitz, Serge and Murphy, Patrick and Ngeow, Chow-Choong and Nordin, Jakob and Nugent, Peter and Ofek, Eran and Patterson, Maria T. and Penprase, Bryan and Porter, Michael and Rauch, Ludwig and Rebbapragada, Umaa and Reiley, Dan and Rigault, Mickael and Rodriguez, Hector and Roestel, Jan van and Rusholme, Ben and Santen, Jakob van and Schulze, S. and Shupe, David L. and Singer, Leo P. and Soumagnac, Maayane T. and Stein, Robert and Surace, Jason and Sollerman, Jesper and Szkody, Paula and Taddia, F. and Terek, Scott and Sistine, Angela Van and Velzen, Sjoert van and Vestrand, W. Thomas and Walters, Richard and Ward, Charlotte and Ye, Quan-Zhi and Yu, Po-Chieh and Yan, Lin and Zolkower, Jeffry},
	month = jan,
	year = {2019},
	note = {arXiv:1902.01932 [astro-ph.IM]},
	pages = {018002},
}

@article{stoppa_snidsage_2026,
	title = {{SNID}─{SAGE}: a modern framework for interactive supernova classification and spectral analysis},
	volume = {549},
	issn = {0035-8711},
	shorttitle = {{SNID}─{SAGE}},
	url = {https://ui.adsabs.harvard.edu/abs/2026MNRAS.549g1066S},
	doi = {10.1093/mnras/stag1066},
	urldate = {2026-07-02},
	journal = {Monthly Notices of the Royal Astronomical Society},
	publisher = {OUP},
	author = {Stoppa, Fiorenzo and Smartt, Stephen J.},
	month = jul,
	year = {2026},
	note = {ADS Bibcode: 2026MNRAS.549g1066S},
	pages = {stag1066},
}

@article{kelsey_archival_2024,
	title = {An archival search for type {Ia} supernova siblings},
	volume = {527},
	issn = {0035-8711},
	url = {https://ui.adsabs.harvard.edu/abs/2024MNRAS.527.8015K},
	doi = {10.1093/mnras/stad3587},
	urldate = {2026-07-02},
	journal = {Monthly Notices of the Royal Astronomical Society},
	publisher = {OUP},
	author = {Kelsey, L.},
	month = jan,
	year = {2024},
	note = {ADS Bibcode: 2024MNRAS.527.8015K},
	pages = {8015--8028},
}

@article{foley_foundation_2018,
	title = {The {Foundation} {Supernova} {Survey}: motivation, design, implementation, and first data release},
	volume = {475},
	issn = {0035-8711},
	shorttitle = {The {Foundation} {Supernova} {Survey}},
	url = {https://ui.adsabs.harvard.edu/abs/2018MNRAS.475..193F},
	doi = {10.1093/mnras/stx3136},
	urldate = {2026-07-02},
	journal = {Monthly Notices of the Royal Astronomical Society},
	publisher = {OUP},
	author = {Foley, Ryan J. and Scolnic, Daniel and Rest, Armin and Jha, S. W. and Pan, Y.-C. and Riess, A. G. and Challis, P. and Chambers, K. C. and Coulter, D. A. and Dettman, K. G. and Foley, M. M. and Fox, O. D. and Huber, M. E. and Jones, D. O. and Kilpatrick, C. D. and Kirshner, R. P. and Schultz, A. S. B. and Siebert, M. R. and Flewelling, H. A. and Gibson, B. and Magnier, E. A. and Miller, J. A. and Primak, N. and Smartt, S. J. and Smith, K. W. and Wainscoat, R. J. and Waters, C. and Willman, M.},
	month = mar,
	year = {2018},
	note = {ADS Bibcode: 2018MNRAS.475..193F},
	pages = {193--219},
}

@article{krisciunas_carnegie_2017,
	title = {The {Carnegie} {Supernova} {Project}. {I}. {Third} {Photometry} {Data} {Release} of {Low}-redshift {Type} {Ia} {Supernovae} and {Other} {White} {Dwarf} {Explosions}},
	volume = {154},
	issn = {0004-6256},
	url = {https://ui.adsabs.harvard.edu/abs/2017AJ....154..211K},
	doi = {10.3847/1538-3881/aa8df0},
	urldate = {2026-07-02},
	journal = {The Astronomical Journal},
	publisher = {IOP},
	author = {Krisciunas, Kevin and Contreras, Carlos and Burns, Christopher R. and Phillips, M. M. and Stritzinger, Maximilian D. and Morrell, Nidia and Hamuy, Mario and Anais, Jorge and Boldt, Luis and Busta, Luis and Campillay, Abdo and Castellón, Sergio and Folatelli, Gastón and Freedman, Wendy L. and González, Consuelo and Hsiao, Eric Y. and Krzeminski, Wojtek and Persson, Sven Eric and Roth, Miguel and Salgado, Francisco and Serón, Jacqueline and Suntzeff, Nicholas B. and Torres, Simón and Filippenko, Alexei V. and Li, Weidong and Madore, Barry F. and DePoy, D. L. and Marshall, Jennifer L. and Rheault, Jean-Philippe and Villanueva, Steven},
	month = nov,
	year = {2017},
	note = {ADS Bibcode: 2017AJ....154..211K},
	pages = {211},
}

@article{riess_precise_1996,
	title = {A {Precise} {Distance} {Indicator}: {Type} {IA} {Supernova} {Multicolor} {Light}-{Curve} {Shapes}},
	volume = {473},
	issn = {0004-637X},
	shorttitle = {A {Precise} {Distance} {Indicator}},
	url = {https://ui.adsabs.harvard.edu/abs/1996ApJ...473...88R},
	doi = {10.1086/178129},
	urldate = {2026-07-02},
	journal = {The Astrophysical Journal},
	publisher = {IOP},
	author = {Riess, Adam G. and Press, William H. and Kirshner, Robert P.},
	month = dec,
	year = {1996},
	note = {ADS Bibcode: 1996ApJ...473...88R},
	pages = {88},
}

@article{burns_carnegie_2011,
	title = {The {Carnegie} {Supernova} {Project}: {Light}-curve {Fitting} with {SNooPy}},
	volume = {141},
	issn = {0004-6256},
	shorttitle = {The {Carnegie} {Supernova} {Project}},
	url = {https://ui.adsabs.harvard.edu/abs/2011AJ....141...19B},
	doi = {10.1088/0004-6256/141/1/19},
	urldate = {2026-07-02},
	journal = {The Astronomical Journal},
	publisher = {IOP},
	author = {Burns, Christopher R. and Stritzinger, Maximilian and Phillips, M. M. and Kattner, ShiAnne and Persson, S. E. and Madore, Barry F. and Freedman, Wendy L. and Boldt, Luis and Campillay, Abdo and Contreras, Carlos and Folatelli, Gaston and Gonzalez, Sergio and Krzeminski, Wojtek and Morrell, Nidia and Salgado, Francisco and Suntzeff, Nicholas B.},
	month = jan,
	year = {2011},
	note = {ADS Bibcode: 2011AJ....141...19B},
	pages = {19},
}

@article{mandel_hierarchical_2022,
	title = {A hierarchical {Bayesian} {SED} model for {Type} {Ia} supernovae in the optical to near-infrared},
	volume = {510},
	issn = {0035-8711},
	url = {https://ui.adsabs.harvard.edu/abs/2022MNRAS.510.3939M},
	doi = {10.1093/mnras/stab3496},
	urldate = {2026-07-02},
	journal = {Monthly Notices of the Royal Astronomical Society},
	publisher = {OUP},
	author = {Mandel, Kaisey S. and Thorp, Stephen and Narayan, Gautham and Friedman, Andrew S. and Avelino, Arturo},
	month = mar,
	year = {2022},
	note = {ADS Bibcode: 2022MNRAS.510.3939M},
	pages = {3939--3966},
}

@article{hayden_fundamental_2013,
	title = {{THE} {FUNDAMENTAL} {METALLICITY} {RELATION} {REDUCES} {TYPE} {Ia} {SN} {HUBBLE} {RESIDUALS} {MORE} {THAN} {HOST} {MASS} {ALONE}},
	volume = {764},
	copyright = {http://iopscience.iop.org/info/page/text-and-data-mining},
	issn = {0004-637X, 1538-4357},
	url = {https://iopscience.iop.org/article/10.1088/0004-637X/764/2/191},
	doi = {10.1088/0004-637X/764/2/191},
	number = {2},
	urldate = {2026-07-03},
	journal = {ApJ},
	author = {Hayden, Brian T. and Gupta, Ravi R. and Garnavich, Peter M. and Mannucci, Filippo and Nichol, Robert C. and Sako, Masao},
	month = feb,
	year = {2013},
	pages = {191},
}

@article{jones_should_2018,
	title = {Should {Type} {Ia} {Supernova} {Distances} {Be} {Corrected} for {Their} {Local} {Environments}?},
	volume = {867},
	issn = {0004-637X, 1538-4357},
	url = {https://iopscience.iop.org/article/10.3847/1538-4357/aae2b9},
	doi = {10.3847/1538-4357/aae2b9},
	number = {2},
	urldate = {2026-07-03},
	journal = {ApJ},
	author = {Jones, D. O. and Riess, A. G. and Scolnic, D. M. and Pan, Y.-C. and Johnson, E. and Coulter, D. A. and Dettman, K. G. and Foley, M. M. and Foley, R. J. and Huber, M. E. and Jha, S. W. and Kilpatrick, C. D. and Kirshner, R. P. and Rest, A. and Schultz, A. S. B. and Siebert, M. R.},
	month = nov,
	year = {2018},
	pages = {108},
}

@article{rigault_strong_2020,
	title = {Strong dependence of {Type} {Ia} supernova standardization on the local specific star formation rate},
	volume = {644},
	issn = {0004-6361},
	url = {https://ui.adsabs.harvard.edu/abs/2020A&A...644A.176R},
	doi = {10.1051/0004-6361/201730404},
	urldate = {2026-07-03},
	journal = {Astronomy and Astrophysics},
	publisher = {EDP},
	author = {Rigault, M. and Brinnel, V. and Aldering, G. and Antilogus, P. and Aragon, C. and Bailey, S. and Baltay, C. and Barbary, K. and Bongard, S. and Boone, K. and Buton, C. and Childress, M. and Chotard, N. and Copin, Y. and Dixon, S. and Fagrelius, P. and Feindt, U. and Fouchez, D. and Gangler, E. and Hayden, B. and Hillebrandt, W. and Howell, D. A. and Kim, A. and Kowalski, M. and Kuesters, D. and Leget, P.-F. and Lombardo, S. and Lin, Q. and Nordin, J. and Pain, R. and Pecontal, E. and Pereira, R. and Perlmutter, S. and Rabinowitz, D. and Runge, K. and Rubin, D. and Saunders, C. and Smadja, G. and Sofiatti, C. and Suzuki, N. and Taubenberger, S. and Tao, C. and Thomas, R. C.},
	month = dec,
	year = {2020},
	note = {ADS Bibcode: 2020A\&A...644A.176R},
	pages = {A176},
}

@article{uddin_carnegie_2020,
	title = {The {Carnegie} {Supernova} {Project}-{I}: {Correlation} between {Type} {Ia} {Supernovae} and {Their} {Host} {Galaxies} from {Optical} to {Near}-infrared {Bands}*},
	volume = {901},
	issn = {0004-637X, 1538-4357},
	shorttitle = {The {Carnegie} {Supernova} {Project}-{I}},
	url = {https://iopscience.iop.org/article/10.3847/1538-4357/abafb7},
	doi = {10.3847/1538-4357/abafb7},
	number = {2},
	urldate = {2026-07-03},
	journal = {ApJ},
	author = {Uddin, Syed A and Burns, Christopher R. and Phillips, M. M. and Suntzeff, Nicholas B. and Contreras, Carlos and Hsiao, Eric Y. and Morrell, Nidia and Galbany, Lluís and Stritzinger, Maximilian and Hoeflich, Peter and Ashall, Chris and Piro, Anthony L. and Freedman, Wendy L. and Persson, S. E. and Krisciunas, Kevin and Brown, Peter},
	month = oct,
	year = {2020},
	pages = {143},
}

@article{kelsey_effect_2021,
	title = {The effect of environment on {Type} {Ia} supernovae in the {Dark} {Energy} {Survey} three-year cosmological sample},
	volume = {501},
	issn = {0035-8711},
	url = {https://ui.adsabs.harvard.edu/abs/2021MNRAS.501.4861K},
	doi = {10.1093/mnras/staa3924},
	urldate = {2026-07-03},
	journal = {Monthly Notices of the Royal Astronomical Society},
	publisher = {OUP},
	author = {Kelsey, L. and Sullivan, M. and Smith, M. and Wiseman, P. and Brout, D. and Davis, T. M. and Frohmaier, C. and Galbany, L. and Grayling, M. and Gutiérrez, C. P. and Hinton, S. R. and Kessler, R. and Lidman, C. and Möller, A. and Sako, M. and Scolnic, D. and Uddin, S. A. and Vincenzi, M. and Abbott, T. M. C. and Aguena, M. and Allam, S. and Annis, J. and Avila, S. and Bacon, D. and Bertin, E. and Brooks, D. and Burke, D. L. and Carnero Rosell, A. and Carrasco Kind, M. and Carretero, J. and Castander, F. J. and Costanzi, M. and da Costa, L. N. and Desai, S. and Diehl, H. T. and Doel, P. and Everett, S. and Ferrero, I. and Ferté, A. and Flaugher, B. and Fosalba, P. and García-Bellido, J. and Gerdes, D. W. and Gruen, D. and Gruendl, R. A. and Gschwend, J. and Gutierrez, G. and Hollowood, D. L. and Honscheid, K. and James, D. J. and Kim, A. G. and Kuehn, K. and Kuropatkin, N. and Lahav, O. and Lima, M. and Marshall, J. L. and Martini, P. and Menanteau, F. and Miquel, R. and Morgan, R. and Ogando, R. L. C. and Palmese, A. and Paz-Chinchón, F. and Plazas, A. A. and Romer, A. K. and Sánchez, C. and Sanchez, E. and Serrano, S. and Sevilla-Noarbe, I. and Suchyta, E. and Tarle, G. and Thomas, D. and To, C. and Varga, T. N. and Walker, A. R. and Wilkinson, R. D. and {DES Collaboration}},
	month = mar,
	year = {2021},
	note = {ADS Bibcode: 2021MNRAS.501.4861K},
	pages = {4861--4876},
}

@article{wiseman_galaxy-driven_2022,
	title = {A galaxy-driven model of type {Ia} supernova luminosity variations},
	volume = {515},
	copyright = {https://creativecommons.org/licenses/by/4.0/},
	issn = {0035-8711, 1365-2966},
	url = {https://academic.oup.com/mnras/article/515/3/4587/6649823},
	doi = {10.1093/mnras/stac1984},
	language = {en},
	number = {3},
	urldate = {2026-07-03},
	journal = {Monthly Notices of the Royal Astronomical Society},
	author = {Wiseman, P and Vincenzi, M and Sullivan, M and Kelsey, L and Popovic, B and Rose, B and Brout, D and Davis, T M and Frohmaier, C and Galbany, L and Lidman, C and Möller, A and Scolnic, D and Smith, M and Aguena, M and Allam, S and Andrade-Oliveira, F and Annis, J and Bertin, E and Bocquet, S and Brooks, D and Burke, D L and Carnero Rosell, A and Carrasco Kind, M and Carretero, J and Castander, F J and Costanzi, M and Pereira, M E S and Desai, S and Diehl, H T and Doel, P and Everett, S and Ferrero, I and Friedel, D and Frieman, J and García-Bellido, J and Gatti, M and Gaztanaga, E and Gruen, D and Gschwend, J and Gutierrez, G and Hinton, S R and Hollowood, D L and Honscheid, K and James, D J and March, M and Menanteau, F and Miquel, R and Morgan, R and Palmese, A and Paz-Chinchón, F and Pieres, A and Plazas Malagón, A A and Romer, A K and Sanchez, E and Scarpine, V and Sevilla-Noarbe, I and Soares-Santos, M and Suchyta, E and Tarle, G and To, C and Varga, T N and {DES Collaboration}},
	month = aug,
	year = {2022},
	pages = {4587--4605},
}

@article{meldorf_dark_2023,
	title = {The {Dark} {Energy} {Survey} {Supernova} {Program} results: type {Ia} supernova brightness correlates with host galaxy dust},
	volume = {518},
	issn = {0035-8711},
	shorttitle = {The {Dark} {Energy} {Survey} {Supernova} {Program} results},
	url = {https://ui.adsabs.harvard.edu/abs/2023MNRAS.518.1985M},
	doi = {10.1093/mnras/stac3056},
	urldate = {2026-07-03},
	journal = {Monthly Notices of the Royal Astronomical Society},
	publisher = {OUP},
	author = {Meldorf, C. and Palmese, A. and Brout, D. and Chen, R. and Scolnic, D. and Kelsey, L. and Galbany, L. and Hartley, W. G. and Davis, T. M. and Drlica-Wagner, A. and Vincenzi, M. and Annis, J. and Dixon, M. and Graur, O. and Lidman, C. and Möller, A. and Nugent, P. and Rose, B. and Smith, M. and Allam, S. and Tucker, D. L. and Asorey, J. and Calcino, J. and Carollo, D. and Glazebrook, K. and Lewis, G. F. and Taylor, G. and Tucker, B. E. and Kim, A. G. and Diehl, H. T. and Aguena, M. and Andrade-Oliveira, F. and Bacon, D. and Bertin, E. and Bocquet, S. and Brooks, D. and Burke, D. L. and Carretero, J. and Carrasco Kind, M. and Castander, F. J. and Costanzi, M. and da Costa, L. N. and Desai, S. and Doel, P. and Everett, S. and Ferrero, I. and Friedel, D. and Frieman, J. and García-Bellido, J. and Gatti, M. and Gruen, D. and Gschwend, J. and Gutierrez, G. and Hinton, S. R. and Hollowood, D. L. and Honscheid, K. and James, D. J. and Kuehn, K. and March, M. and Marshall, J. L. and Menanteau, F. and Miquel, R. and Morgan, R. and Paz-Chinchón, F. and Pereira, M. E. S. and Plazas Malagón, A. A. and Sanchez, E. and Scarpine, V. and Sevilla-Noarbe, I. and Suchyta, E. and Tarle, G. and Varga, T. N. and {DES Collaboration}},
	month = jan,
	year = {2023},
	note = {ADS Bibcode: 2023MNRAS.518.1985M},
	pages = {1985--2004},
}

@article{popovic_pantheon_2023,
	title = {The {Pantheon}+ {Analysis}: {Forward} {Modeling} the {Dust} and {Intrinsic} {Color} {Distributions} of {Type} {Ia} {Supernovae}, and {Quantifying} {Their} {Impact} on {Cosmological} {Inferences}},
	volume = {945},
	issn = {0004-637X},
	shorttitle = {The {Pantheon}+ {Analysis}},
	url = {https://ui.adsabs.harvard.edu/abs/2023ApJ...945...84P},
	doi = {10.3847/1538-4357/aca273},
	urldate = {2026-07-03},
	journal = {The Astrophysical Journal},
	publisher = {IOP},
	author = {Popovic, Brodie and Brout, Dillon and Kessler, Richard and Scolnic, Daniel},
	month = mar,
	year = {2023},
	note = {ADS Bibcode: 2023ApJ...945...84P},
	pages = {84},
}

@misc{toy_reduction_2025,
	title = {Reduction of the type {Ia} supernova host galaxy step in the outer regions of galaxies},
	url = {http://arxiv.org/abs/2408.03749},
	doi = {10.48550/arXiv.2408.03749},
	urldate = {2026-07-03},
	publisher = {arXiv},
	author = {Toy, M. and Wiseman, P. and Sullivan, M. and Scolnic, D. and Vincenzi, M. and Brout, D. and Davis, T. M. and Frohmaier, C. and Galbany, L. and Lidman, C. and Lee, J. and Kelsey, L. and Kessler, R. and Möller, A. and Popovic, B. and Sánchez, B. O. and Shah, P. and Smith, M. and Allam, S. and Aguena, M. and Alves, O. and Bacon, D. and Brooks, D. and Burke, D. L. and Rosell, A. Carnero and Carretero, J. and Costa, L. N. da and Pereira, M. E. S. and Desai, S. and Diehl, H. T. and Doel, P. and Drlica-Wagner, A. and Everett, S. and Ferrero, I. and Flaugher, B. and Frieman, J. and García-Bellido, J. and Gatti, M. and Gaztanaga, E. and Giannini, G. and Gruendl, R. A. and Gutierrez, G. and Hinton, S. R. and Hollowood, D. L. and Honscheid, K. and James, D. J. and Lahav, O. and Lee, S. and Marshall, J. L. and Mena-Fernández, J. and Miquel, R. and Palmese, A. and Pieres, A. and Malagón, A. A. Plazas and Romer, A. K. and Samuroff, S. and Sanchez, E. and Cid, D. Sanchez and Schubnell, M. and Suchyta, E. and Swanson, M. E. C. and Tarle, G. and Tucker, D. L. and Vikram, V. and Walker, A. R. and Weaverdyck, N.},
	month = mar,
	year = {2025},
	note = {arXiv:2408.03749 [astro-ph.CO]},
}

@misc{murakami_old_2026,
	title = {Old {Universe}, {Young} {SNe} {Ia}: {A} {Statistical} {Analysis} of {Type} {Ia} {Supernova} {Progenitor} {Age} from 6,983 {TITAN} {Host} {Galaxies}, and {Implications} for {Cosmology}},
	shorttitle = {Old {Universe}, {Young} {SNe} {Ia}},
	url = {http://arxiv.org/abs/2604.16597},
	doi = {10.48550/arXiv.2604.16597},
	urldate = {2026-07-03},
	publisher = {arXiv},
	author = {Murakami, Yukei and Tweddle, Jack and Wiseman, Phil and Jha, Saurabh and Riess, Adam and Smartt, Stephen and Vincenzi, Maria and Pallathadka, Gautham Adamane and Brout, Dillon and Jones, David and Scolnic, Daniel and Marlin, Elijah and Popovic, Brodie and Galbany, Lluís and Schmidt, Brian and Zhang, Keto and Dixon, Mitchell and Larison, Conor and Ferguson, Henry and Filippenko, Alexei},
	month = apr,
	year = {2026},
	note = {arXiv:2604.16597 [astro-ph.CO]},
}

@article{boruah_cosmic_2020,
	title = {Cosmic flows in the nearby {Universe}: new peculiar velocities from {SNe} and cosmological constraints},
	volume = {498},
	issn = {0035-8711},
	shorttitle = {Cosmic flows in the nearby {Universe}},
	url = {https://ui.adsabs.harvard.edu/abs/2020MNRAS.498.2703B},
	doi = {10.1093/mnras/staa2485},
	urldate = {2026-07-03},
	journal = {Monthly Notices of the Royal Astronomical Society},
	publisher = {OUP},
	author = {Boruah, Supranta S. and Hudson, Michael J. and Lavaux, Guilhem},
	month = oct,
	year = {2020},
	note = {ADS Bibcode: 2020MNRAS.498.2703B},
	pages = {2703--2718},
}

@article{stahl_peculiar-velocity_2021,
	title = {Peculiar-velocity cosmology with {Types} {Ia} and {II} supernovae},
	copyright = {arXiv.org perpetual, non-exclusive license},
	url = {https://arxiv.org/abs/2105.05185},
	doi = {10.48550/ARXIV.2105.05185},
	urldate = {2026-07-03},
	publisher = {arXiv},
	author = {Stahl, Benjamin E. and de Jaeger, Thomas and Boruah, Supranta S. and Zheng, WeiKang and Filippenko, Alexei V. and Hudson, Michael J.},
	year = {2021},
	note = {Version Number: 2},
}

@ARTICLE{dixon_peculiar_2026,
       author = {{Dixon}, Mitchell and {Jones}, David O. and {Drakos}, Nicole E. and {Carreres}, Bastien and {Tweddle}, Jack W. and {Murakami}, Yukei S. and {Ashall}, Chris and {Brout}, Dillon and {de Jaeger}, Thomas and {Do}, Aaron and {Marlin}, Elijah G. and {Rubin}, David and {Smartt}, Stephen J. and {Smith}, Ken W.},
        title = "{The Peculiar Growth of Structure: Validating $fσ_8$ Measurements from TITAN Type Ia Supernovae and the Uchuu Simulations}",
      journal = {arXiv e-prints},
         year = 2026,
        month = jun,
          eid = {arXiv:2606.14962},
        pages = {arXiv:2606.14962},
          doi = {10.48550/arXiv.2606.14962},
archivePrefix = {arXiv},
       eprint = {2606.14962},
 primaryClass = {astro-ph.CO},
       adsurl = {https://ui.adsabs.harvard.edu/abs/2026arXiv260614962D}
}

@article{zhang_drafts_2025,
	title = {{DRAFTS}: {A} {Deep}-learning-based {Radio} {Fast} {Transient} {Search} {Pipeline}},
	volume = {276},
	issn = {0067-0049},
	shorttitle = {{DRAFTS}},
	url = {https://ui.adsabs.harvard.edu/abs/2025ApJS..276...20Z},
	doi = {10.3847/1538-4365/ad8f31},
	urldate = {2026-07-03},
	journal = {The Astrophysical Journal Supplement Series},
	publisher = {IOP},
	author = {Zhang, Yong-Kun and Li, Di and Feng, Yi and Tsai, Chao-Wei and Wang, Pei and Niu, Chen-Hui and Chen, Hua-Xi and Zhu, Yu-Hao},
	month = jan,
	year = {2025},
	note = {ADS Bibcode: 2025ApJS..276...20Z},
	pages = {20},
}

@article{koribalski_wallaby_2020,
	title = {{WALLABY} ─ an {SKA} {Pathfinder} {H} {I} survey},
	volume = {365},
	issn = {0004-640X},
	url = {https://ui.adsabs.harvard.edu/abs/2020Ap&SS.365..118K},
	doi = {10.1007/s10509-020-03831-4},
	urldate = {2026-07-03},
	journal = {Astrophysics and Space Science},
	publisher = {Springer},
	author = {Koribalski, Bärbel S. and Staveley-Smith, L. and Westmeier, T. and Serra, P. and Spekkens, K. and Wong, O. I. and Lee-Waddell, K. and Lagos, C. D. P. and Obreschkow, D. and Ryan-Weber, E. V. and Zwaan, M. and Kilborn, V. and Bekiaris, G. and Bekki, K. and Bigiel, F. and Boselli, A. and Bosma, A. and Catinella, B. and Chauhan, G. and Cluver, M. E. and Colless, M. and Courtois, H. M. and Crain, R. A. and de Blok, W. J. G. and Dénes, H. and Duffy, A. R. and Elagali, A. and Fluke, C. J. and For, B.-Q. and Heald, G. and Henning, P. A. and Hess, K. M. and Holwerda, B. W. and Howlett, C. and Jarrett, T. and Jones, D. H. and Jones, M. G. and Józsa, G. I. G. and Jurek, R. and Jütte, E. and Kamphuis, P. and Karachentsev, I. and Kerp, J. and Kleiner, D. and Kraan-Korteweg, R. C. and López-Sánchez, Á. R. and Madrid, J. and Meyer, M. and Mould, J. and Murugeshan, C. and Norris, R. P. and Oh, S.-H. and Oosterloo, T. A. and Popping, A. and Putman, M. and Reynolds, T. N. and Rhee, J. and Robotham, A. S. G. and Ryder, S. and Schröder, A. C. and Shao, Li and Stevens, A. R. H. and Taylor, E. N. and vanÂ der Hulst, J. M. and Verdes-Montenegro, L. and Wakker, B. P. and Wang, J. and Whiting, M. and Winkel, B. and Wolf, C.},
	month = jul,
	year = {2020},
	note = {ADS Bibcode: 2020Ap\&SS.365..118K},
	pages = {118},
}

@misc{desi_collaboration_desi_2016,
	title = {The {DESI} {Experiment} {Part} {I}: {Science},{Targeting}, and {Survey} {Design}},
	shorttitle = {The {DESI} {Experiment} {Part} {I}},
	url = {https://ui.adsabs.harvard.edu/abs/2016arXiv161100036D},
	doi = {10.48550/arXiv.1611.00036},
	urldate = {2026-07-03},
	publisher = {arXiv},
	author = {{DESI Collaboration} and Aghamousa, Amir and Aguilar, Jessica and Ahlen, Steve and Alam, Shadab and Allen, Lori E. and Allende Prieto, Carlos and Annis, James and Bailey, Stephen and Balland, Christophe and Ballester, Otger and Baltay, Charles and Beaufore, Lucas and Bebek, Chris and Beers, Timothy C. and Bell, Eric F. and Bernal, José Luis and Besuner, Robert and Beutler, Florian and Blake, Chris and Bleuler, Hannes and Blomqvist, Michael and Blum, Robert and Bolton, Adam S. and Briceno, Cesar and Brooks, David and Brownstein, Joel R. and Buckley-Geer, Elizabeth and Burden, Angela and Burtin, Etienne and Busca, Nicolas G. and Cahn, Robert N. and Cai, Yan-Chuan and Cardiel-Sas, Laia and Carlberg, Raymond G. and Carton, Pierre-Henri and Casas, Ricard and Castander, Francisco J. and Cervantes-Cota, Jorge L. and Claybaugh, Todd M. and Close, Madeline and Coker, Carl T. and Cole, Shaun and Comparat, Johan and Cooper, Andrew P. and Cousinou, M.-C. and Crocce, Martin and Cuby, Jean-Gabriel and Cunningham, Daniel P. and Davis, Tamara M. and Dawson, Kyle S. and de la Macorra, Axel and De Vicente, Juan and Delubac, Timothée and Derwent, Mark and Dey, Arjun and Dhungana, Govinda and Ding, Zhejie and Doel, Peter and Duan, Yutong T. and Ealet, Anne and Edelstein, Jerry and Eftekharzadeh, Sarah and Eisenstein, Daniel J. and Elliott, Ann and Escoffier, Stéphanie and Evatt, Matthew and Fagrelius, Parker and Fan, Xiaohui and Fanning, Kevin and Farahi, Arya and Farihi, Jay and Favole, Ginevra and Feng, Yu and Fernandez, Enrique and Findlay, Joseph R. and Finkbeiner, Douglas P. and Fitzpatrick, Michael J. and Flaugher, Brenna and Flender, Samuel and Font-Ribera, Andreu and Forero-Romero, Jaime E. and Fosalba, Pablo and Frenk, Carlos S. and Fumagalli, Michele and Gaensicke, Boris T. and Gallo, Giuseppe and Garcia-Bellido, Juan and Gaztanaga, Enrique and Pietro Gentile Fusillo, Nicola and Gerard, Terry and Gershkovich, Irena and Giannantonio, Tommaso and Gillet, Denis and Gonzalez-de-Rivera, Guillermo and Gonzalez-Perez, Violeta and Gott, Shelby and Graur, Or and Gutierrez, Gaston and Guy, Julien and Habib, Salman and Heetderks, Henry and Heetderks, Ian and Heitmann, Katrin and Hellwing, Wojciech A. and Herrera, David A. and Ho, Shirley and Holland, Stephen and Honscheid, Klaus and Huff, Eric and Hutchinson, Timothy A. and Huterer, Dragan and Hwang, Ho Seong and Illa Laguna, Joseph Maria and Ishikawa, Yuzo and Jacobs, Dianna and Jeffrey, Niall and Jelinsky, Patrick and Jennings, Elise and Jiang, Linhua and Jimenez, Jorge and Johnson, Jennifer and Joyce, Richard and Jullo, Eric and Juneau, Stéphanie and Kama, Sami and Karcher, Armin and Karkar, Sonia and Kehoe, Robert and Kennamer, Noble and Kent, Stephen and Kilbinger, Martin and Kim, Alex G. and Kirkby, David and Kisner, Theodore and Kitanidis, Ellie and Kneib, Jean-Paul and Koposov, Sergey and Kovacs, Eve and Koyama, Kazuya and Kremin, Anthony and Kron, Richard and Kronig, Luzius and Kueter-Young, Andrea and Lacey, Cedric G. and Lafever, Robin and Lahav, Ofer and Lambert, Andrew and Lampton, Michael and Landriau, Martin and Lang, Dustin and Lauer, Tod R. and Le Goff, Jean-Marc and Le Guillou, Laurent and Le Van Suu, Auguste and Lee, Jae Hyeon and Lee, Su-Jeong and Leitner, Daniela and Lesser, Michael and Levi, Michael E. and L'Huillier, Benjamin and Li, Baojiu and Liang, Ming and Lin, Huan and Linder, Eric and Loebman, Sarah R. and Lukić, Zarija and Ma, Jun and MacCrann, Niall and Magneville, Christophe and Makarem, Laleh and Manera, Marc and Manser, Christopher J. and Marshall, Robert and Martini, Paul and Massey, Richard and Matheson, Thomas and McCauley, Jeremy and McDonald, Patrick and McGreer, Ian D. and Meisner, Aaron and Metcalfe, Nigel and Miller, Timothy N. and Miquel, Ramon and Moustakas, John and Myers, Adam and Naik, Milind and Newman, Jeffrey A. and Nichol, Robert C. and Nicola, Andrina and Nicolati da Costa, Luiz and Nie, Jundan and Niz, Gustavo and Norberg, Peder and Nord, Brian and Norman, Dara and Nugent, Peter and O'Brien, Thomas and Oh, Minji and Olsen, Knut A. G.},
	month = oct,
	year = {2016},
	note = {ADS Bibcode: 2016arXiv161100036D},
}

@article{boone_twins_2021,
	title = {The {Twins} {Embedding} of {Type} {Ia} {Supernovae}. {I}. {The} {Diversity} of {Spectra} at {Maximum} {Light}},
	volume = {912},
	issn = {0004-637X, 1538-4357},
	url = {https://iopscience.iop.org/article/10.3847/1538-4357/abec3c},
	doi = {10.3847/1538-4357/abec3c},
	number = {1},
	urldate = {2026-07-03},
	journal = {ApJ},
	author = {Boone, K. and Aldering, G. and Antilogus, P. and Aragon, C. and Bailey, S. and Baltay, C. and Bongard, S. and Buton, C. and Copin, Y. and Dixon, S. and Fouchez, D. and Gangler, E. and Gupta, R. and Hayden, B. and Hillebrandt, W. and Kim, A. G. and Kowalski, M. and Küsters, D. and Léget, P.-F. and Mondon, F. and Nordin, J. and Pain, R. and Pecontal, E. and Pereira, R. and Perlmutter, S. and Ponder, K. A. and Rabinowitz, D. and Rigault, M. and Rubin, D. and Runge, K. and Saunders, C. and Smadja, G. and Suzuki, N. and Tao, C. and Taubenberger, S. and Thomas, R. C. and Vincenzi, M.},
	month = may,
	year = {2021},
	pages = {70},
}

@article{fakhouri_improving_2015,
	title = {{IMPROVING} {COSMOLOGICAL} {DISTANCE} {MEASUREMENTS} {USING} {TWIN} {TYPE} {IA} {SUPERNOVAE}},
	volume = {815},
	copyright = {http://iopscience.iop.org/info/page/text-and-data-mining},
	issn = {1538-4357},
	url = {https://iopscience.iop.org/article/10.1088/0004-637X/815/1/58},
	doi = {10.1088/0004-637X/815/1/58},
	number = {1},
	urldate = {2026-07-03},
	journal = {ApJ},
	author = {Fakhouri, H. K. and Boone, K. and Aldering, G. and Antilogus, P. and Aragon, C. and Bailey, S. and Baltay, C. and Barbary, K. and Baugh, D. and Bongard, S. and Buton, C. and Chen, J. and Childress, M. and Chotard, N. and Copin, Y. and Fagrelius, P. and Feindt, U. and Fleury, M. and Fouchez, D. and Gangler, E. and Hayden, B. and Kim, A. G. and Kowalski, M. and Leget, P.-F. and Lombardo, S. and Nordin, J. and Pain, R. and Pecontal, E. and Pereira, R. and Perlmutter, S. and Rabinowitz, D. and Ren, J. and Rigault, M. and Rubin, D. and Runge, K. and Saunders, C. and Scalzo, R. and Smadja, G. and Sofiatti, C. and Strovink, M. and Suzuki, N. and Tao, C. and Thomas, R. C. and Weaver, B. A. and {The Nearby Supernova Factory}},
	month = dec,
	year = {2015},
	pages = {58},
}

@article{ward_relative_2023,
	title = {Relative {Intrinsic} {Scatter} in {Hierarchical} {Type} {Ia} {Supernova} {Sibling} {Analyses}: {Application} to {SNe} 2021hpr, 1997bq, and 2008fv in {NGC} 3147},
	volume = {956},
	issn = {0004-637X, 1538-4357},
	shorttitle = {Relative {Intrinsic} {Scatter} in {Hierarchical} {Type} {Ia} {Supernova} {Sibling} {Analyses}},
	url = {https://iopscience.iop.org/article/10.3847/1538-4357/acf7bb},
	doi = {10.3847/1538-4357/acf7bb},
	number = {2},
	urldate = {2026-07-03},
	journal = {ApJ},
	author = {Ward, Sam M. and Thorp, Stephen and Mandel, Kaisey S. and Dhawan, Suhail and Jones, David O. and Taggart, Kirsty and Foley, Ryan J. and Narayan, Gautham and Chambers, Kenneth C. and Coulter, David A. and Davis, Kyle W. and De Boer, Thomas and De Soto, Kaylee and Earl, Nicholas and Gagliano, Alex and Gao, Hua and Hjorth, Jens and Huber, Mark E. and Izzo, Luca and Langeroodi, Danial and Magnier, Eugene A. and McGill, Peter and Rest, Armin and Rojas-Bravo, César and Wojtak, Radosław and {Young Supernova Experiment}},
	month = oct,
	year = {2023},
	pages = {111},
}

@article{nugent_characterizing_2026,
	title = {Characterizing {Supernova} {Host} {Galaxies} with {FrankenBlast} : {A} {Scalable} {Tool} for {Transient} {Host} {Galaxy} {Association}, {Photometry}, and {Stellar} {Population} {Modeling}},
	volume = {997},
	issn = {0004-637X, 1538-4357},
	shorttitle = {Characterizing {Supernova} {Host} {Galaxies} with {FrankenBlast}},
	url = {https://iopscience.iop.org/article/10.3847/1538-4357/ae247b},
	doi = {10.3847/1538-4357/ae247b},
	number = {1},
	urldate = {2026-07-03},
	journal = {ApJ},
	author = {Nugent, Anya E. and Villar, V. Ashley and Gagliano, Alex and Jones, David O. and Horowicz, Asaf and De Soto, Kaylee and Wang, Bingjie and Margalit, Ben},
	month = jan,
	year = {2026},
	pages = {38},
}

@article{peterson_dehvils_2024,
	title = {The {DEHVILS} in the details: {Type} {Ia} supernova {Hubble} residual comparisons and mass step analysis in the near-infrared},
	volume = {690},
	issn = {0004-6361},
	shorttitle = {The {DEHVILS} in the details},
	url = {https://ui.adsabs.harvard.edu/abs/2024A&A...690A..56P},
	doi = {10.1051/0004-6361/202450052},
	urldate = {2026-07-27},
	journal = {Astronomy and Astrophysics},
	publisher = {EDP},
	author = {Peterson, E. R. and Scolnic, D. and Jones, D. O. and Do, A. and Popovic, B. and Riess, A. G. and Dwomoh, A. and Johansson, J. and Rubin, D. and Sánchez, B. O. and Shappee, B. J. and Tonry, J. L. and Tully, R. B. and Vincenzi, M.},
	month = oct,
	year = {2024},
	note = {ADS Bibcode: 2024A\&A...690A..56P},
	pages = {A56},
}

@article{lavaux_2m_2011,
	title = {The {2M}++ galaxy redshift catalogue},
	volume = {416},
	issn = {0035-8711},
	url = {https://ui.adsabs.harvard.edu/abs/2011MNRAS.416.2840L},
	doi = {10.1111/j.1365-2966.2011.19233.x},
	urldate = {2026-07-28},
	journal = {Monthly Notices of the Royal Astronomical Society},
	publisher = {OUP},
	author = {Lavaux, Guilhem and Hudson, Michael J.},
	month = oct,
	year = {2011},
	note = {ADS Bibcode: 2011MNRAS.416.2840L},
	pages = {2840--2856},
}

@book{Hoaglin00,
  asin = {0471384917},
  author = {Hoaglin, David C. and Mosteller, Frederick and (Editor), John W. Tukey},
  description = {Amazon.com: Understanding Robust and Exploratory Data Analysis (9780471384915): David C. Hoaglin, Frederick Mosteller, John W. Tukey: Books},
  dewey = {001.422},
  ean = {9780471384915},
  edition = 1,
  isbn = {0471384917},
  publisher = {Wiley-Interscience},
  title = {Understanding Robust and Exploratory Data Analysis},
  year = 2000
}

@ARTICLE{Sullivan2006,
       author = {{Sullivan}, M. and {Le Borgne}, D. and {Pritchet}, C.~J. and {Hodsman}, A. and {Neill}, J.~D. and {Howell}, D.~A. and {Carlberg}, R.~G. and {Astier}, P. and {Aubourg}, E. and {Balam}, D. and {Basa}, S. and {Conley}, A. and {Fabbro}, S. and {Fouchez}, D. and {Guy}, J. and {Hook}, I. and {Pain}, R. and {Palanque-Delabrouille}, N. and {Perrett}, K. and {Regnault}, N. and {Rich}, J. and {Taillet}, R. and {Baumont}, S. and {Bronder}, J. and {Ellis}, R.~S. and {Filiol}, M. and {Lusset}, V. and {Perlmutter}, S. and {Ripoche}, P. and {Tao}, C.},
        title = "{Rates and Properties of Type Ia Supernovae as a Function of Mass and Star Formation in Their Host Galaxies}",
      journal = {\apj},
         year = 2006,
        month = sep,
       volume = {648},
       number = {2},
        pages = {868-883},
          doi = {10.1086/506137},
archivePrefix = {arXiv},
       eprint = {astro-ph/0605455},
 primaryClass = {astro-ph},
       adsurl = {https://ui.adsabs.harvard.edu/abs/2006ApJ...648..868S}
}

@ARTICLE{Gupta2016,
       author = {{Gupta}, Ravi R. and {Kuhlmann}, Steve and {Kovacs}, Eve and {Spinka}, Harold and {Kessler}, Richard and {Goldstein}, Daniel A. and {Liotine}, Camille and {Pomian}, Katarzyna and {D'Andrea}, Chris B. and {Sullivan}, Mark and {Carretero}, Jorge and {Castander}, Francisco J. and {Nichol}, Robert C. and {Finley}, David A. and {Fischer}, John A. and {Foley}, Ryan J. and {Kim}, Alex G. and {Papadopoulos}, Andreas and {Sako}, Masao and {Scolnic}, Daniel M. and {Smith}, Mathew and {Tucker}, Brad E. and {Uddin}, Syed and {Wolf}, Rachel C. and {Yuan}, Fang and {Abbott}, Tim M.~C. and {Abdalla}, Filipe B. and {Benoit-L{\'e}vy}, Aur{\'e}lien and {Bertin}, Emmanuel and {Brooks}, David and {Carnero Rosell}, Aurelio and {Carrasco Kind}, Matias and {Cunha}, Carlos E. and {da Costa}, Luiz N. and {Desai}, Shantanu and {Doel}, Peter and {Eifler}, Tim F. and {Evrard}, August E. and {Flaugher}, Brenna and {Fosalba}, Pablo and {Gazta{\~n}aga}, Enrique and {Gruen}, Daniel and {Gruendl}, Robert and {James}, David J. and {Kuehn}, Kyler and {Kuropatkin}, Nikolay and {Maia}, Marcio A.~G. and {Marshall}, Jennifer L. and {Miquel}, Ramon and {Plazas}, Andr{\'e}s A. and {Romer}, A. Kathy and {S{\'a}nchez}, Eusebio and {Schubnell}, Michael and {Sevilla-Noarbe}, Ignacio and {Sobreira}, Fl{\'a}via and {Suchyta}, Eric and {Swanson}, Molly E.~C. and {Tarle}, Gregory and {Walker}, Alistair R. and {Wester}, William},
        title = "{Host Galaxy Identification for Supernova Surveys}",
      journal = {\aj},
         year = 2016,
        month = dec,
       volume = {152},
       number = {6},
          eid = {154},
        pages = {154},
          doi = {10.3847/0004-6256/152/6/154},
archivePrefix = {arXiv},
       eprint = {1604.06138},
 primaryClass = {astro-ph.CO},
       adsurl = {https://ui.adsabs.harvard.edu/abs/2016AJ....152..154G}
}

@ARTICLE{Bagpipes,
       author = {{Carnall}, A.~C. and {McLure}, R.~J. and {Dunlop}, J.~S. and {Dav{\'e}}, R.},
        title = "{Inferring the star formation histories of massive quiescent galaxies with BAGPIPES: evidence for multiple quenching mechanisms}",
      journal = {\mnras},
         year = 2018,
        month = nov,
       volume = {480},
       number = {4},
        pages = {4379-4401},
          doi = {10.1093/mnras/sty2169},
archivePrefix = {arXiv},
       eprint = {1712.04452},
 primaryClass = {astro-ph.GA},
       adsurl = {https://ui.adsabs.harvard.edu/abs/2018MNRAS.480.4379C}
}

@ARTICLE{HostPhot,
       author = {{M{\"u}ller-Bravo}, Tom{\'a}s and {Galbany}, Llu{\'\i}s},
        title = "{HostPhot: global and local photometry of galaxies hosting supernovae or other transients}",
      journal = {The Journal of Open Source Software},
         year = 2022,
        month = aug,
       volume = {7},
       number = {76},
          eid = {4508},
        pages = {4508},
          doi = {10.21105/joss.04508},
archivePrefix = {arXiv},
       eprint = {2208.08117},
 primaryClass = {astro-ph.CO},
       adsurl = {https://ui.adsabs.harvard.edu/abs/2022JOSS....7.4508M}
}
\bibliographystyle{aasjournalv7.1}



\end{document}